\documentclass[acmtog,nonacm]{acmart}

\usepackage[htt]{hyphenat}
\usepackage{multirow}
\usepackage{tikz}
\usepackage{xcolor}
\usepackage{tabularx}
\usepackage{booktabs}

\newcommand{\makecropbox}[6]
{{\begin{tikzpicture}
			\node[anchor=south west,inner sep=0] (image) at (0,0) {\includegraphics[height=#2]{#1}};
			\begin{scope}[x={(image.south east)},y={(image.north west)}]
				\draw[cyan,very thick] (#3, #4) rectangle (#5, #6);
			\end{scope}
\end{tikzpicture}}}

\AtBeginDocument{%
  }

\setcopyright{rightsretained} 
\acmJournal{TOG}
\acmYear{2025}
\acmDOI{XXXXXXX.XXXXXXX}

\begin{document}

\title{VideoMat: Extracting PBR Materials from Video Diffusion Models}

\author{Jacob Munkberg}
\orcid{0009-0004-0451-7442}
\affiliation{%
  \institution{NVIDIA}
  \country{Sweden}
}

\author{Zian Wang}
\orcid{0000-0003-4166-3807}
\affiliation{%
  \institution{NVIDIA},
  \institution{University of Toronto},
  \institution{Vector Institute}
  \country{Canada}
}

\author{Ruofan Liang}
\orcid{0009-0005-7667-1809}
\affiliation{%
  \institution{NVIDIA},
  \institution{University of Toronto},
  \institution{Vector Institute}
  \country{Canada}
}

\author{Tianchang Shen}
\orcid{0000-0002-7133-2761}
\affiliation{%
  \institution{NVIDIA},
  \institution{University of Toronto},
  \institution{Vector Institute}
  \country{Canada}
}

\author{Jon Hasselgren}
\orcid{0009-0002-3423-190X}
\affiliation{%
  \institution{NVIDIA}
  \country{Sweden}
}

\renewcommand{\shortauthors}{Munkberg et al.}

\begin{abstract}
We leverage finetuned video diffusion models, intrinsic decomposition of videos, 
and physically-based differentiable rendering to generate high quality materials for 
3D models given a text prompt or a single image. We condition a video diffusion model to 
respect the input geometry and lighting condition. This model produces multiple views of a 
given 3D model with coherent material properties. Secondly, we use a recent model to extract 
intrinsics (base color, roughness, metallic) from the generated video. Finally, we use the 
intrinsics alongside the generated video in a differentiable path tracer to robustly extract 
PBR materials directly compatible with common content creation tools.
\end{abstract}

\begin{CCSXML}
<ccs2012>
<concept>
<concept_id>10010147.10010178.10010224.10010240.10010243</concept_id>
<concept_desc>Computing methodologies~Appearance and texture representations</concept_desc>
<concept_significance>300</concept_significance>
</concept>
<concept>
<concept_id>10010147.10010371.10010372.10010374</concept_id>
<concept_desc>Computing methodologies~Ray tracing</concept_desc>
<concept_significance>100</concept_significance>
</concept>
<concept>
<concept_id>10010147.10010371.10010372.10010376</concept_id>
<concept_desc>Computing methodologies~Reflectance modeling</concept_desc>
<concept_significance>100</concept_significance>
</concept>
</ccs2012>
\end{CCSXML}

\ccsdesc[300]{Computing methodologies~Appearance and texture representations}
\ccsdesc[100]{Computing methodologies~Ray tracing}
\ccsdesc[100]{Computing methodologies~Reflectance modeling}

\keywords{Text-to-material, Material generation, Differentiable Rendering, Video diffusion models}

%\received{1 June 2025}
%\received[revised]{1 June 2025}
%\received[accepted]{1 June 2025}

\begin{teaserfigure}
   \centering
   \includegraphics[width=0.99\textwidth]{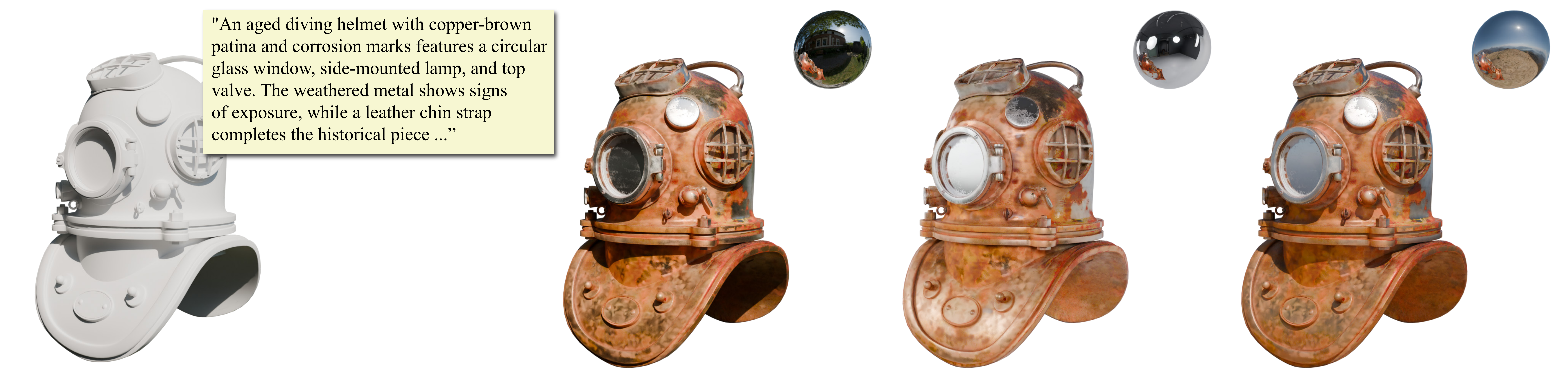}
   \vspace*{-2mm}
   \caption{Given a 3D model and a prompt, we extract high quality PBR materials
   from finetuned video diffusion models.}
   \label{fig:teaser}
\end{teaserfigure}

%%%%%%%%%%%%%%%%%%%%%%%%%%%%%%%%%%
% System overview
%%%%%%%%%%%%%%%%%%%%%%%%%%%%%%%%%%

\newcommand{\figSystem}{
\begin{figure*}
    \centering
        \includegraphics[width=0.99\textwidth]{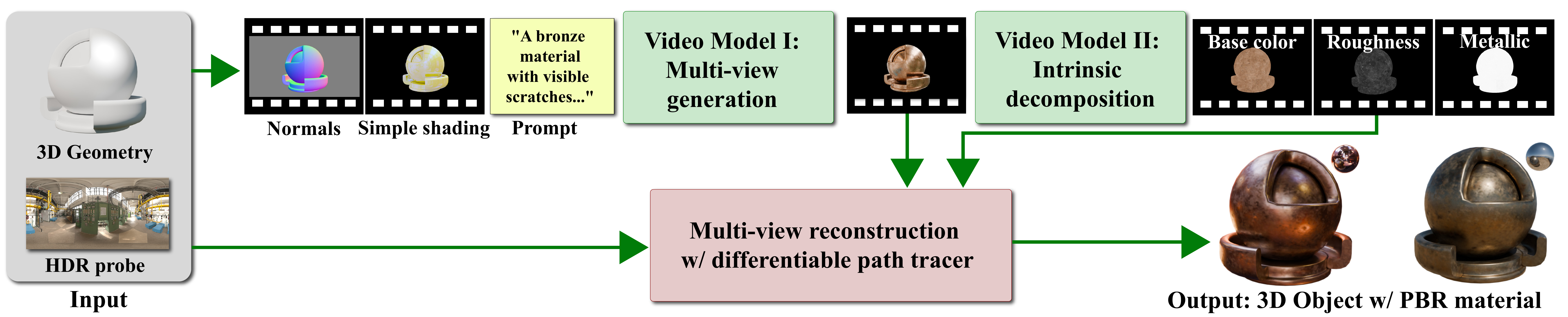}
    \vspace*{-2mm}
    \caption{Our method starts from a known 3D model and an HDR environment map. We first render videos of normal maps and 
    three simple uniform shading conditions (diffuse, semi-specular, fully specular) lit with the provided probe. 
    Next, these conditions, together with a text prompt, are passed to our finetuned video model, which generates a coherent 
    video of the object with a novel material, while respecting the given lighting condition. We then pass this video into 
    a second video model, which performs intrinsic decomposition, and generate per-frame G-buffers for the material properties.
    Finally, the output from the two video models, alongside the given geometry and lighting, are passed to a differentiable path tracer, 
    which performs multi-view reconstruction to extract high quality PBR materials from the generated views.}
    \label{fig:system}
\end{figure*}
}

%%%%%%%%%%%%%%%%%%%%%%%%%%%%%%%%%%
% Impact of multiple bounces
%%%%%%%%%%%%%%%%%%%%%%%%%%%%%%%%%%

\newcommand{\figRayDepth}{
\begin{figure}
    \centering
    \setlength{\tabcolsep}{1pt}
    \begin{tabular}{ccc}
       \includegraphics[width=0.3\columnwidth]{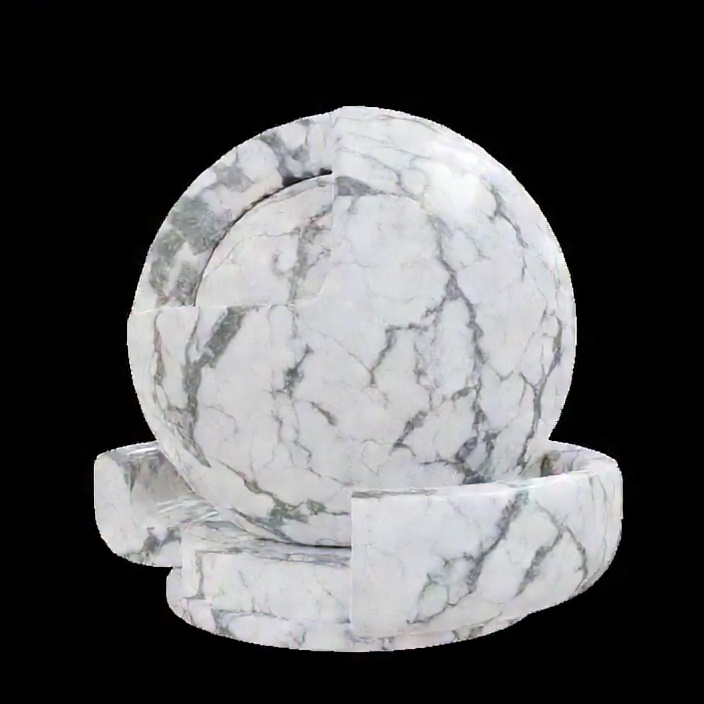} &
       \includegraphics[width=0.3\columnwidth]{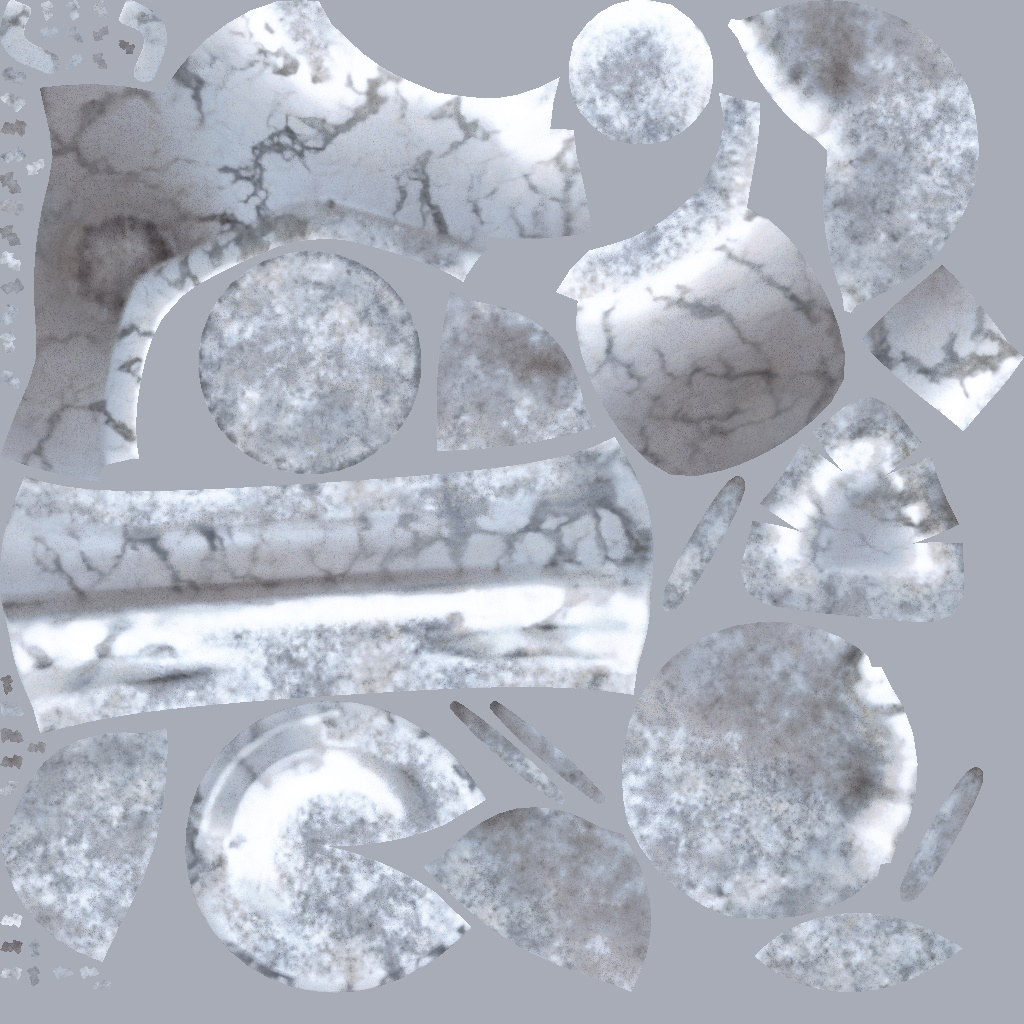} &
       \includegraphics[width=0.3\columnwidth]{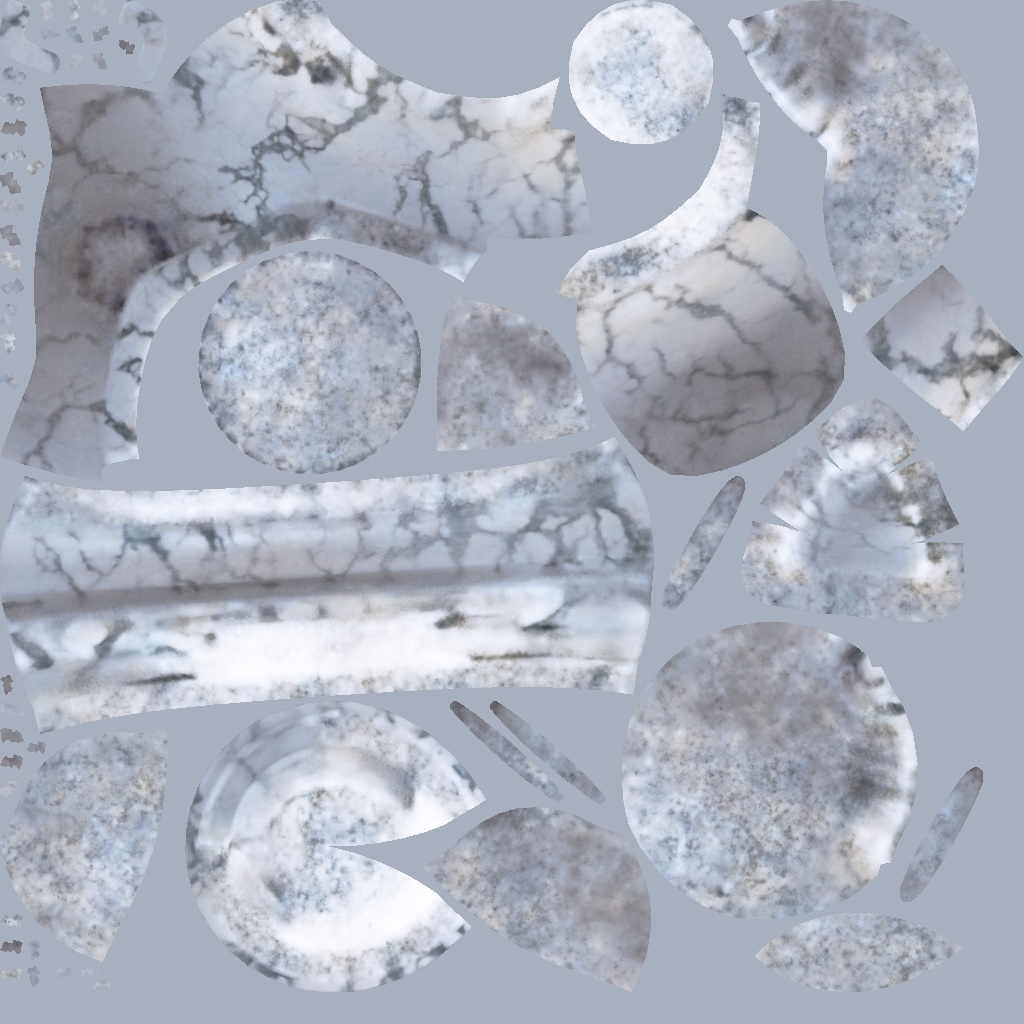} \\   
       \small{Frame from video model} & \small{Base color $d=1$} & \small{Base color $d=3$} 
    \end{tabular}
    \caption{Impact of physically-based rendering. In this example, a significant portion of the lighting comes from indirect bounces. We visualize the reconstructions from differentiable path tracing with one ($d=1$) vs. three bounces ($d=3$). 
    By more accurately simulating global illumination, we obtain a more uniform base 
    color with less baked lighting.}
    \label{fig:raydepth}
 \end{figure}
}

%%%%%%%%%%%%%%%%%%%%%%%%%%%%%%%%%%
% Main results
%%%%%%%%%%%%%%%%%%%%%%%%%%%%%%%%%%

\newcommand{\figMainQualityResults}{
\begin{figure*}
\centering
\small
\setlength{\tabcolsep}{1pt}
\begin{tabular}{cccccccccccc}
	\multirow{2}{*}{\rotatebox{90}{\makebox[0.16\textwidth]{\centering \textsc{Diver}}}} &
	\multirow{2}{*}{\includegraphics[width=0.16\textwidth]{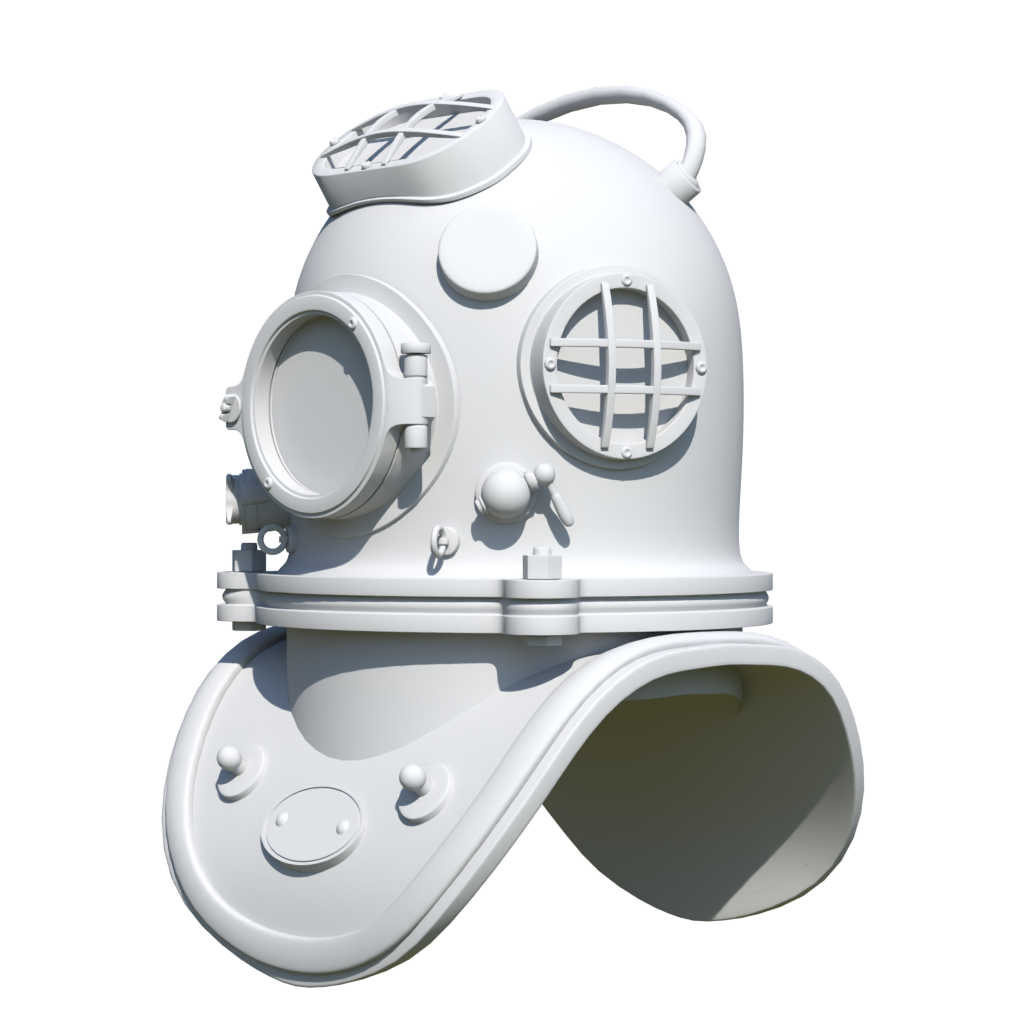}} & 
    \rotatebox[origin=c]{90}{Our} &
	\raisebox{-0.5\height}{\includegraphics[width=0.093\textwidth]{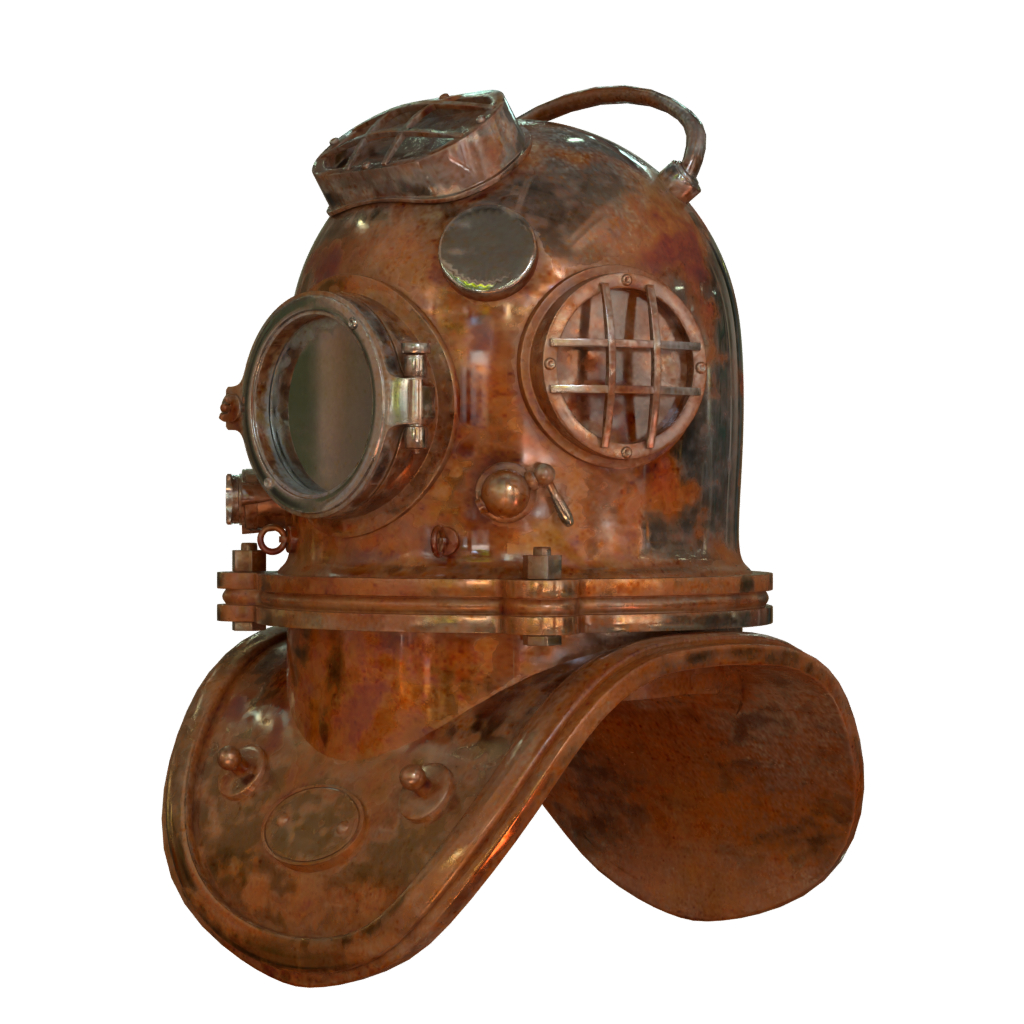}} &
	\raisebox{-0.5\height}{\includegraphics[width=0.093\textwidth]{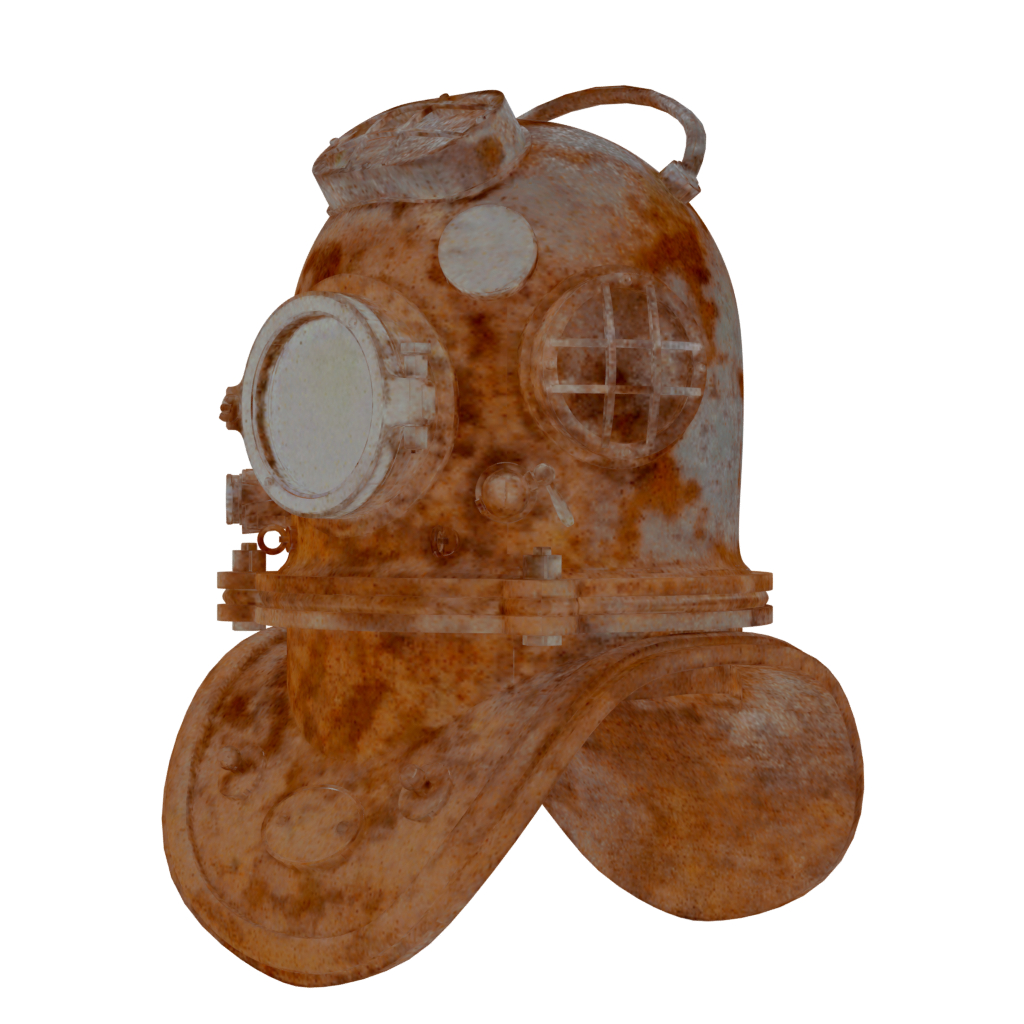}} &
	\raisebox{-0.5\height}{\includegraphics[width=0.093\textwidth]{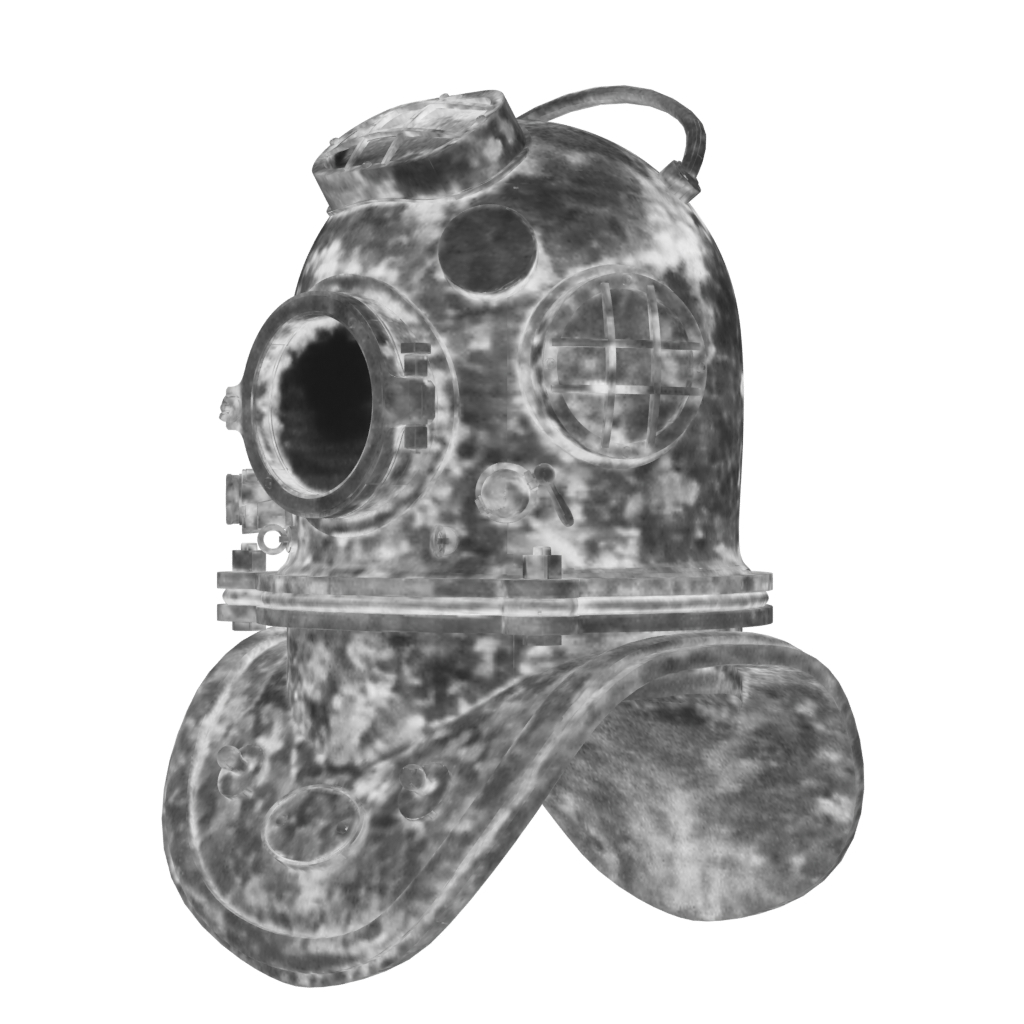}} &
	\raisebox{-0.5\height}{\includegraphics[width=0.093\textwidth]{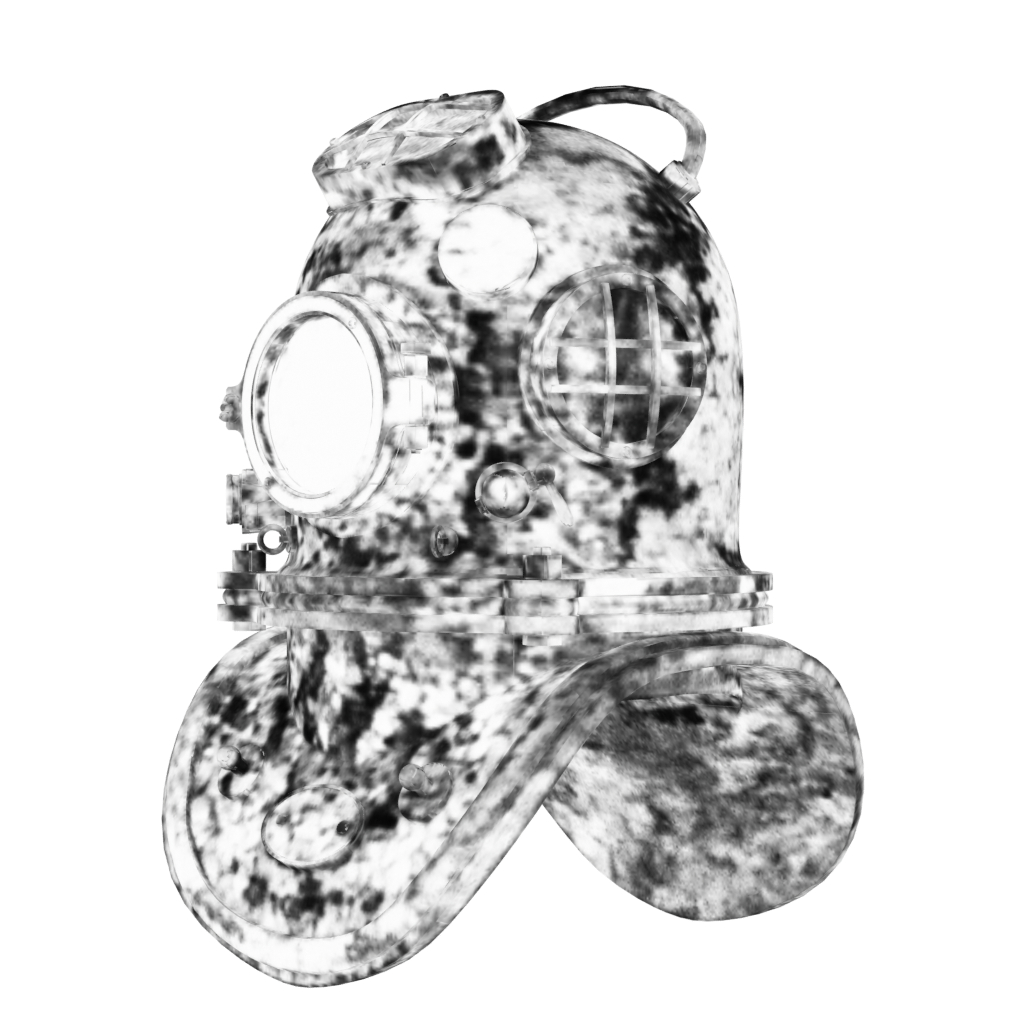}} &
    \rotatebox[origin=c]{90}{Paint-it} &
	\raisebox{-0.5\height}{\includegraphics[width=0.093\textwidth]{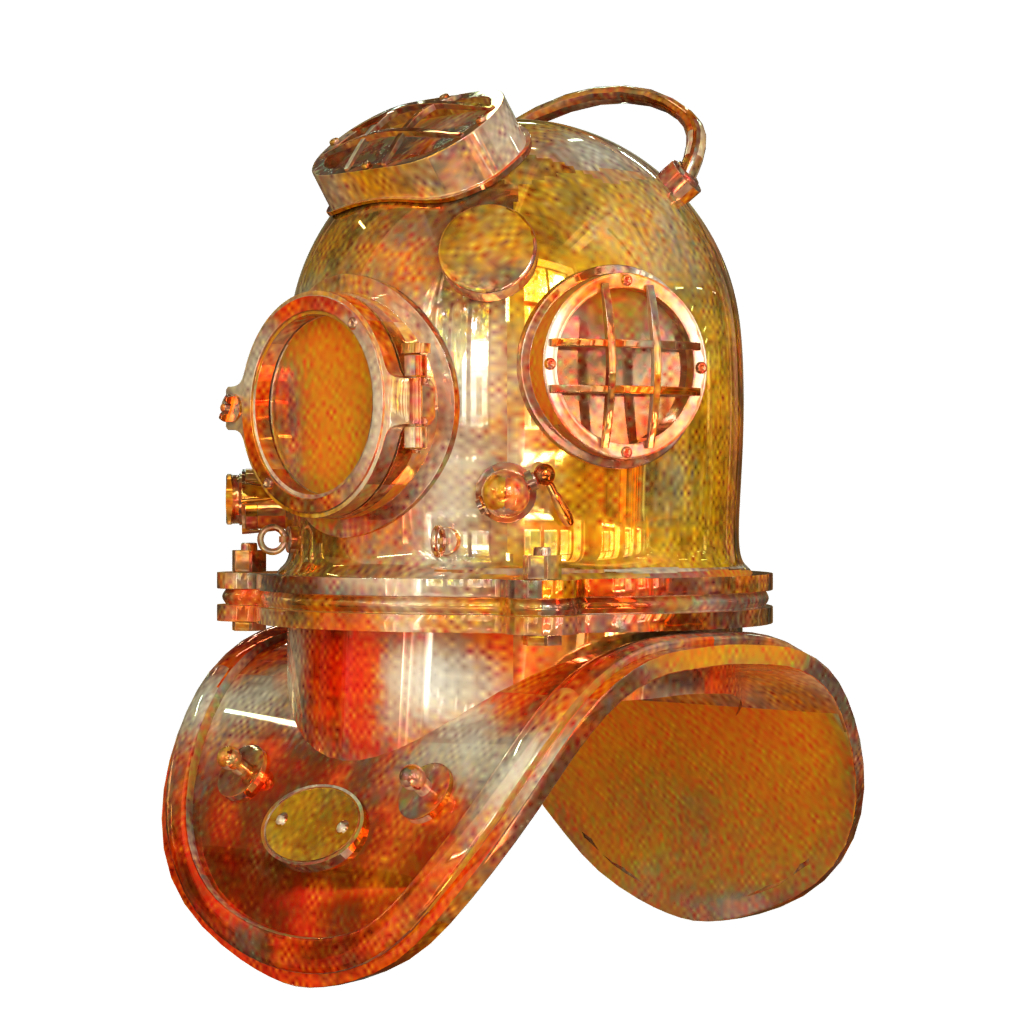}} &
	\raisebox{-0.5\height}{\includegraphics[width=0.093\textwidth]{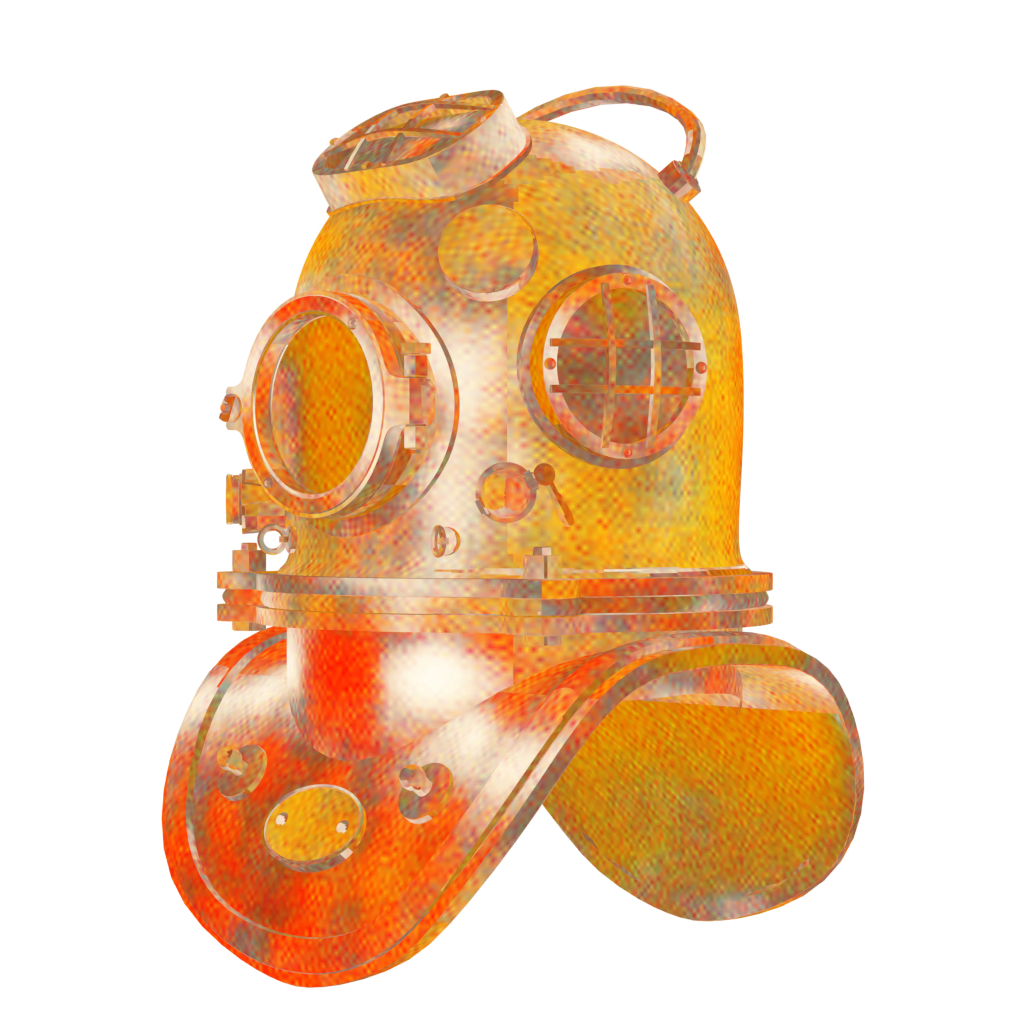}} &
	\raisebox{-0.5\height}{\includegraphics[width=0.093\textwidth]{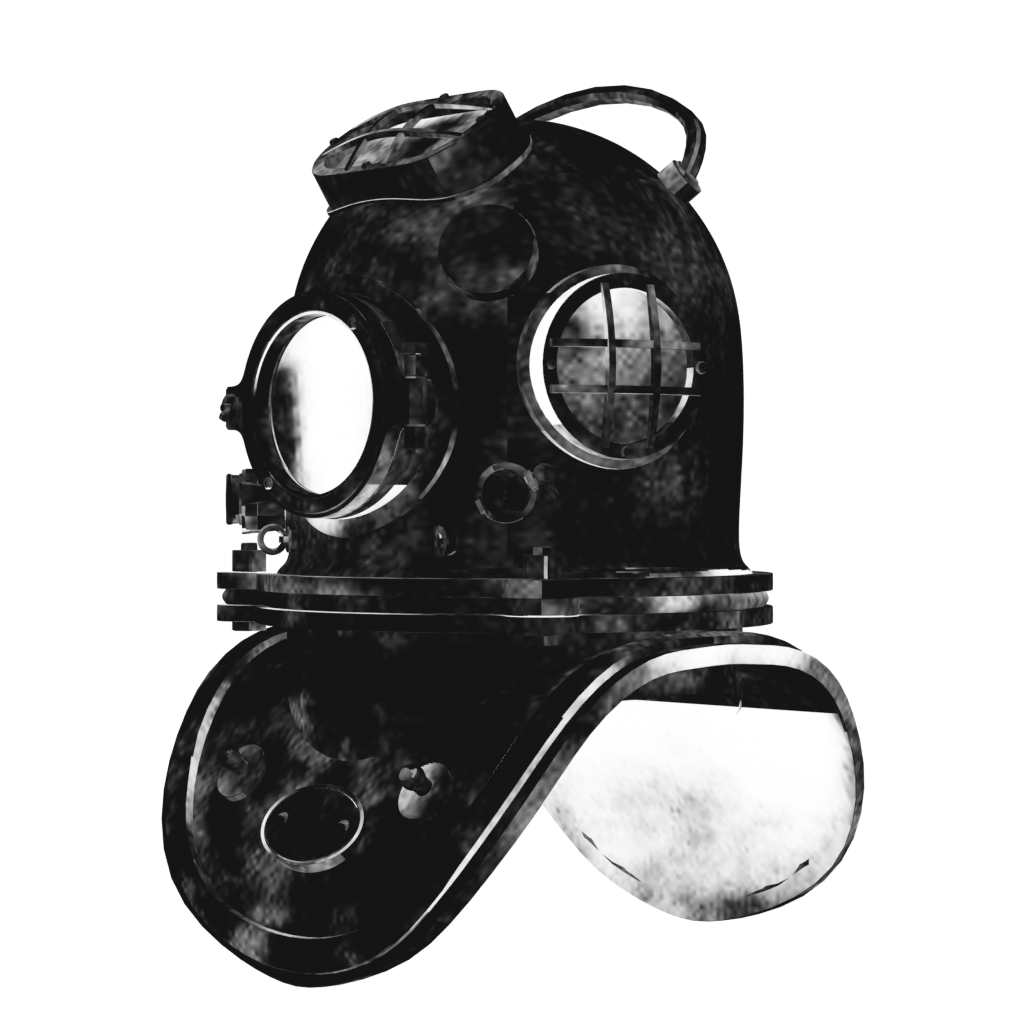}} &
	\raisebox{-0.5\height}{\includegraphics[width=0.093\textwidth]{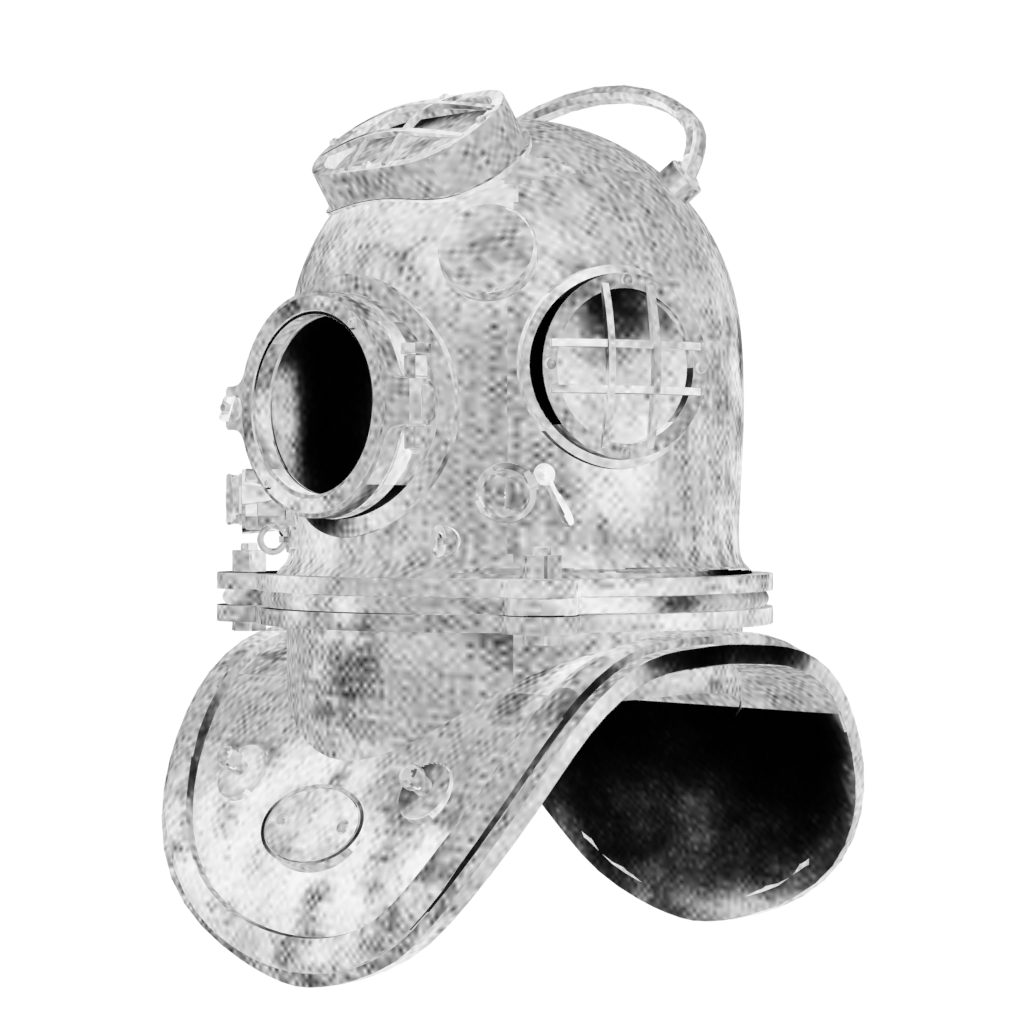}} \\
	& & 
    \rotatebox[origin=c]{90}{Reference} &
	\raisebox{-0.5\height}{\includegraphics[width=0.093\textwidth]{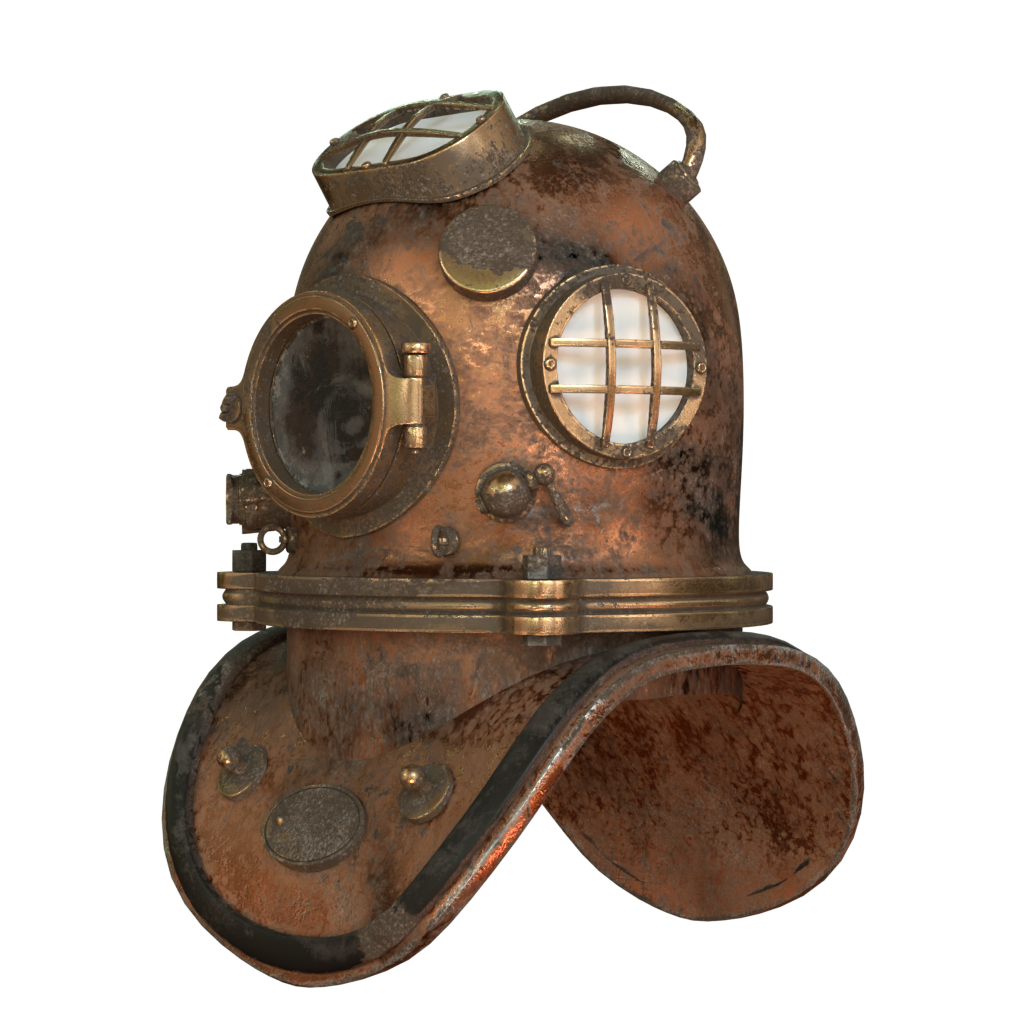}} &
	\raisebox{-0.5\height}{\includegraphics[width=0.093\textwidth]{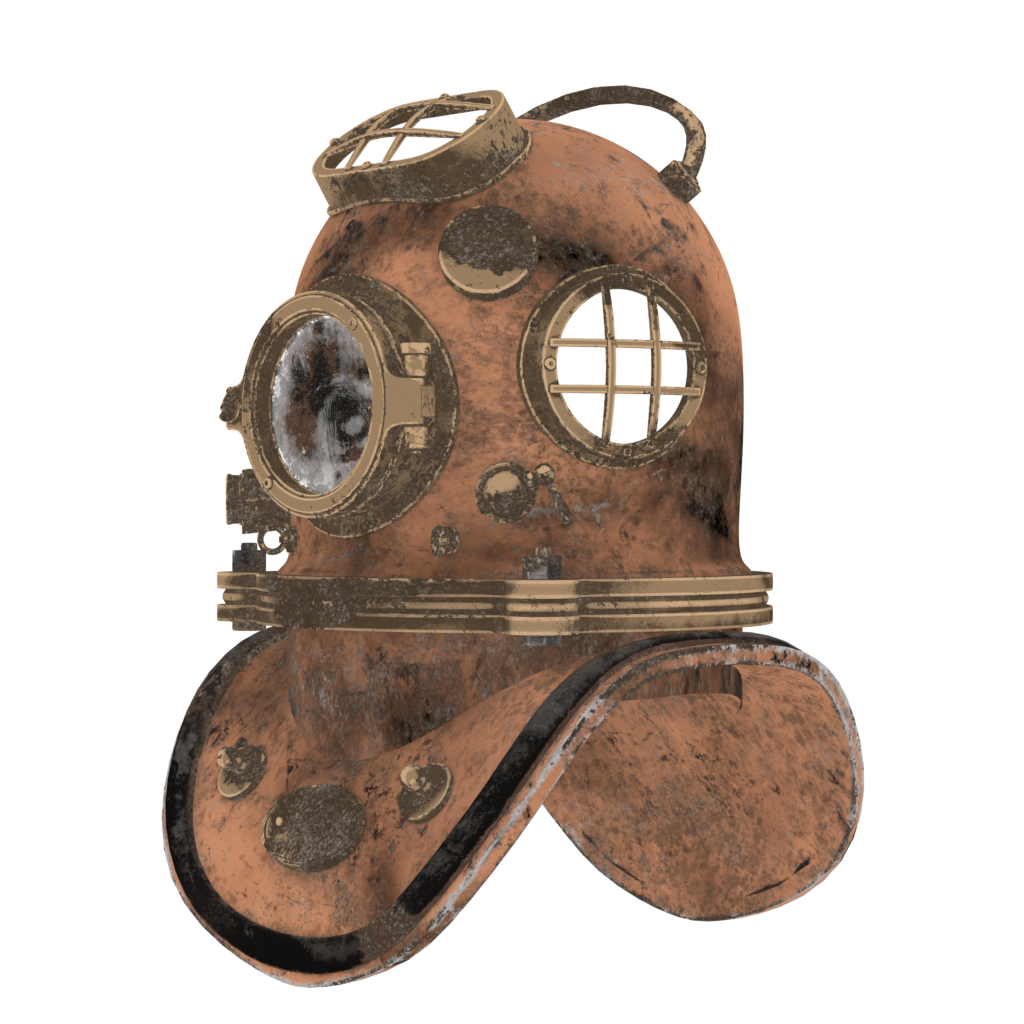}} &
	\raisebox{-0.5\height}{\includegraphics[width=0.093\textwidth]{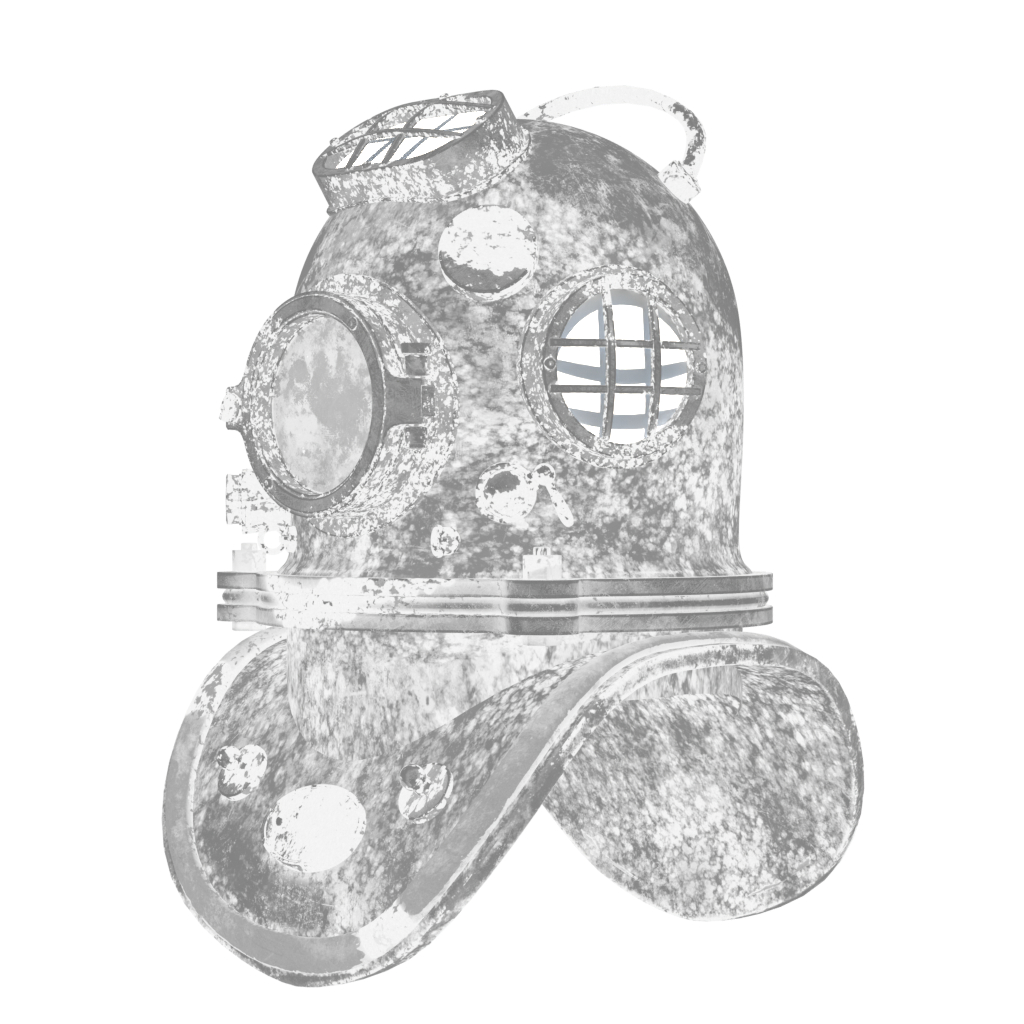}} &
	\raisebox{-0.5\height}{\includegraphics[width=0.093\textwidth]{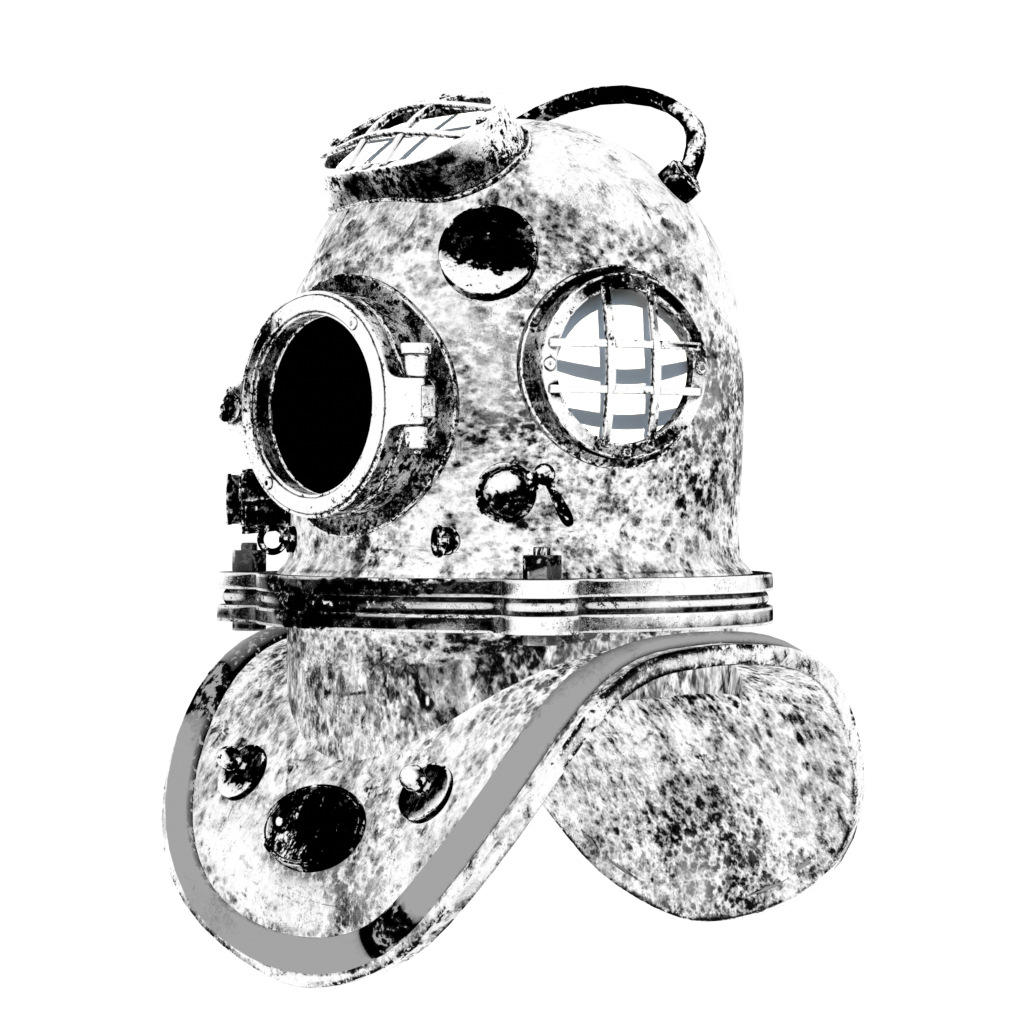}} &
    \rotatebox[origin=c]{90}{DreamMat} &
	\raisebox{-0.5\height}{\includegraphics[width=0.093\textwidth]{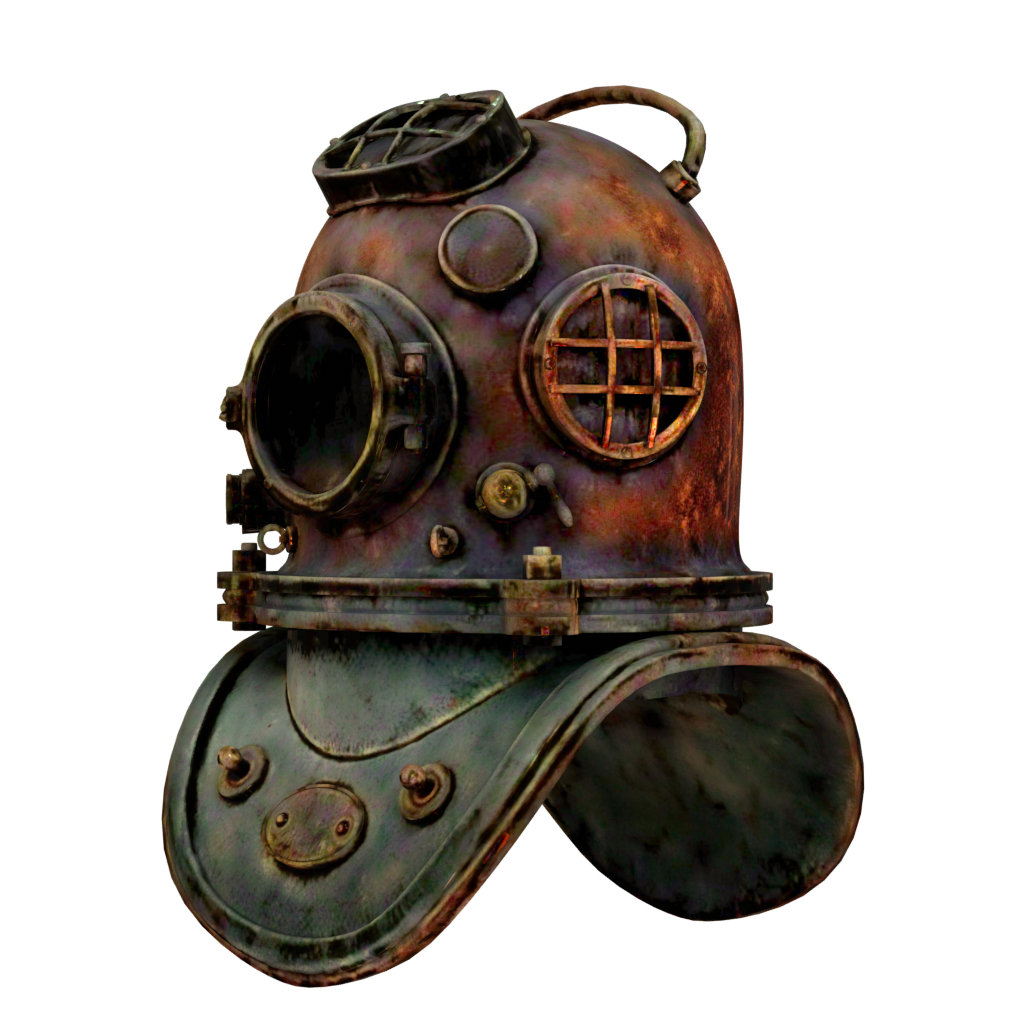}} &
	\raisebox{-0.5\height}{\includegraphics[width=0.093\textwidth]{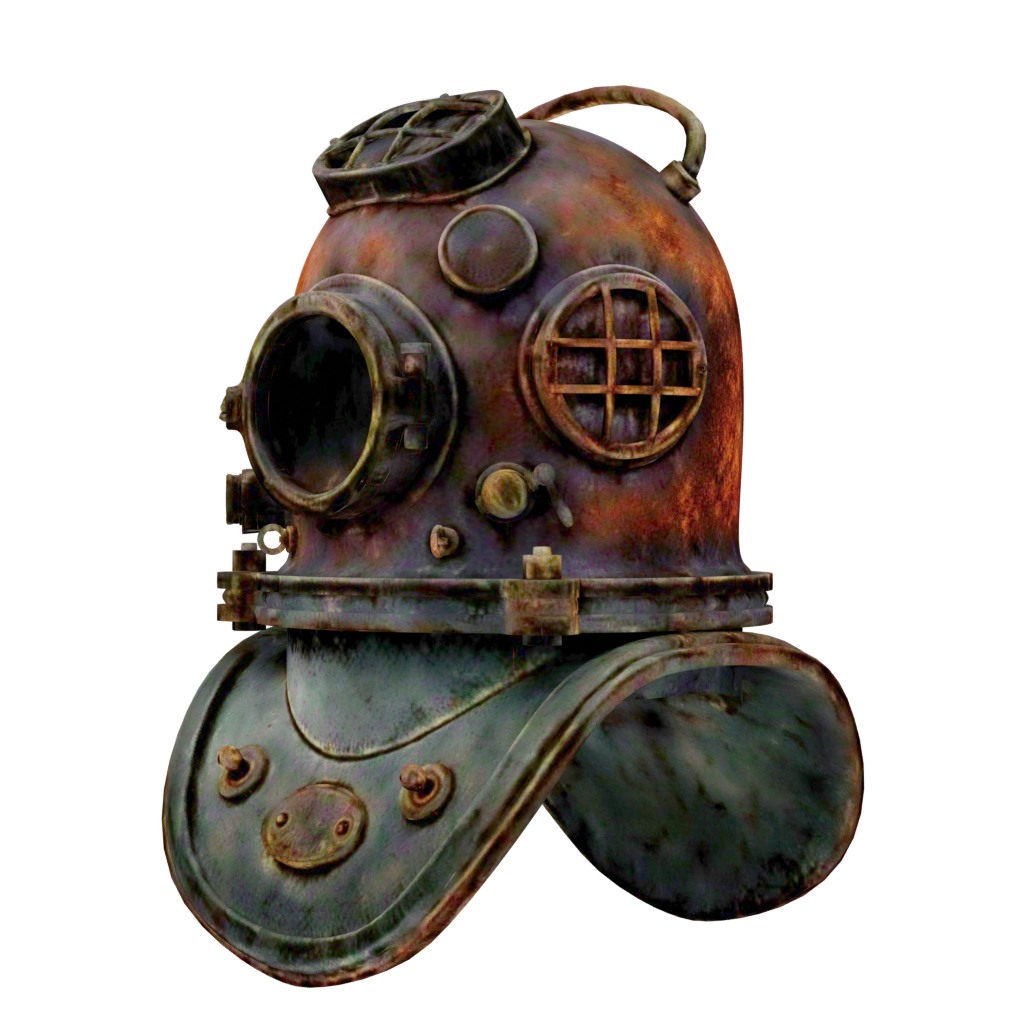}} &
	\raisebox{-0.5\height}{\includegraphics[width=0.093\textwidth]{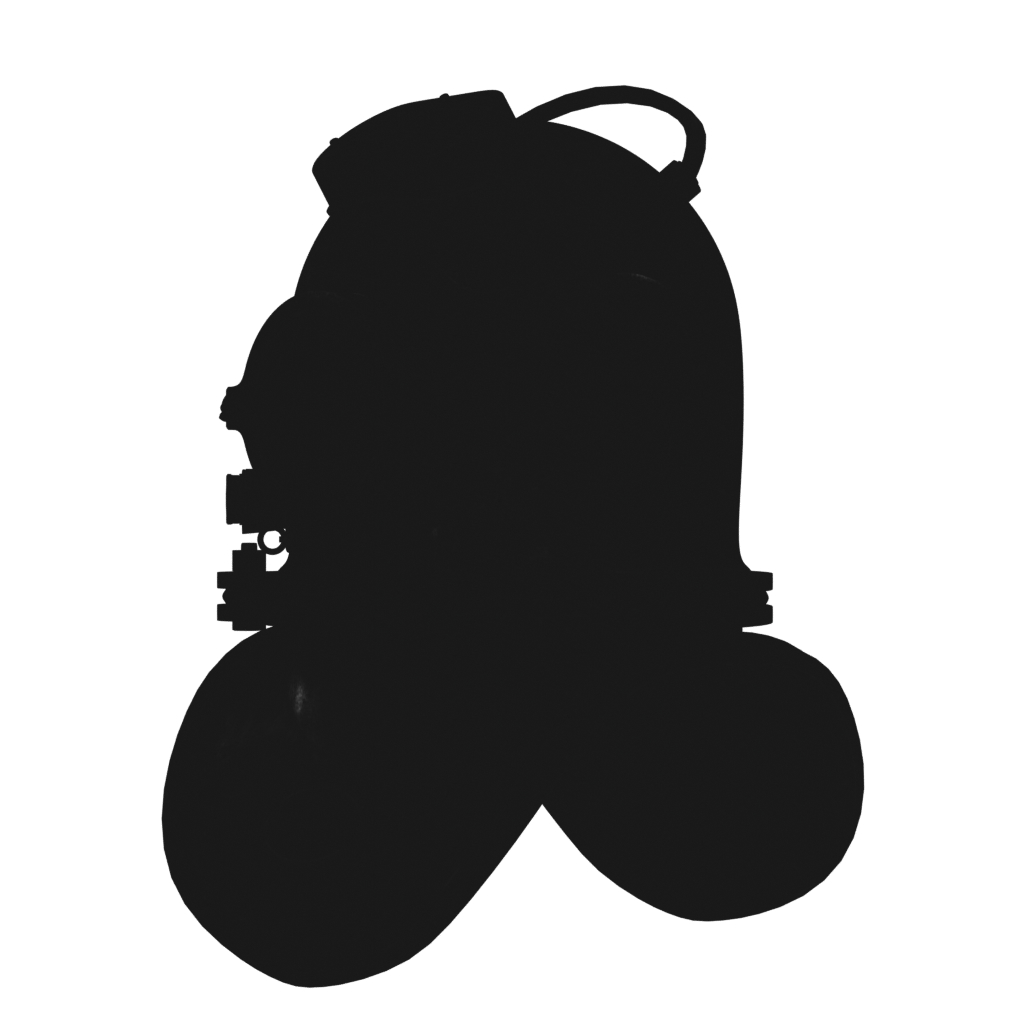}} &
	\raisebox{-0.5\height}{\includegraphics[width=0.093\textwidth]{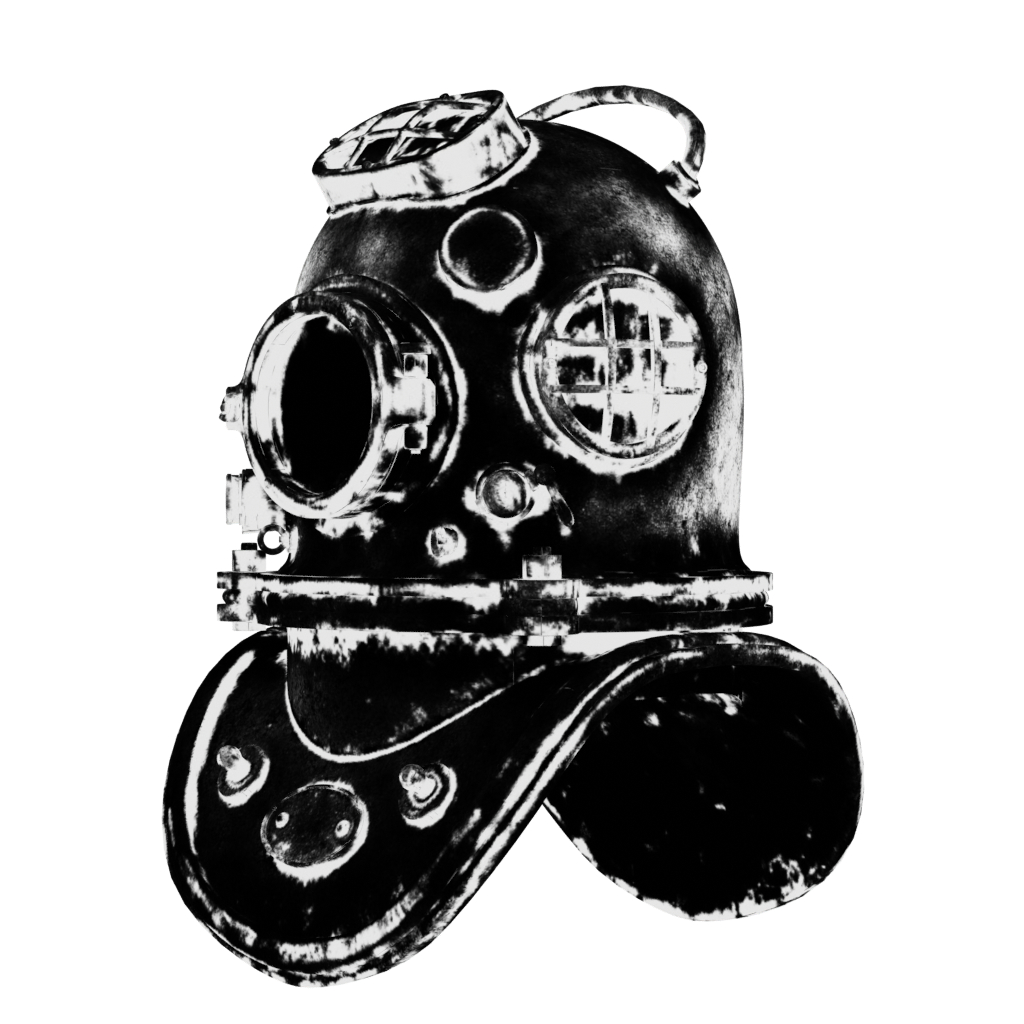}} \\

	\multirow{2}{*}{\rotatebox{90}{\makebox[0.16\textwidth]{\centering \textsc{Gramophone}}}} &
	\multirow{2}{*}{\includegraphics[width=0.16\textwidth]{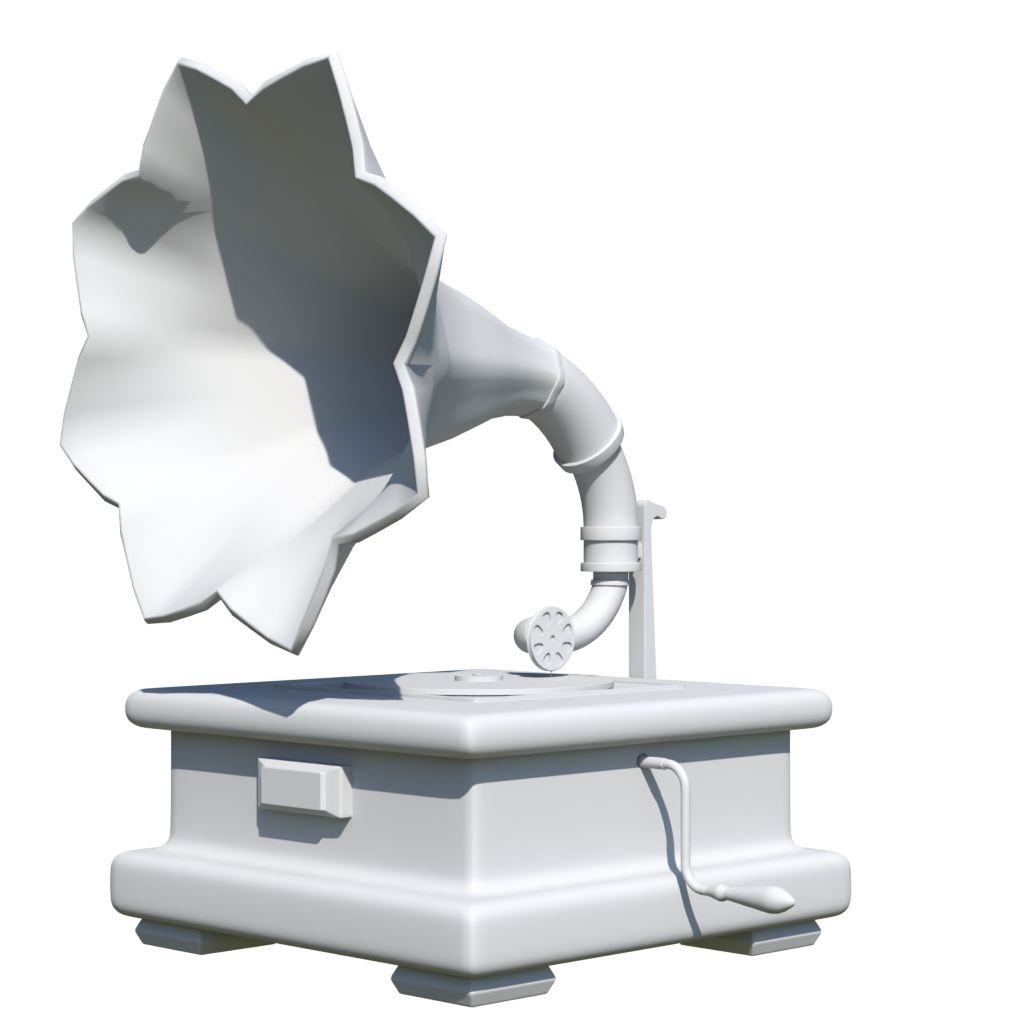}} & 
    \rotatebox[origin=c]{90}{Our} &
	\raisebox{-0.5\height}{\includegraphics[width=0.093\textwidth]{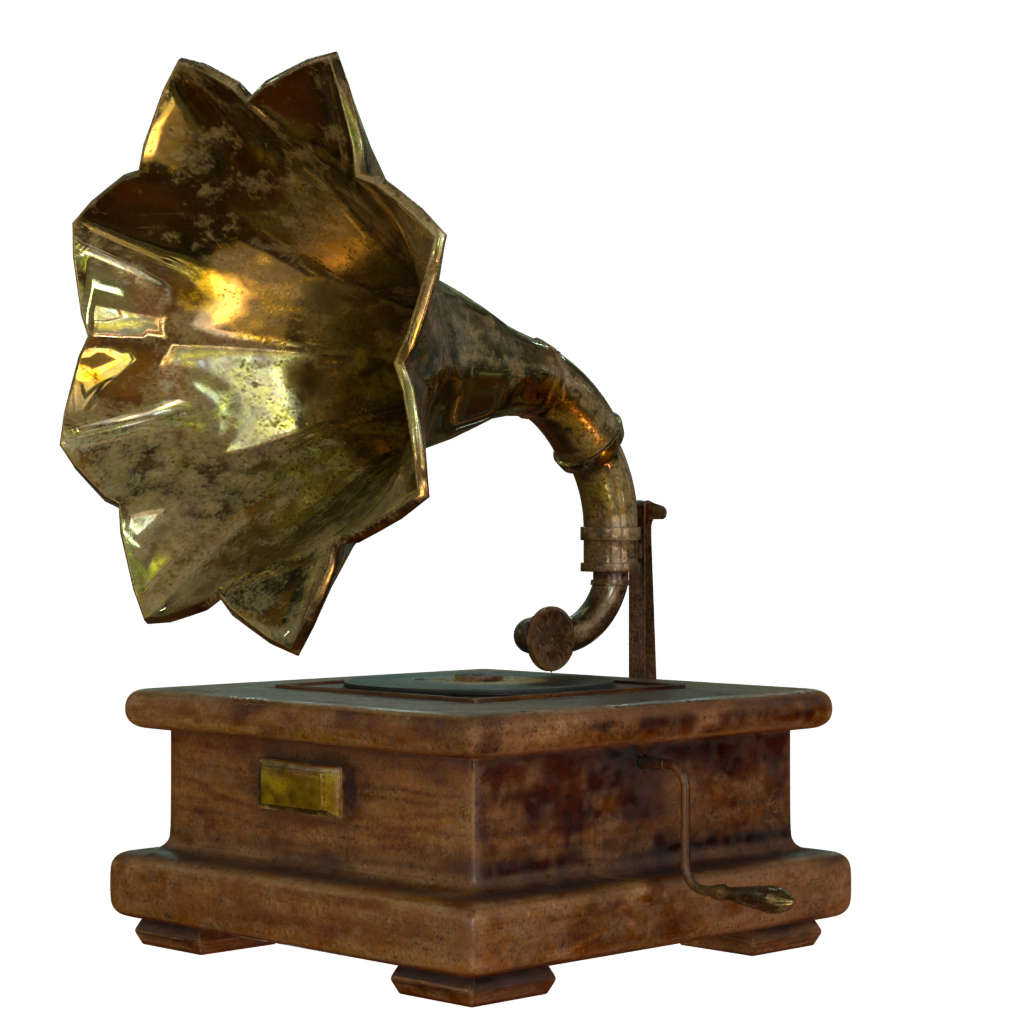}} &
	\raisebox{-0.5\height}{\includegraphics[width=0.093\textwidth]{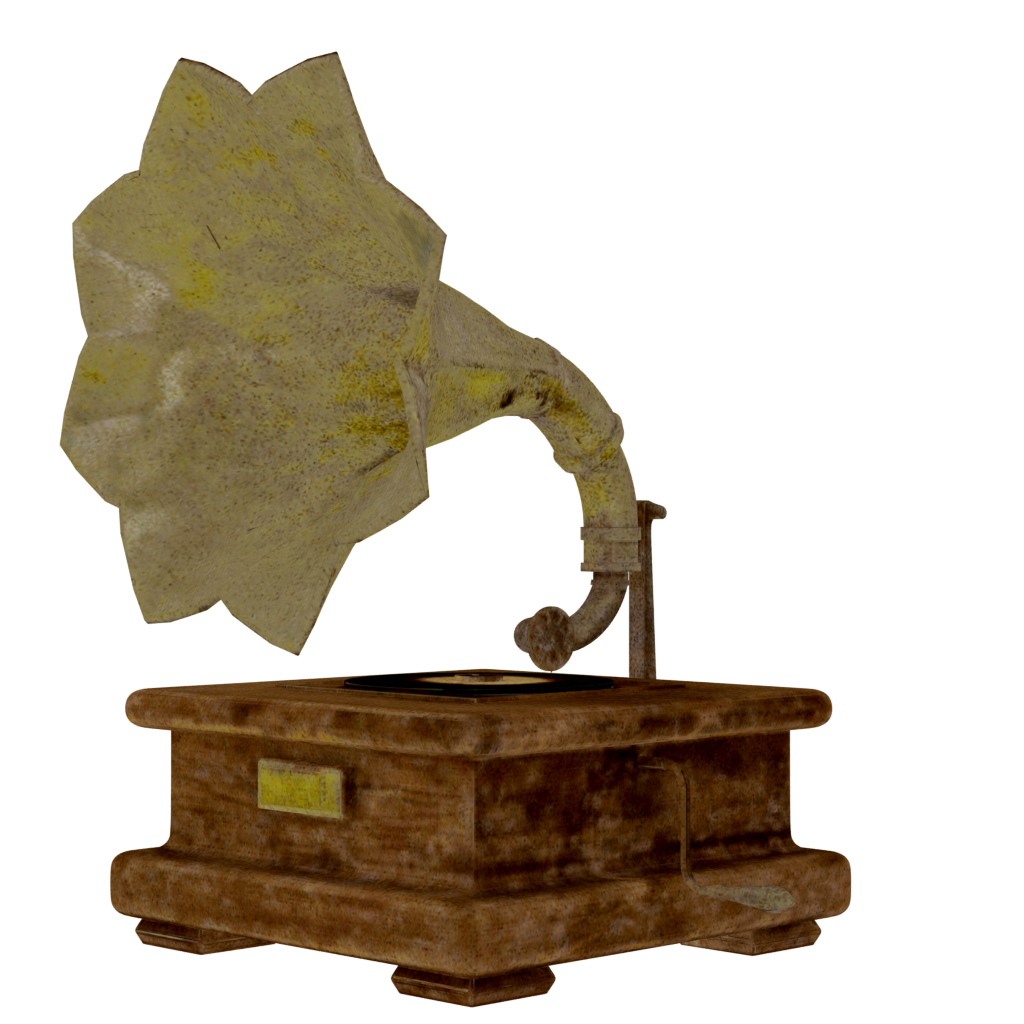}} &
	\raisebox{-0.5\height}{\includegraphics[width=0.093\textwidth]{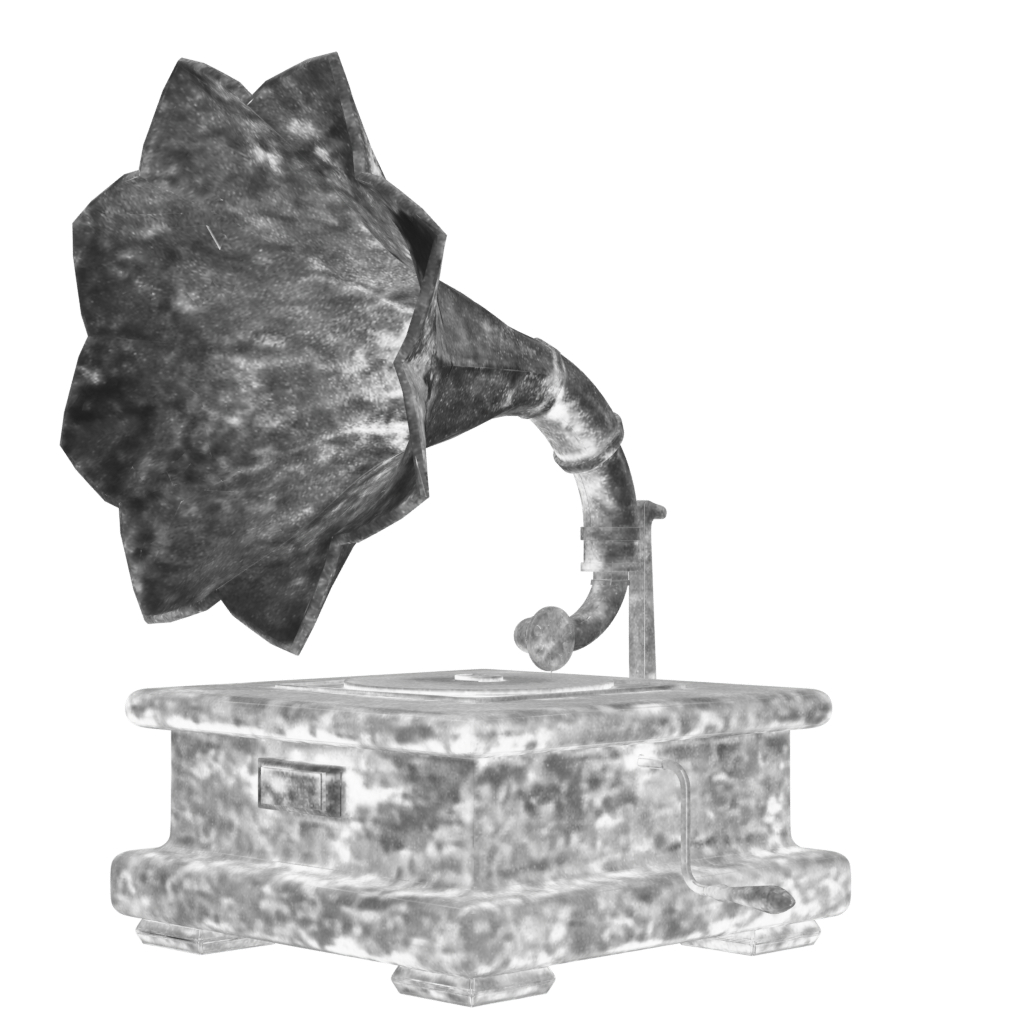}} &
	\raisebox{-0.5\height}{\includegraphics[width=0.093\textwidth]{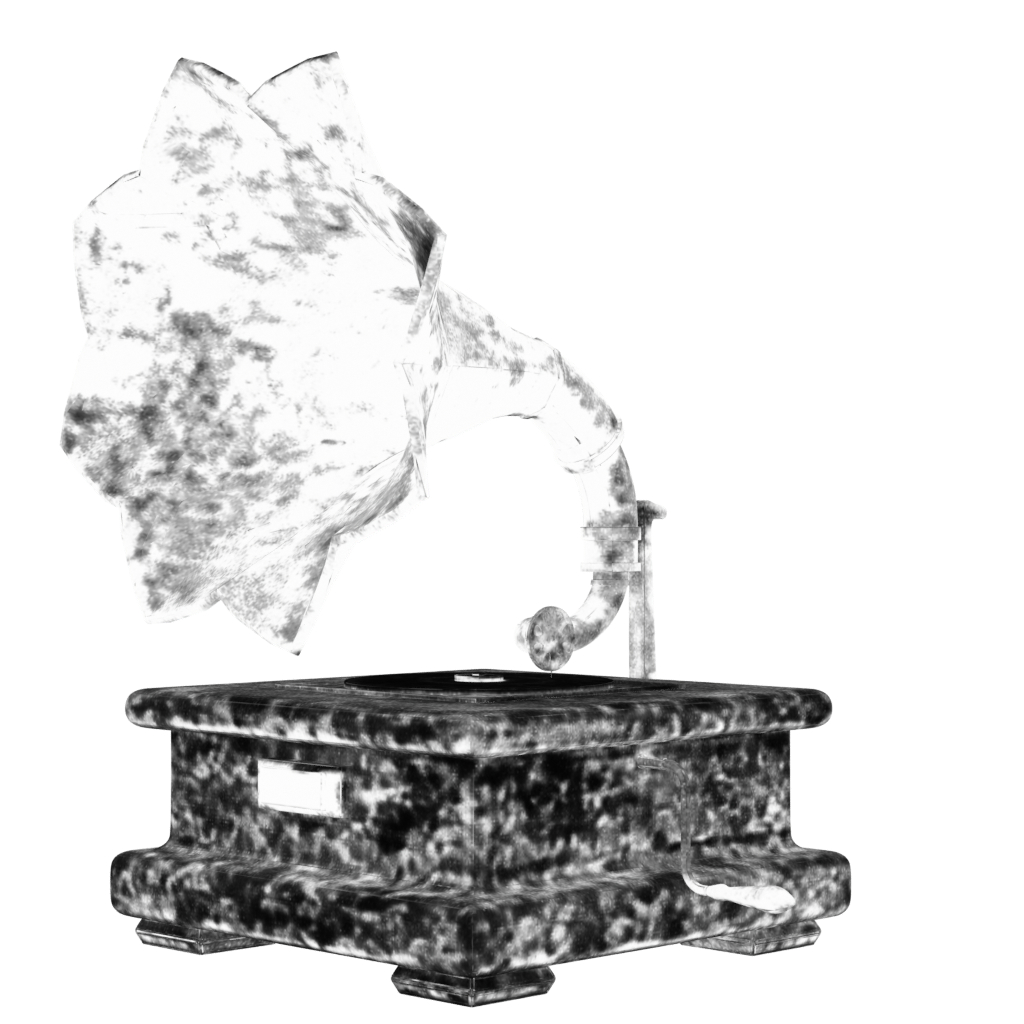}} &
    \rotatebox[origin=c]{90}{Paint-it} &
	\raisebox{-0.5\height}{\includegraphics[width=0.093\textwidth]{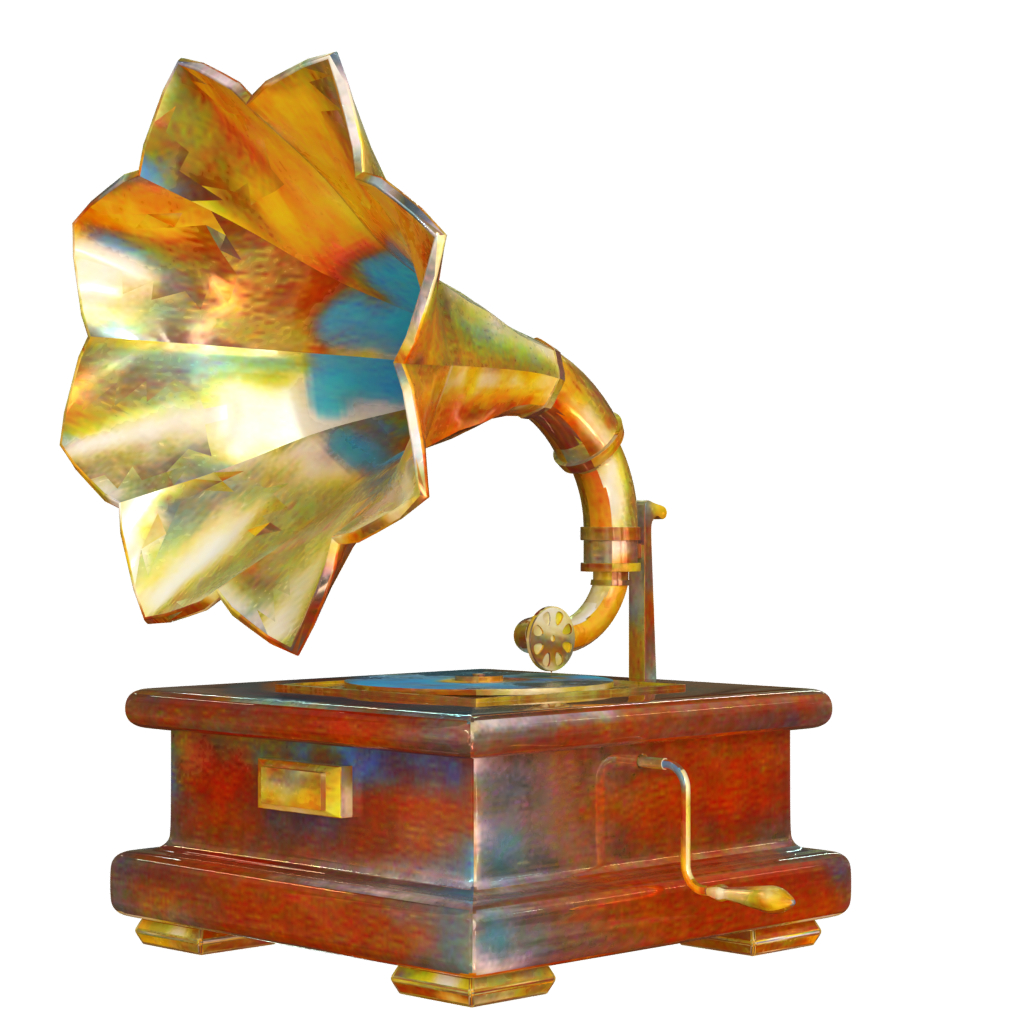}} &
	\raisebox{-0.5\height}{\includegraphics[width=0.093\textwidth]{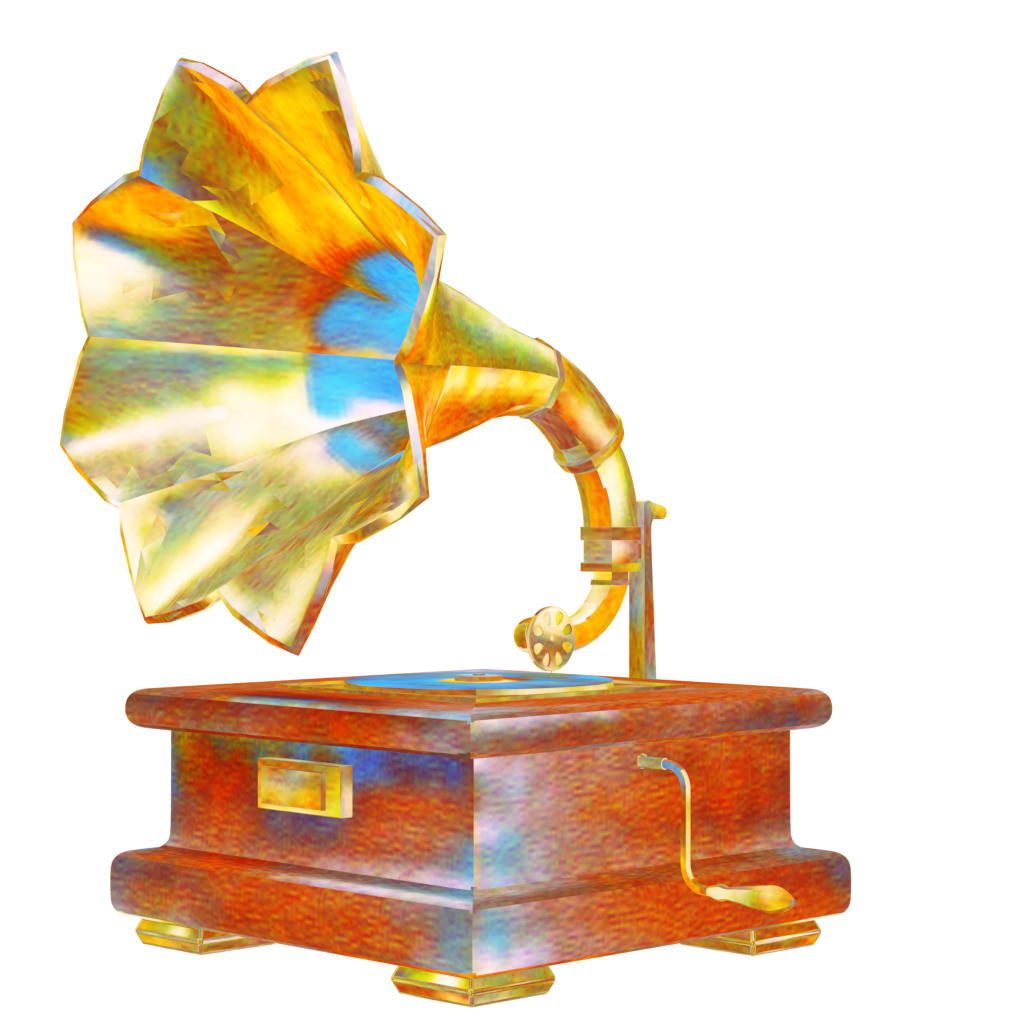}} &
	\raisebox{-0.5\height}{\includegraphics[width=0.093\textwidth]{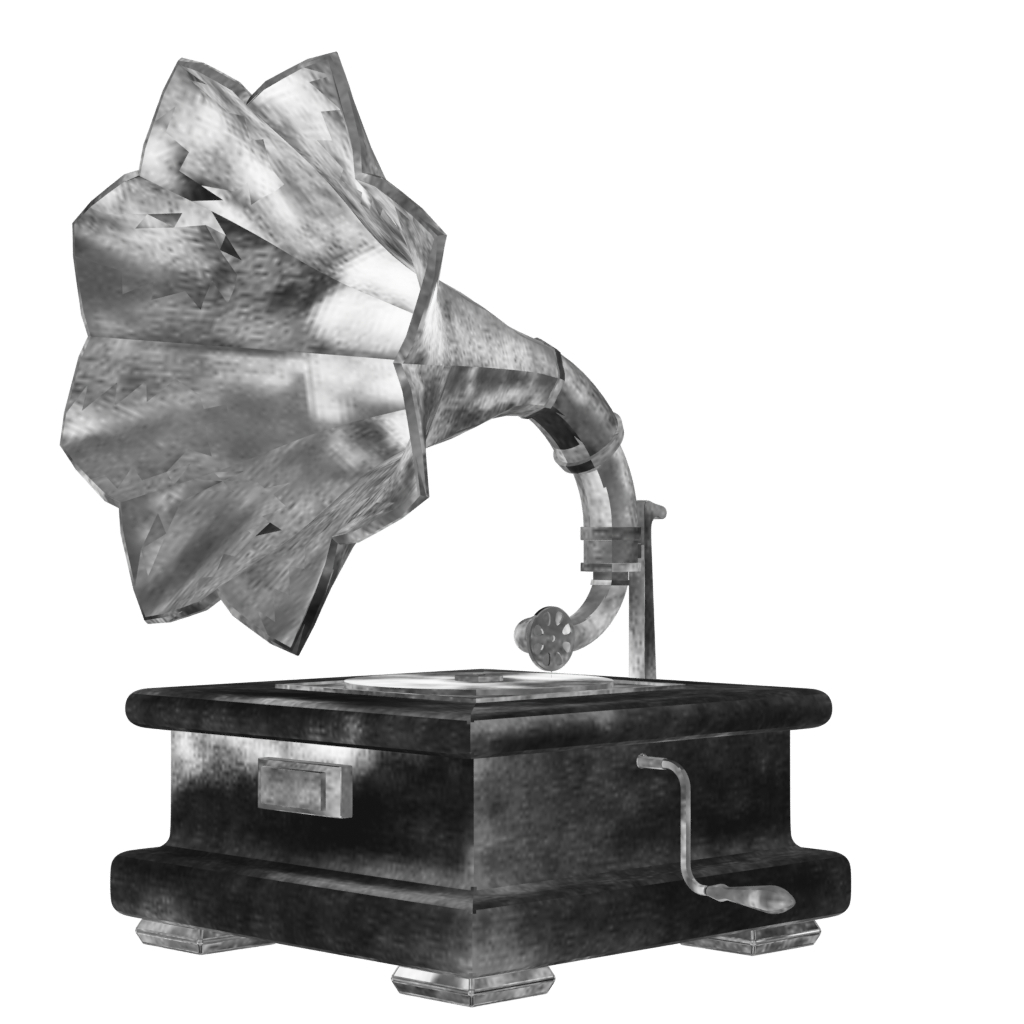}} &
	\raisebox{-0.5\height}{\includegraphics[width=0.093\textwidth]{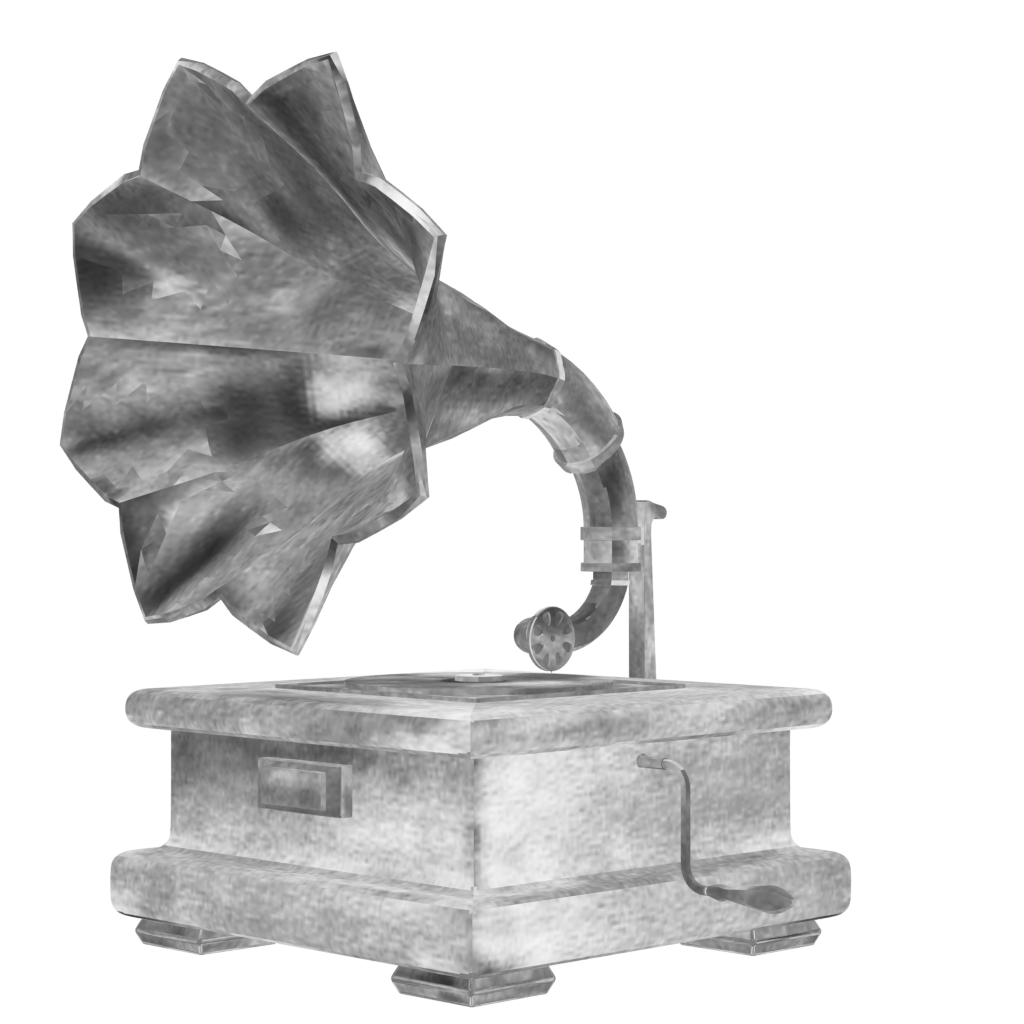}} \\
	& & 
    \rotatebox[origin=c]{90}{Reference} &
	\raisebox{-0.5\height}{\includegraphics[width=0.093\textwidth]{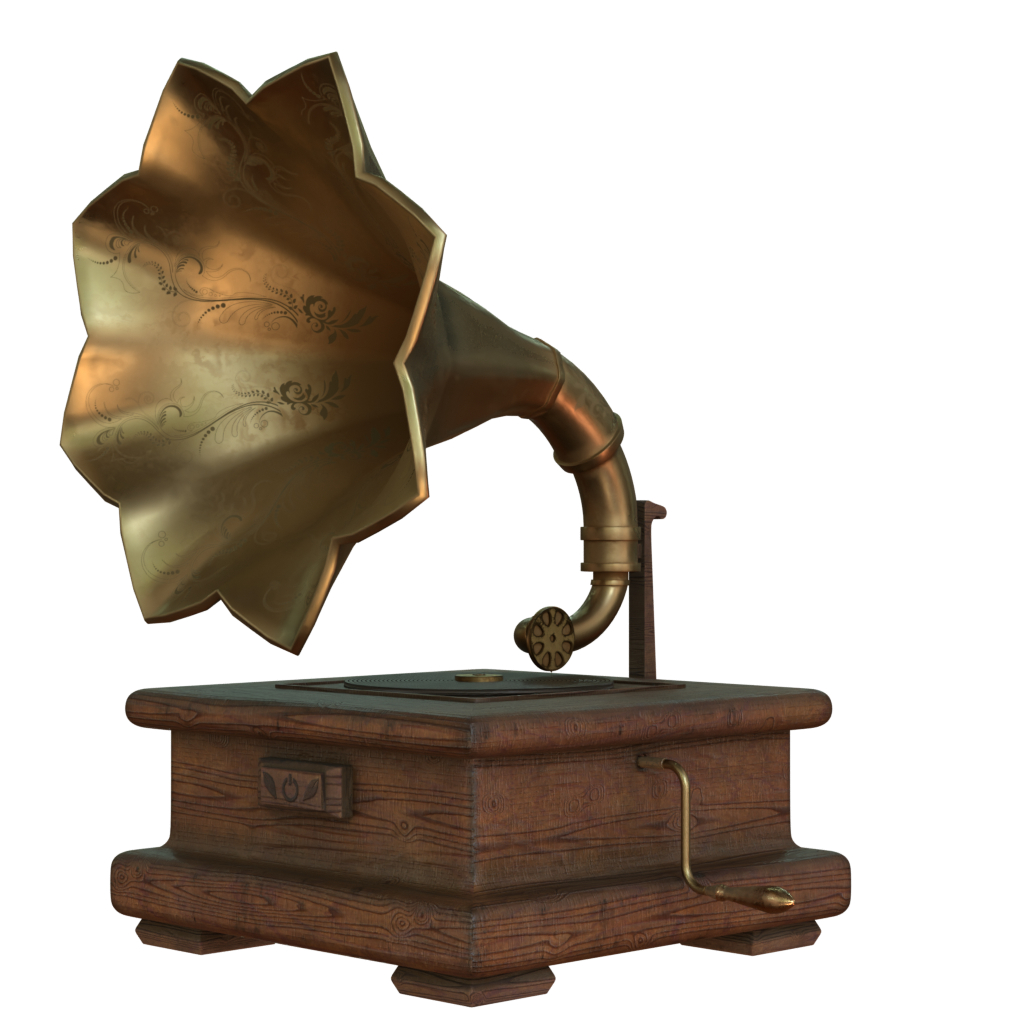}} &
	\raisebox{-0.5\height}{\includegraphics[width=0.093\textwidth]{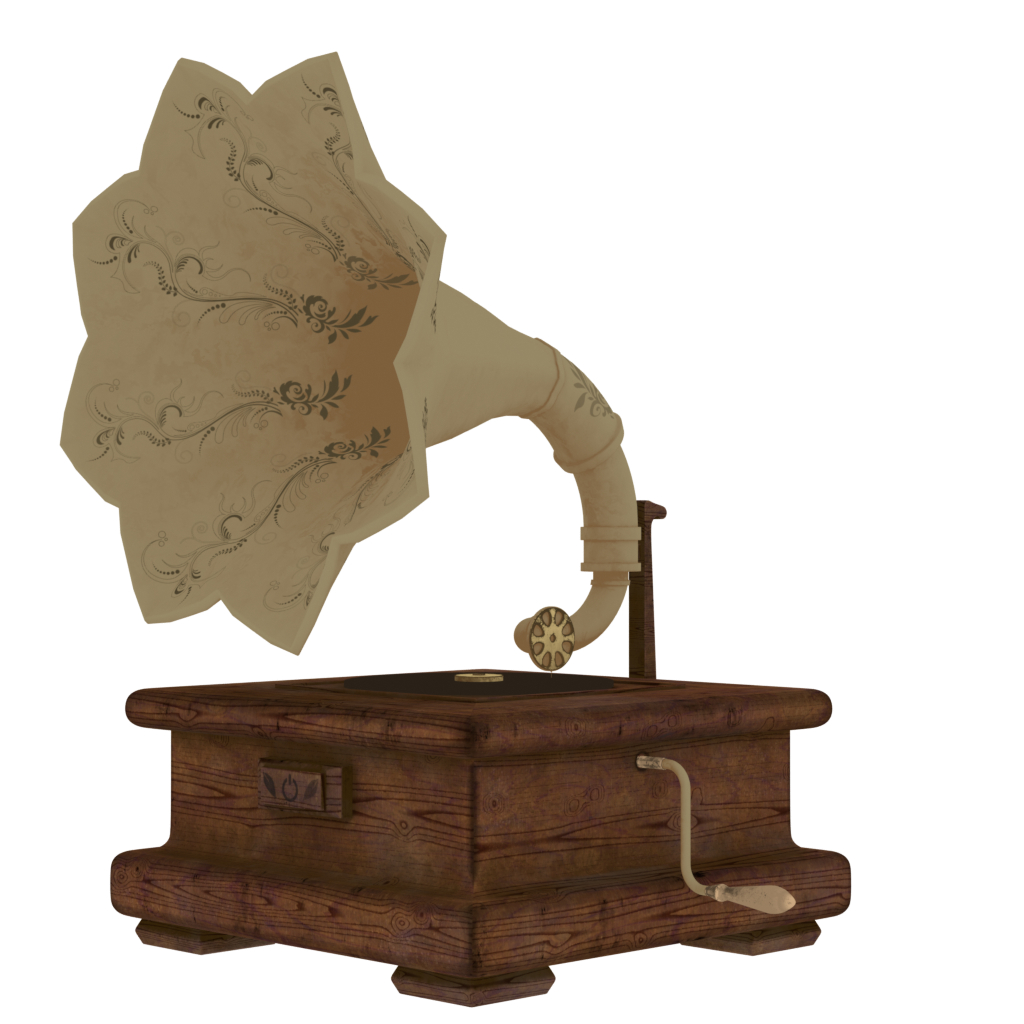}} &
	\raisebox{-0.5\height}{\includegraphics[width=0.093\textwidth]{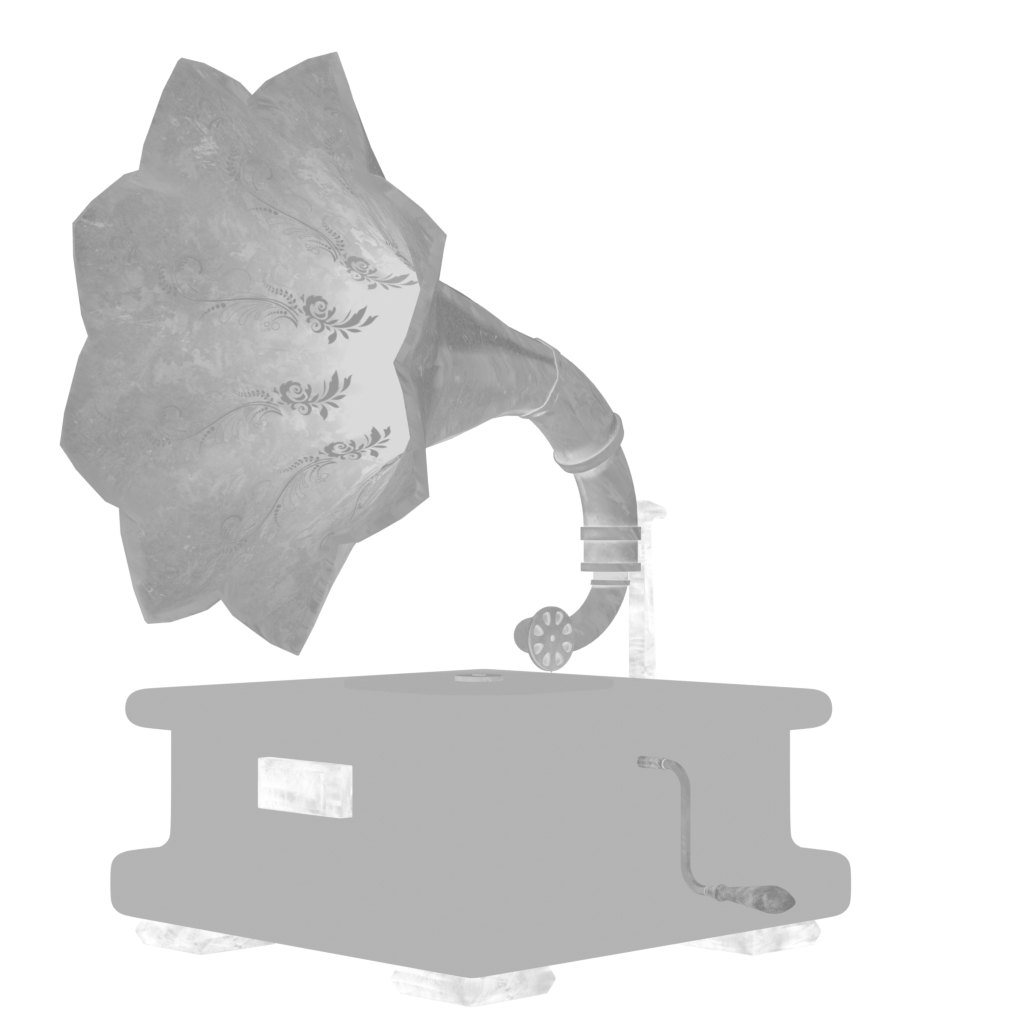}} &
	\raisebox{-0.5\height}{\includegraphics[width=0.093\textwidth]{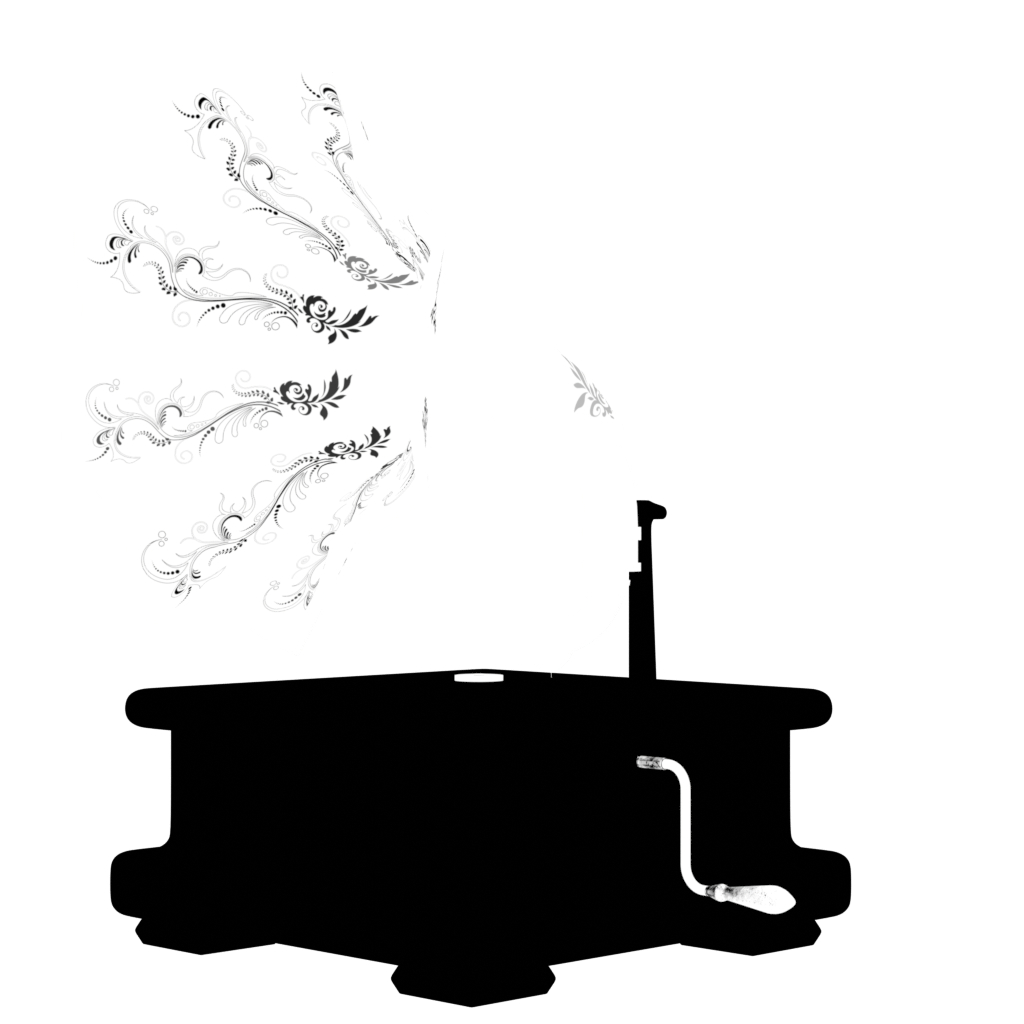}} &
    \rotatebox[origin=c]{90}{DreamMat} &
	\raisebox{-0.5\height}{\includegraphics[width=0.093\textwidth]{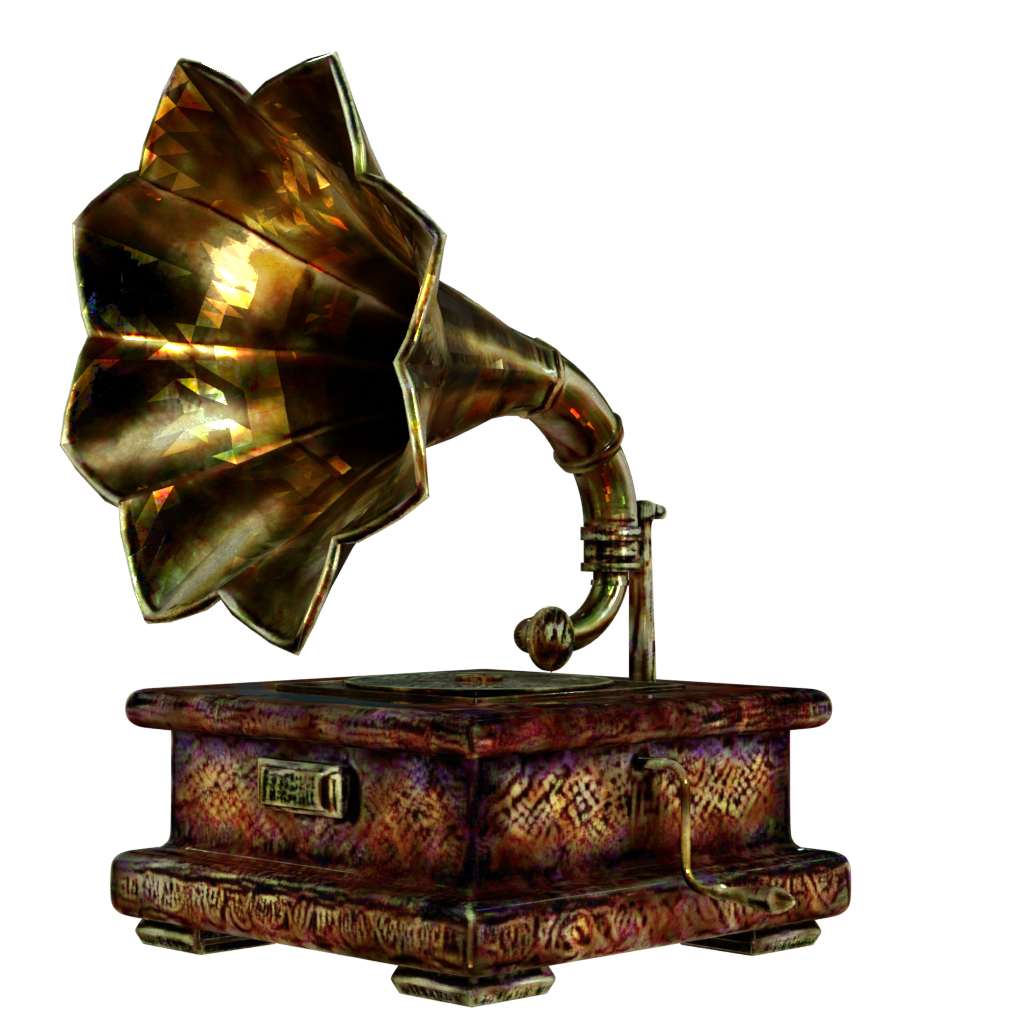}} &
	\raisebox{-0.5\height}{\includegraphics[width=0.093\textwidth]{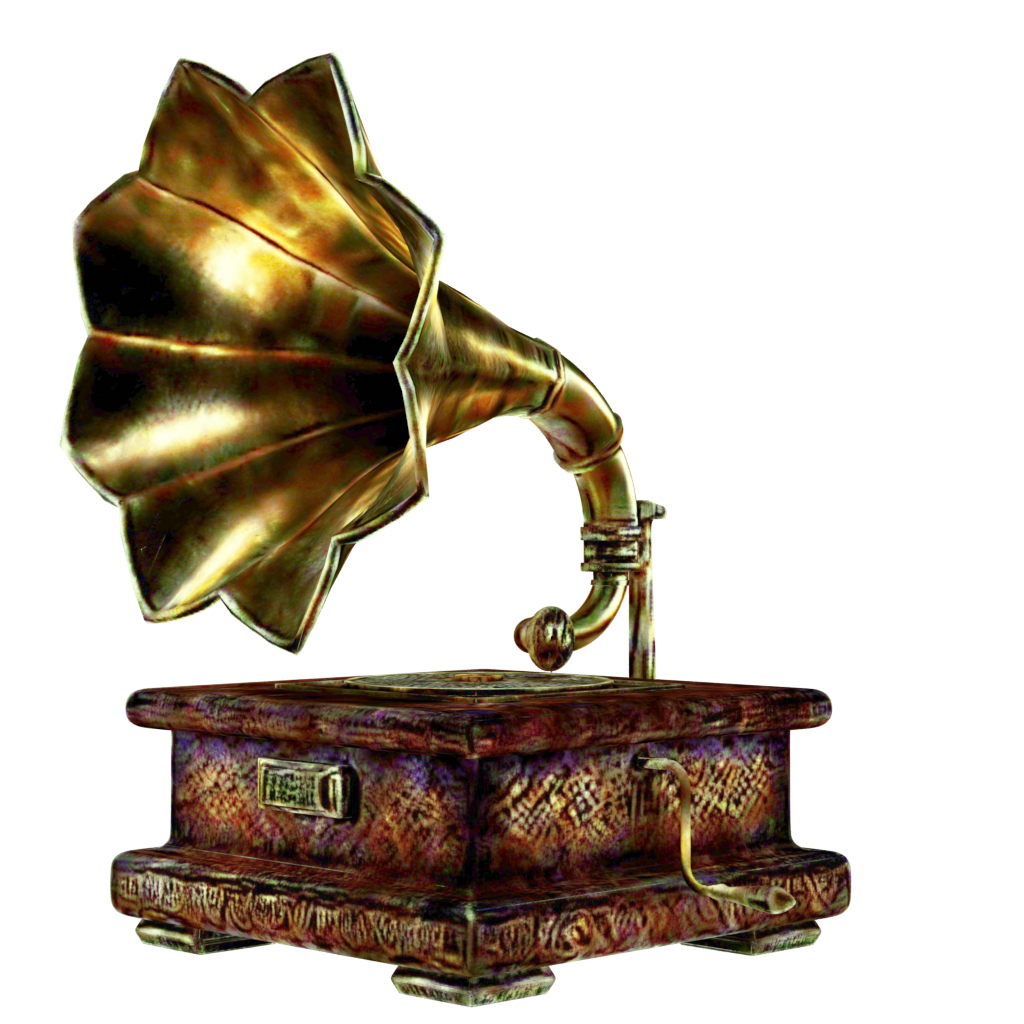}} &
	\raisebox{-0.5\height}{\includegraphics[width=0.093\textwidth]{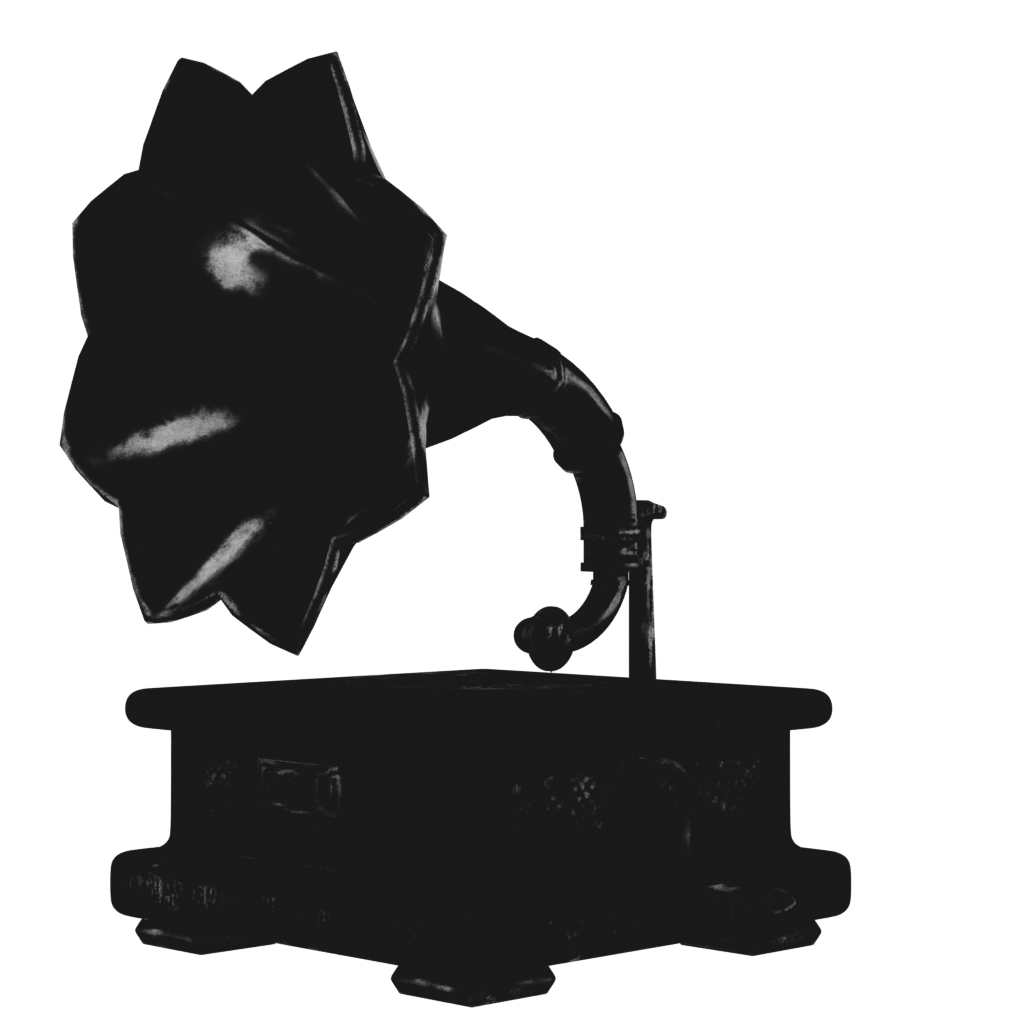}} &
	\raisebox{-0.5\height}{\includegraphics[width=0.093\textwidth]{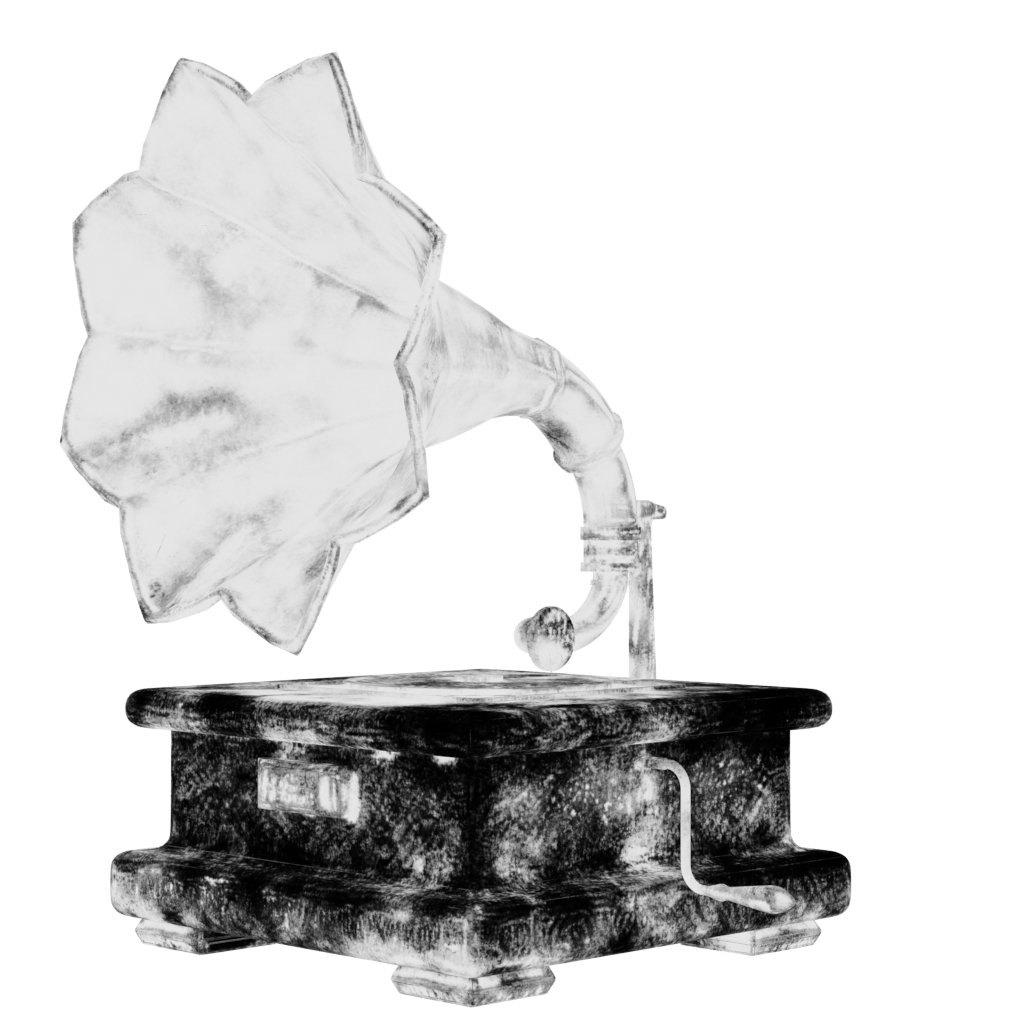}} \\

	\multirow{2}{*}{\rotatebox{90}{\makebox[0.16\textwidth]{\centering \textsc{Shoes}}}} &
	\multirow{2}{*}{\includegraphics[width=0.16\textwidth]{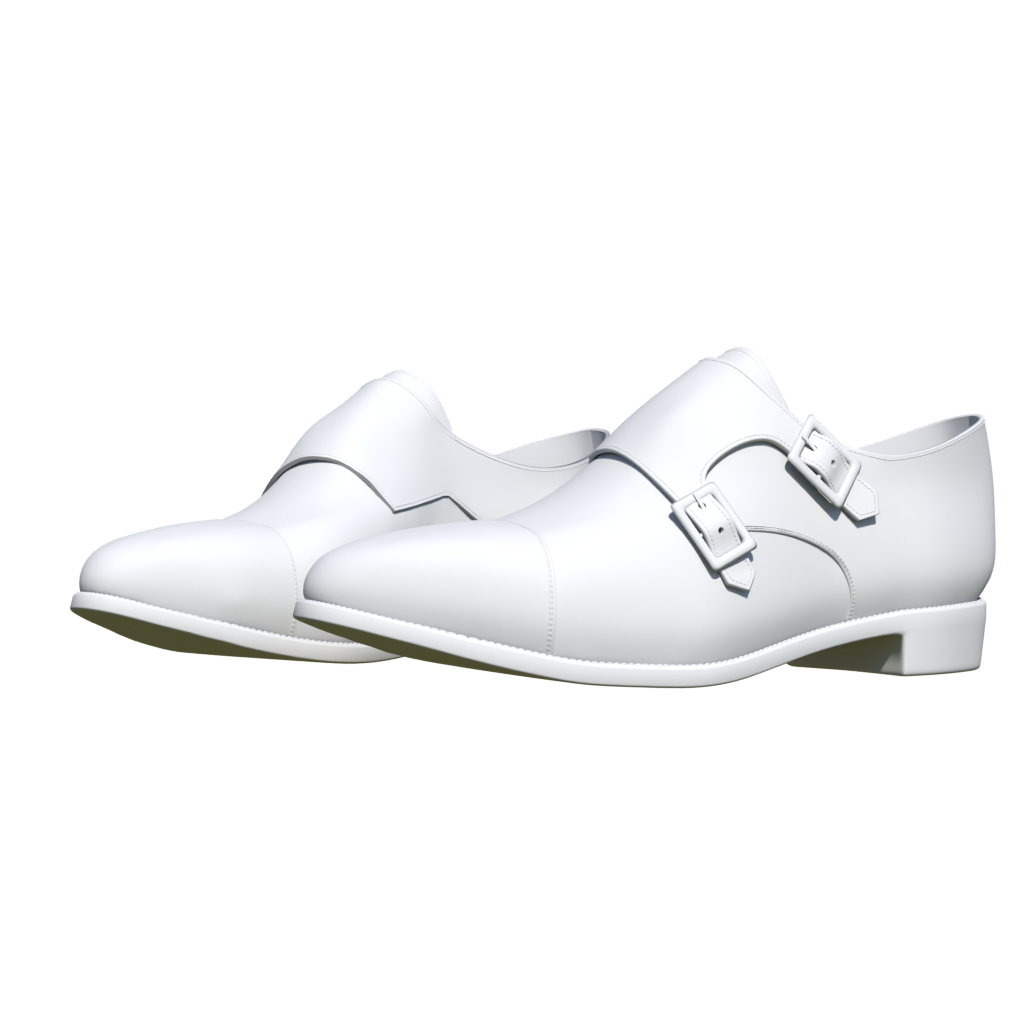}} &
    \rotatebox[origin=c]{90}{Our} &
	\raisebox{-0.5\height}{\includegraphics[width=0.093\textwidth]{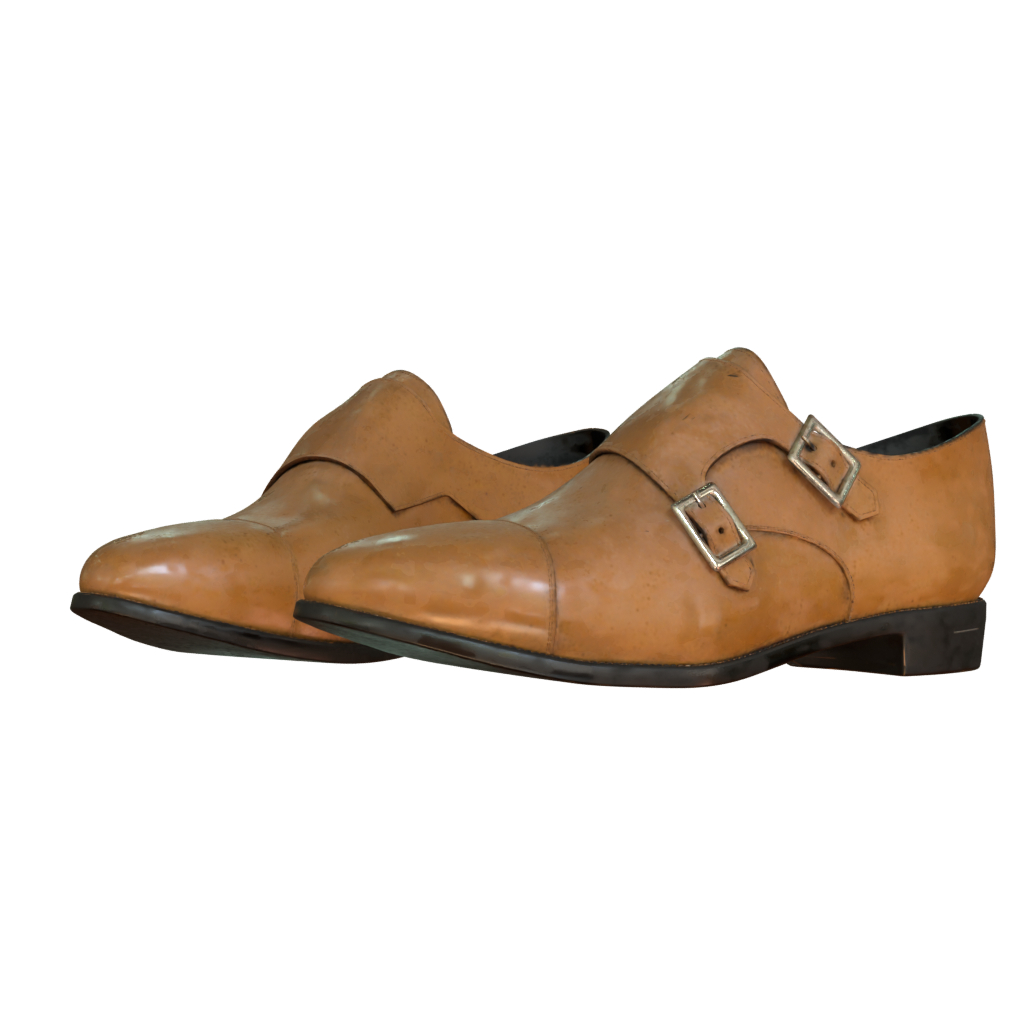}} &
	\raisebox{-0.5\height}{\includegraphics[width=0.093\textwidth]{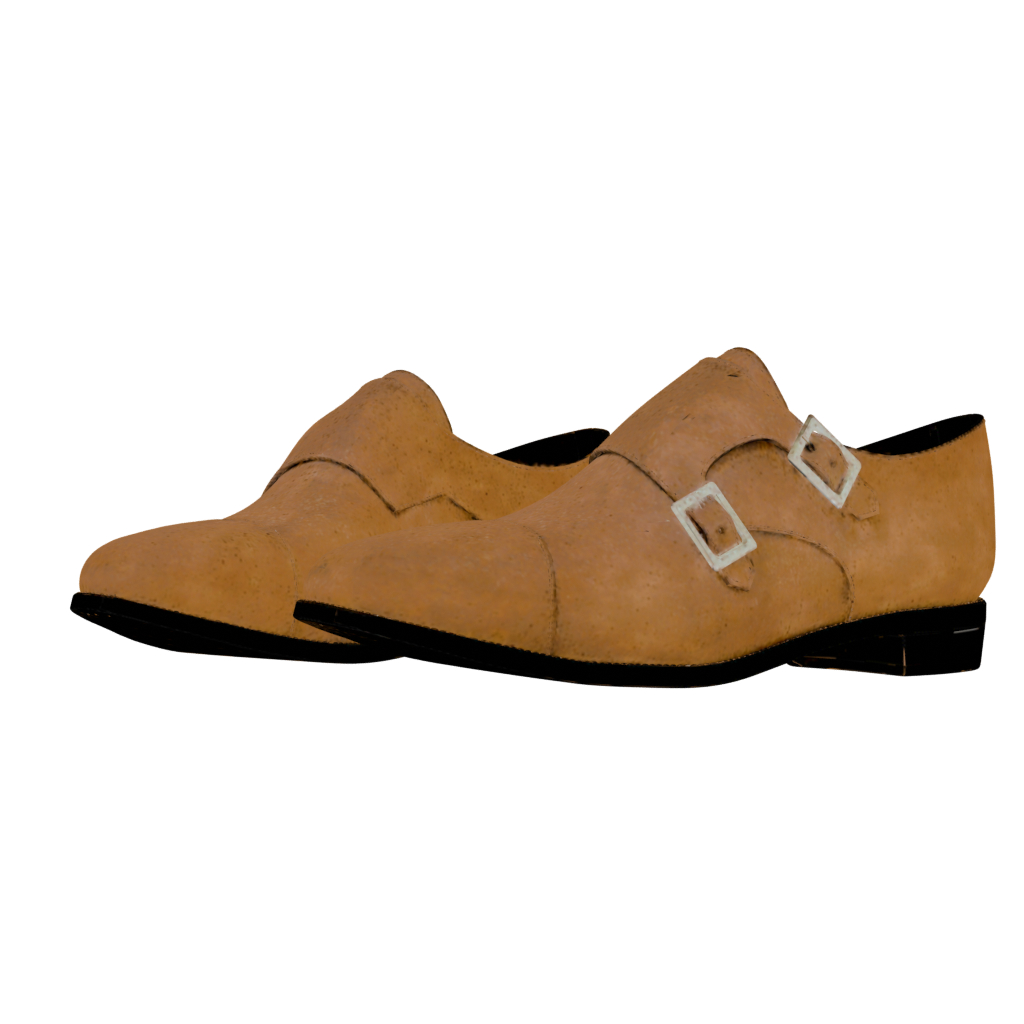}} &
	\raisebox{-0.5\height}{\includegraphics[width=0.093\textwidth]{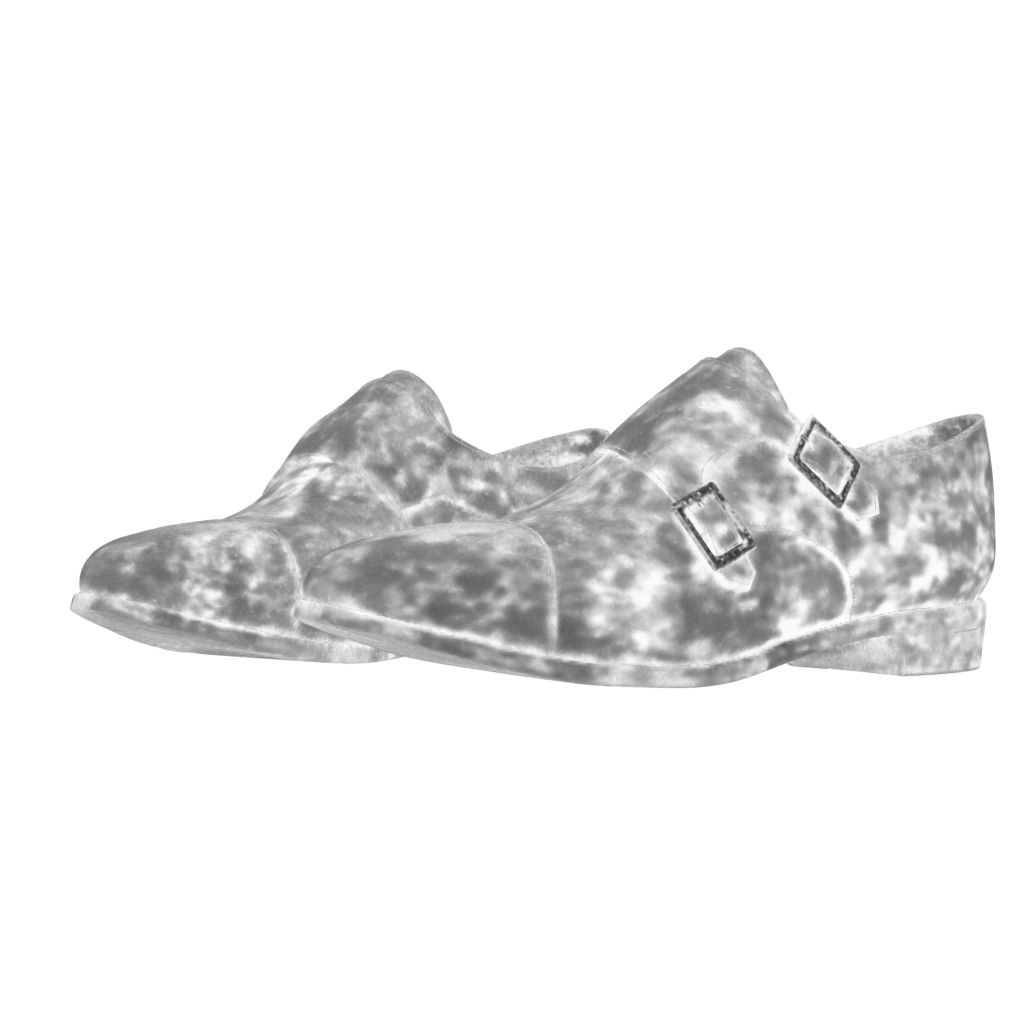}} &
	\raisebox{-0.5\height}{\includegraphics[width=0.093\textwidth]{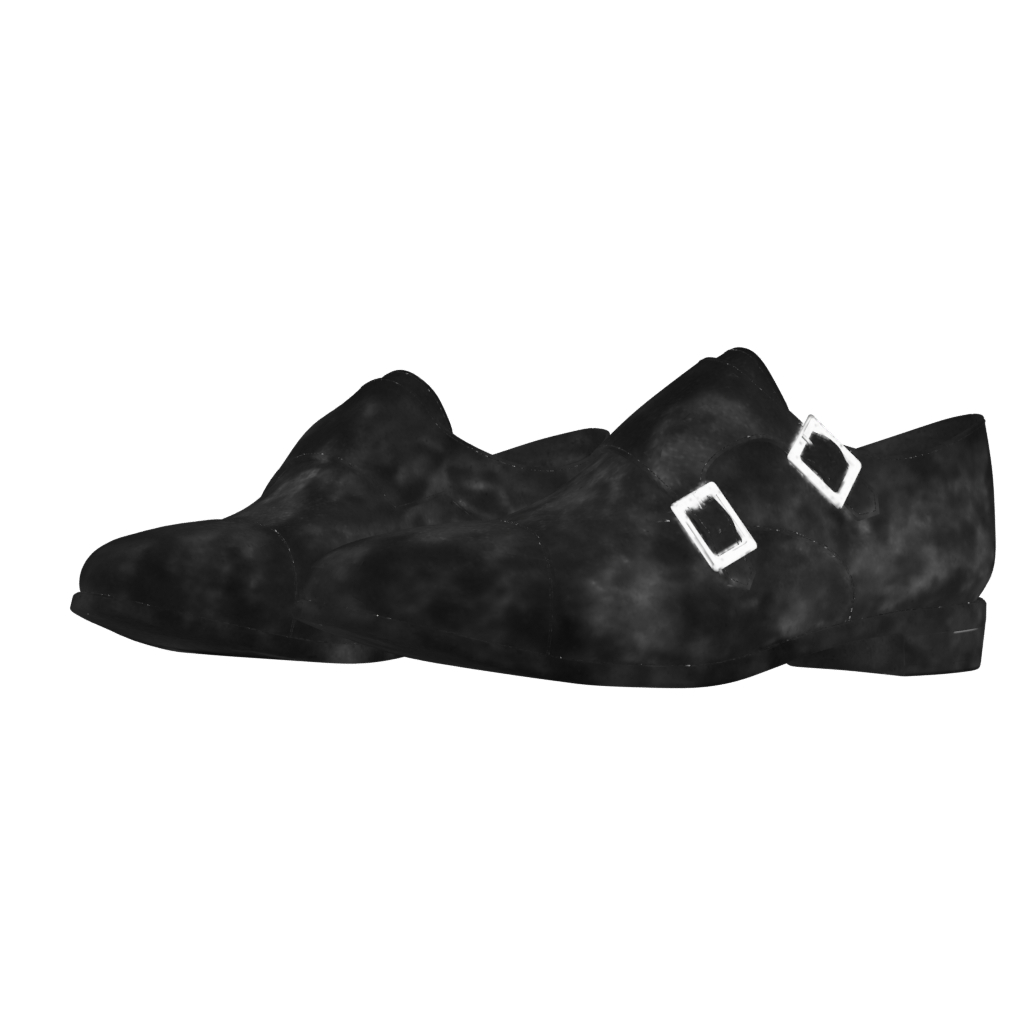}} &
    \rotatebox[origin=c]{90}{Paint-it} &
	\raisebox{-0.5\height}{\includegraphics[width=0.093\textwidth]{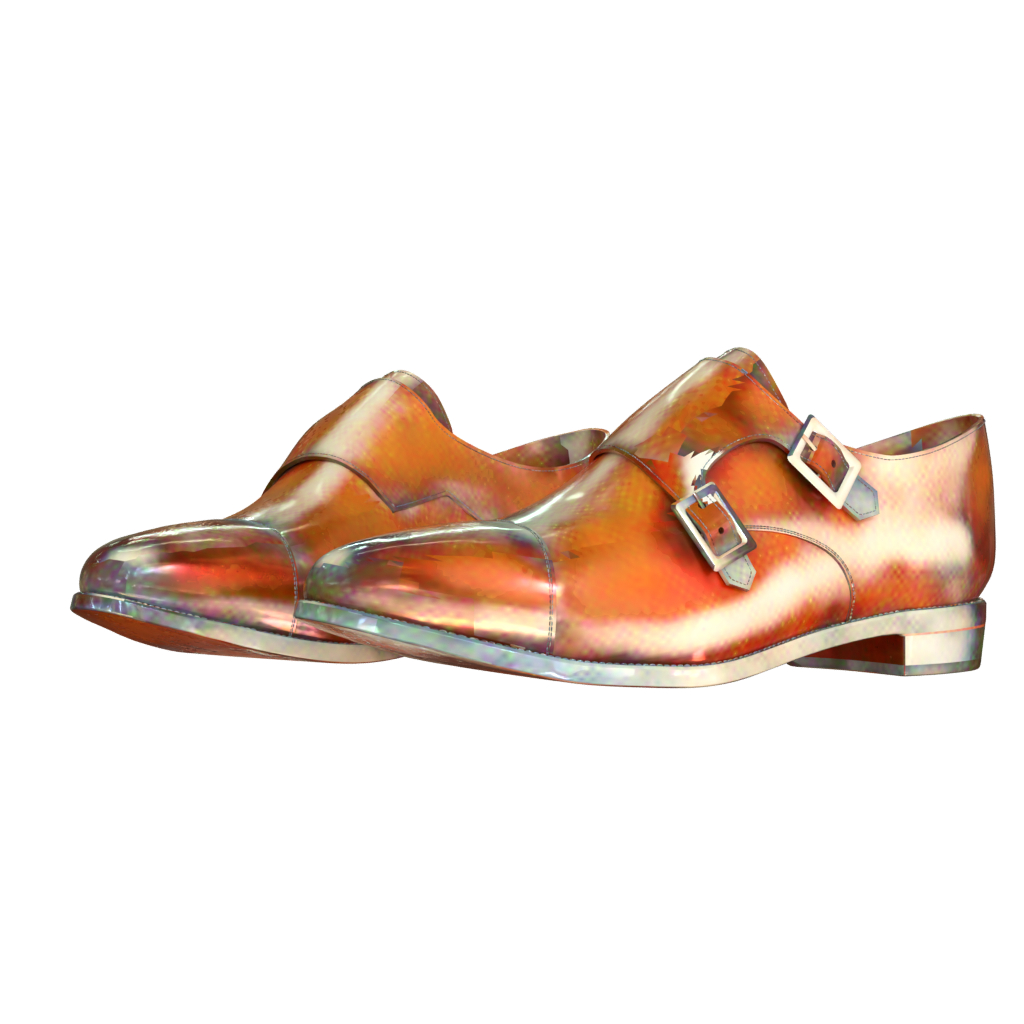}} &
	\raisebox{-0.5\height}{\includegraphics[width=0.093\textwidth]{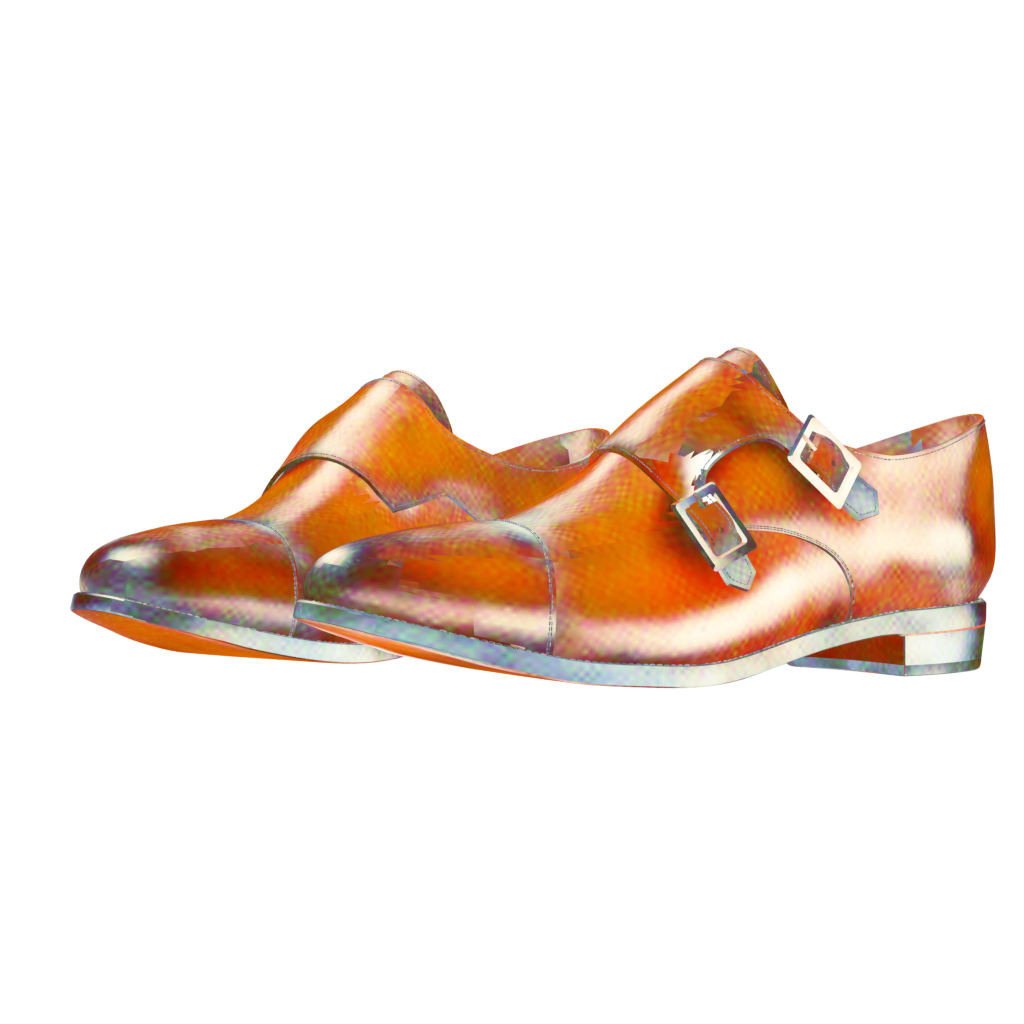}} &
	\raisebox{-0.5\height}{\includegraphics[width=0.093\textwidth]{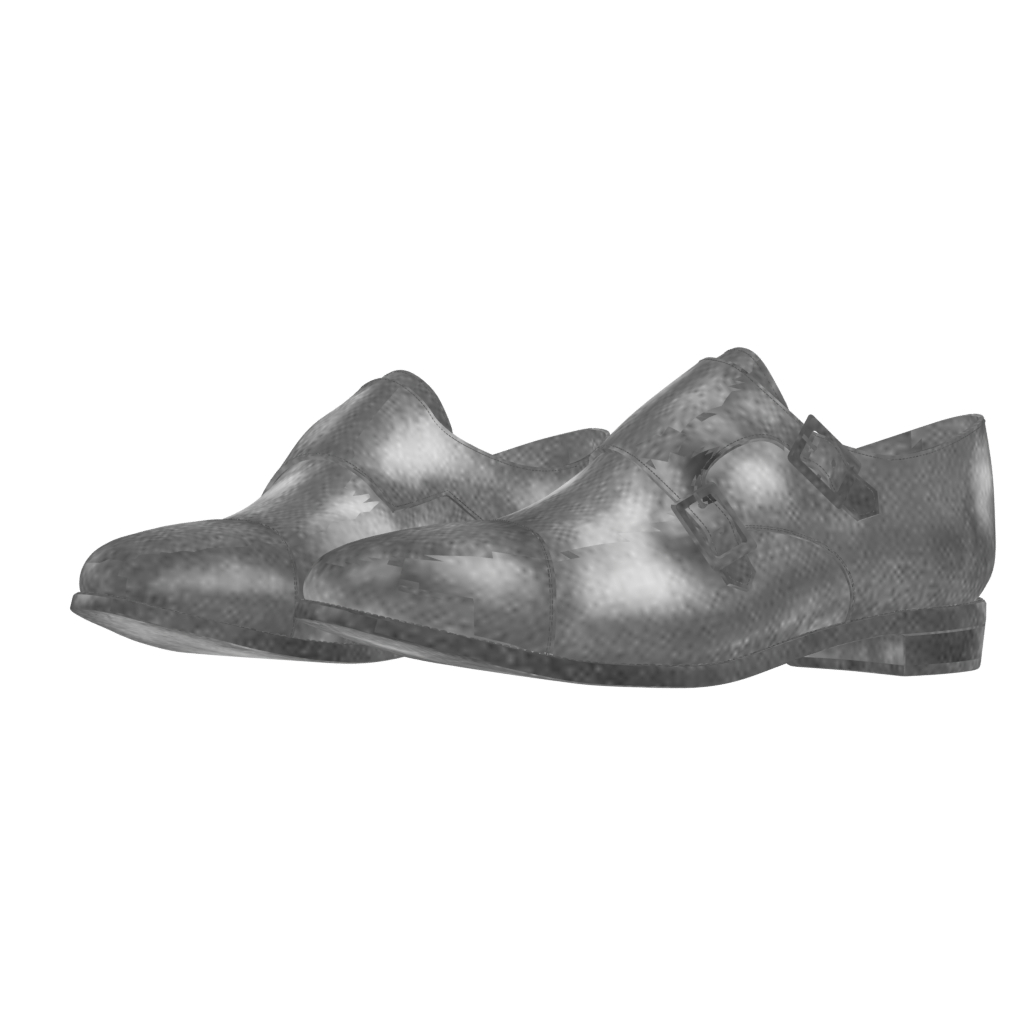}} &
	\raisebox{-0.5\height}{\includegraphics[width=0.093\textwidth]{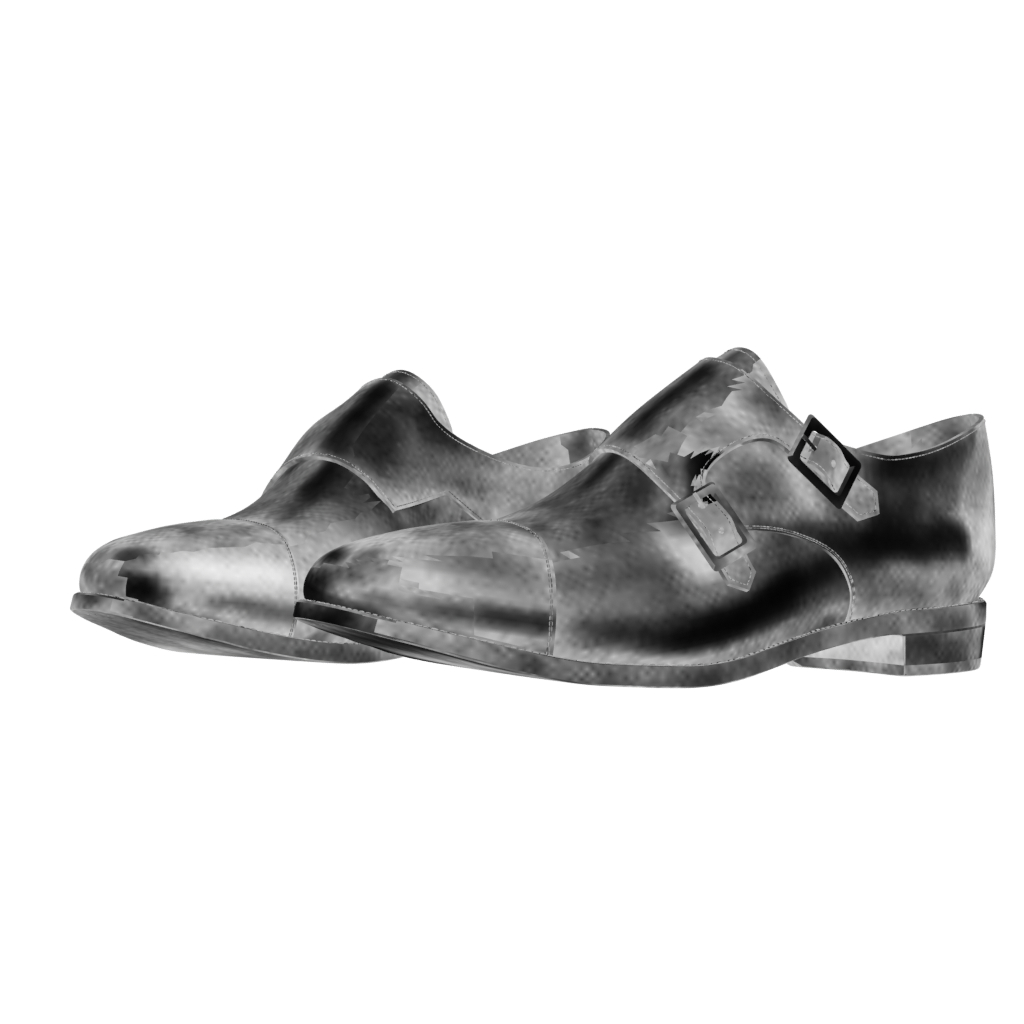}} \\
	& & 
    \rotatebox[origin=c]{90}{Reference} &
	\raisebox{-0.5\height}{\includegraphics[width=0.093\textwidth]{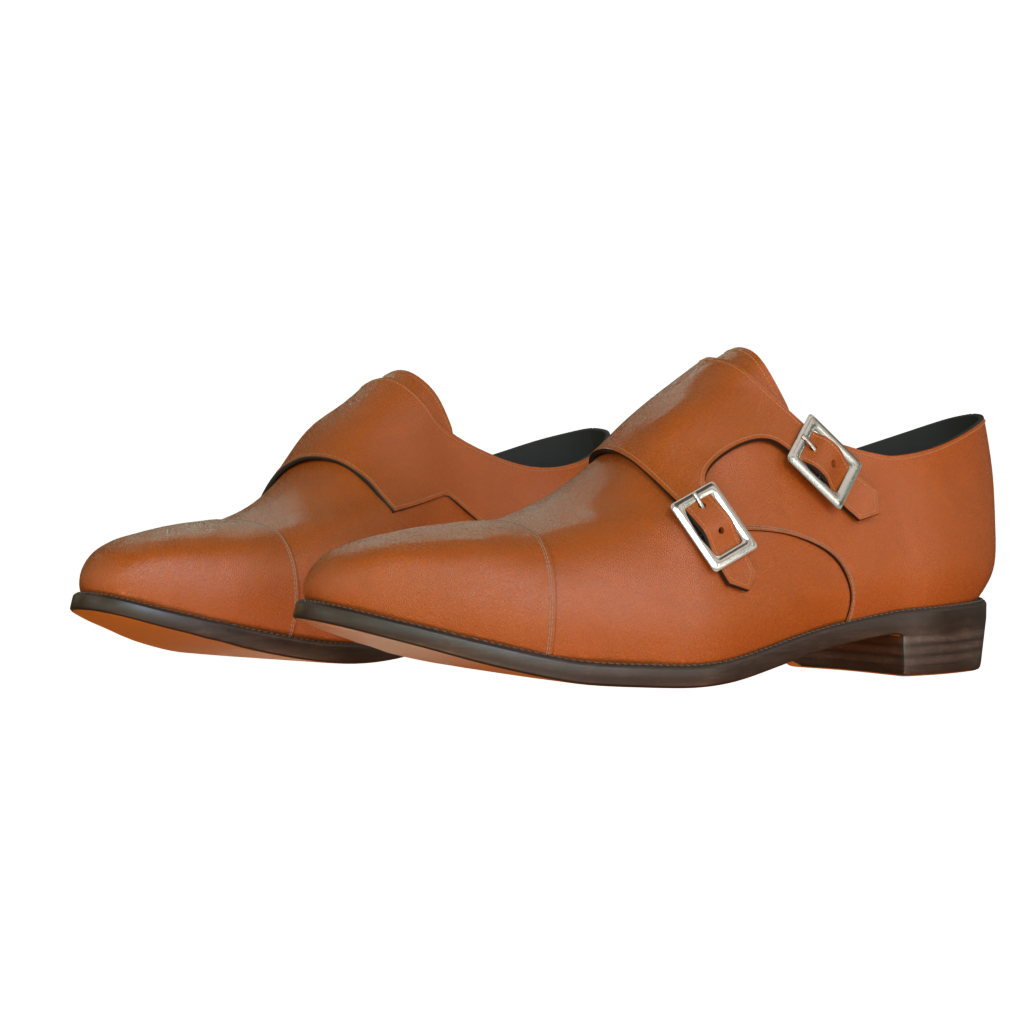}} &
	\raisebox{-0.5\height}{\includegraphics[width=0.093\textwidth]{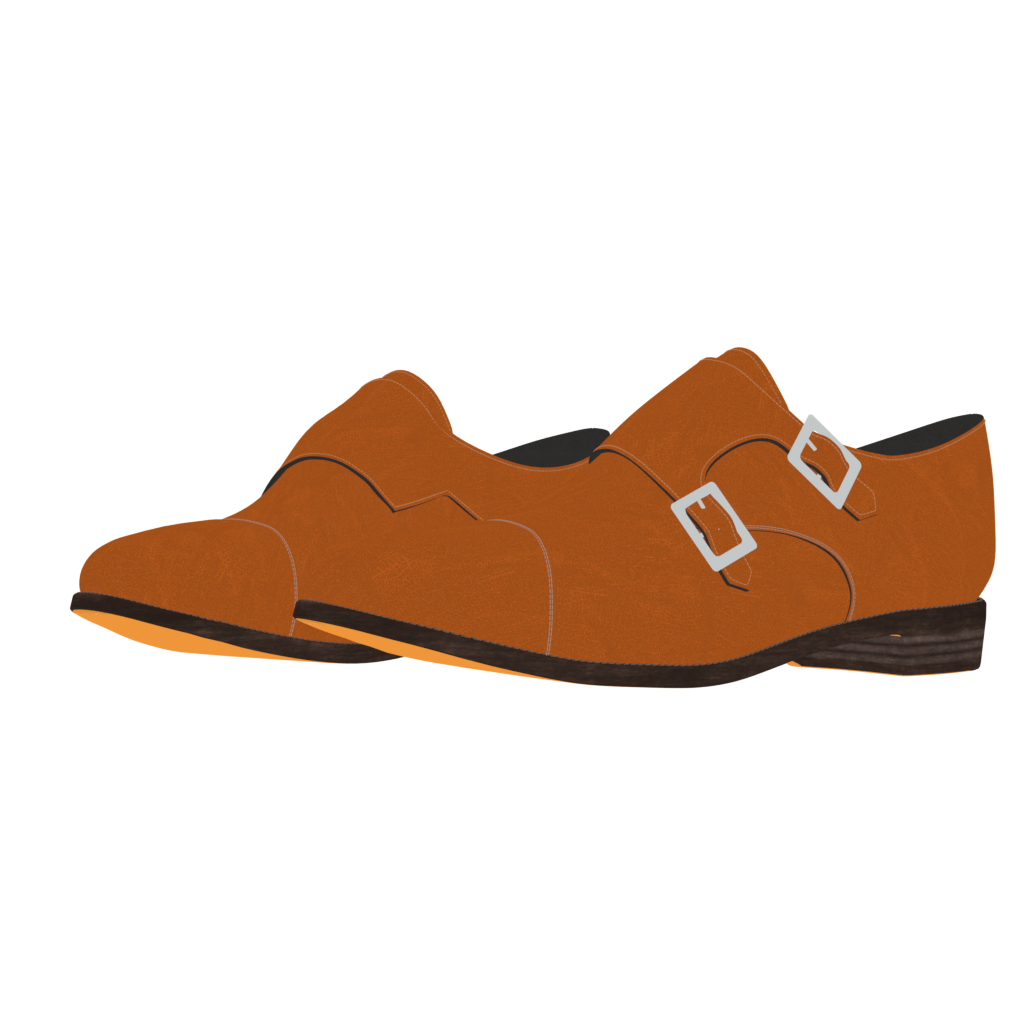}} &
	\raisebox{-0.5\height}{\includegraphics[width=0.093\textwidth]{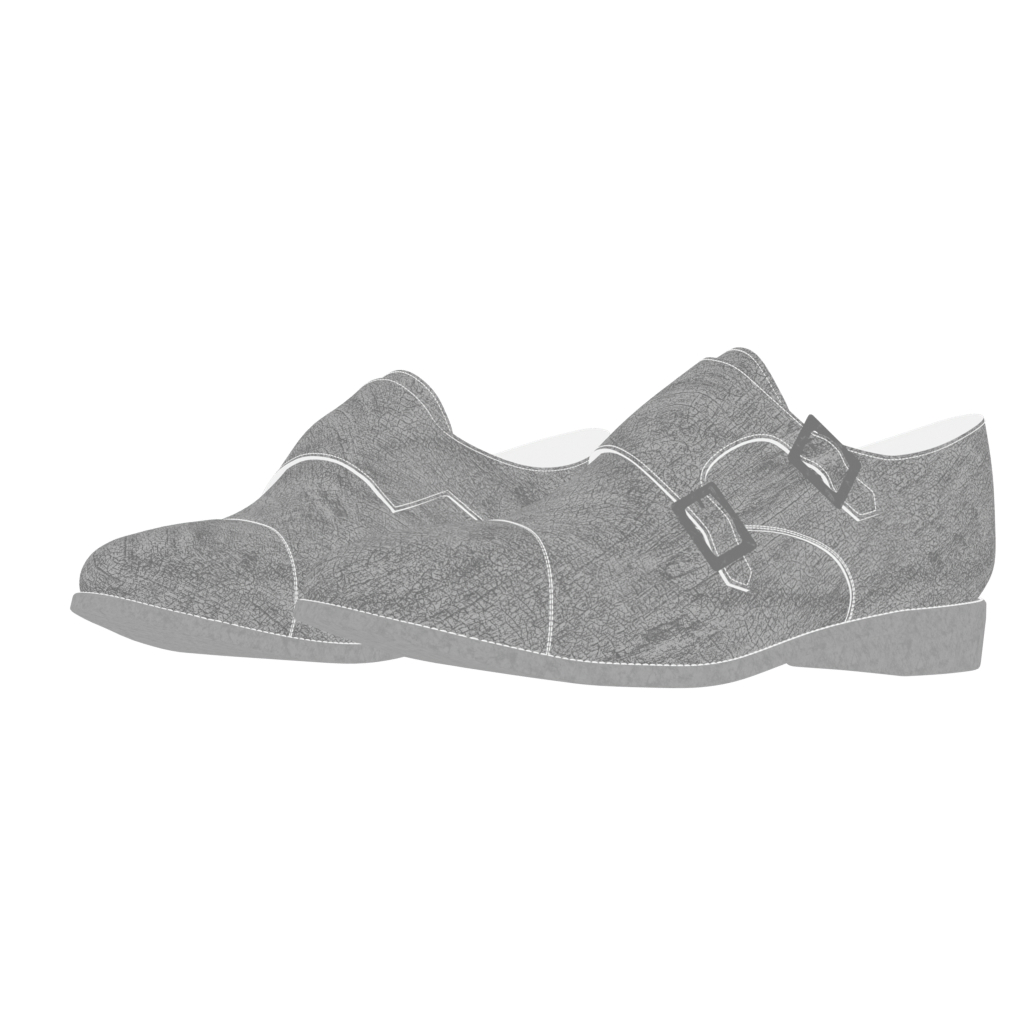}} &
	\raisebox{-0.5\height}{\includegraphics[width=0.093\textwidth]{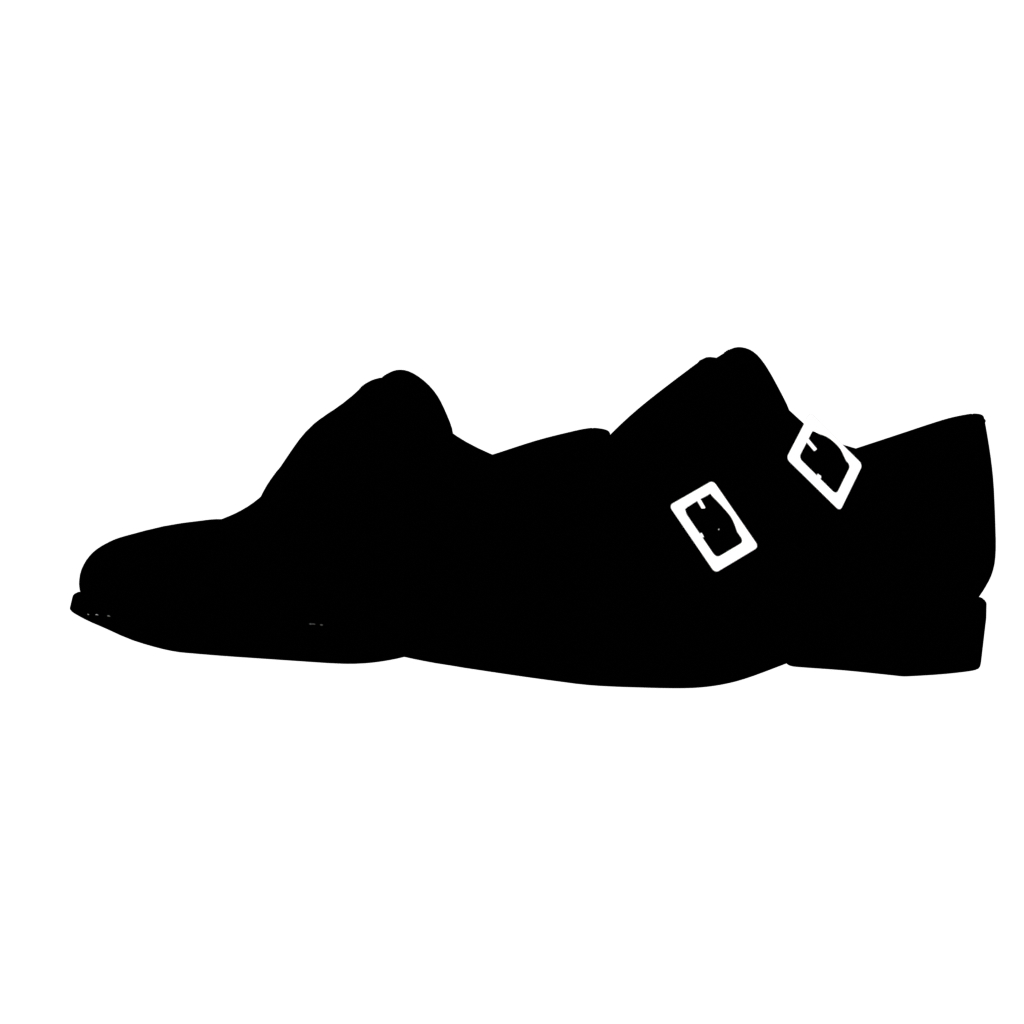}} &
    \rotatebox[origin=c]{90}{DreamMat} &
	\raisebox{-0.5\height}{\includegraphics[width=0.093\textwidth]{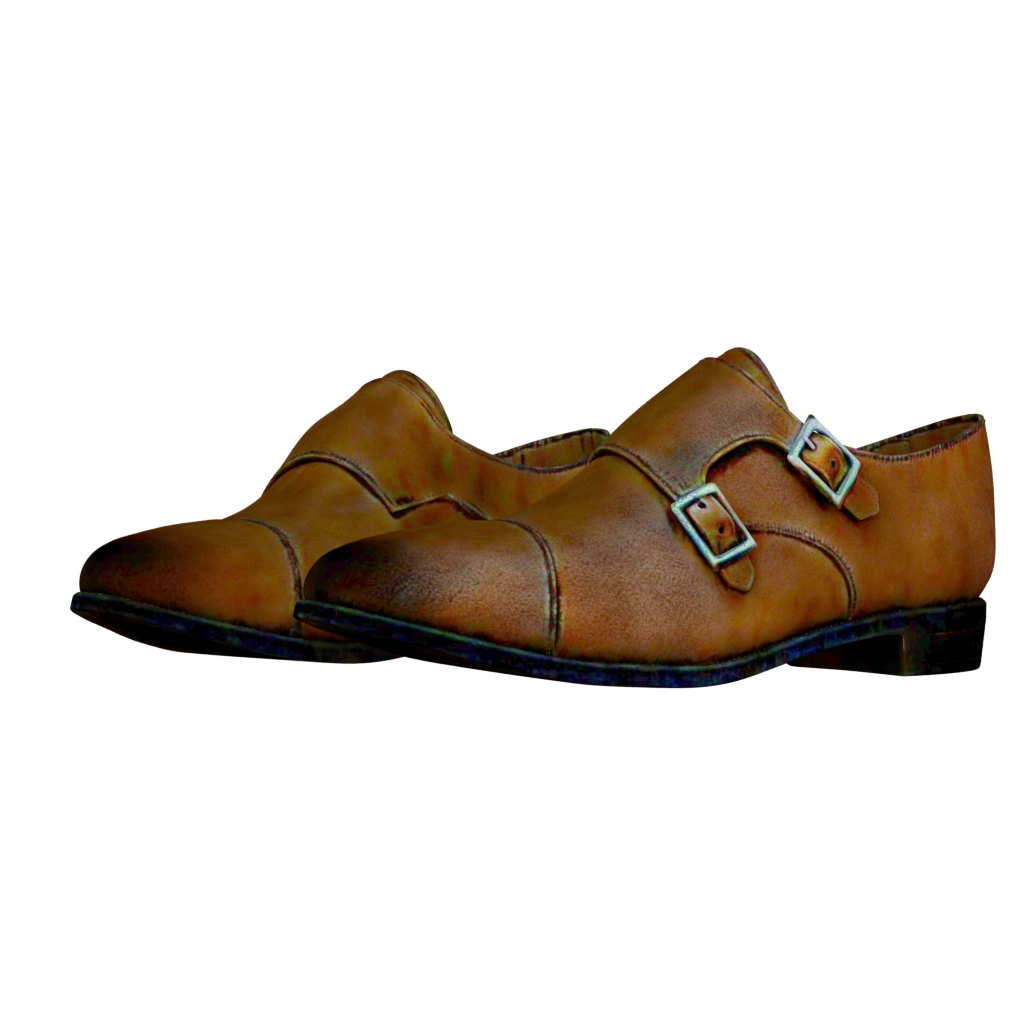}} &
	\raisebox{-0.5\height}{\includegraphics[width=0.093\textwidth]{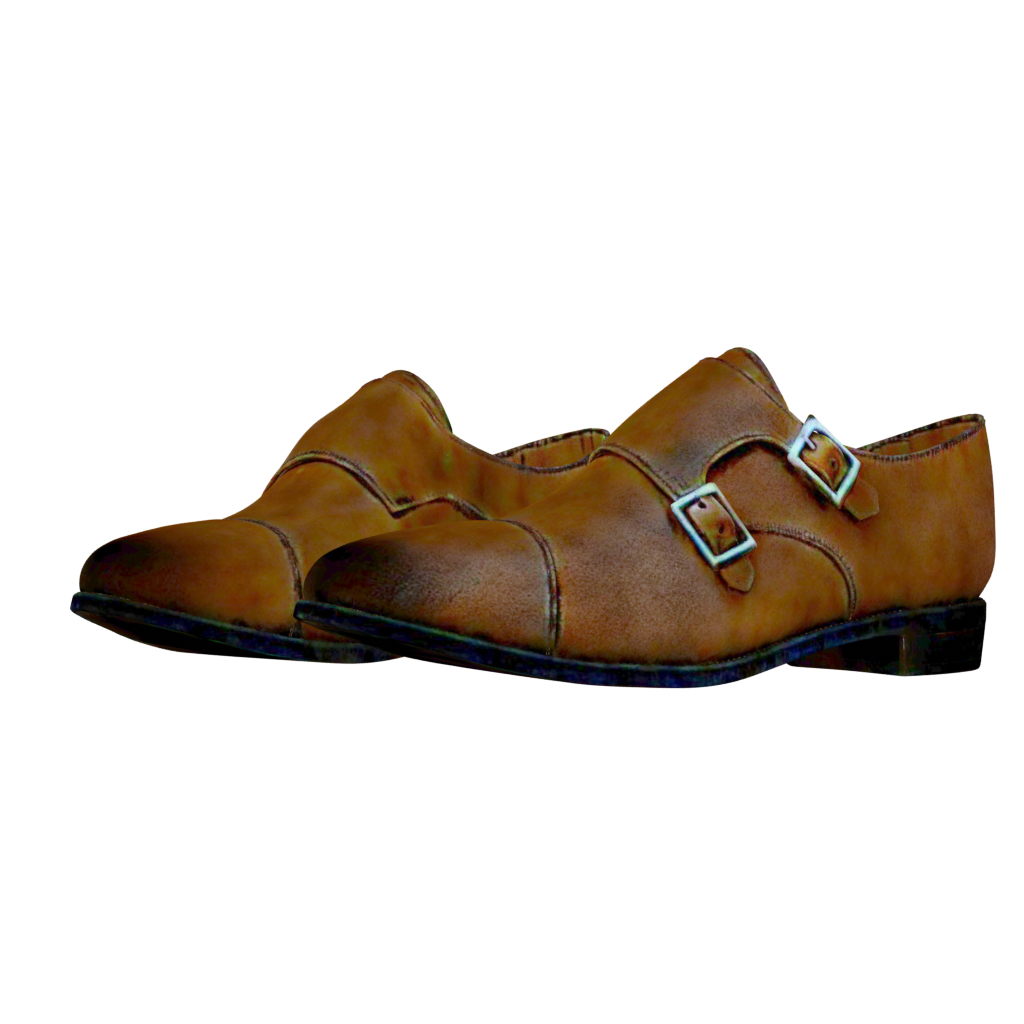}} &
	\raisebox{-0.5\height}{\includegraphics[width=0.093\textwidth]{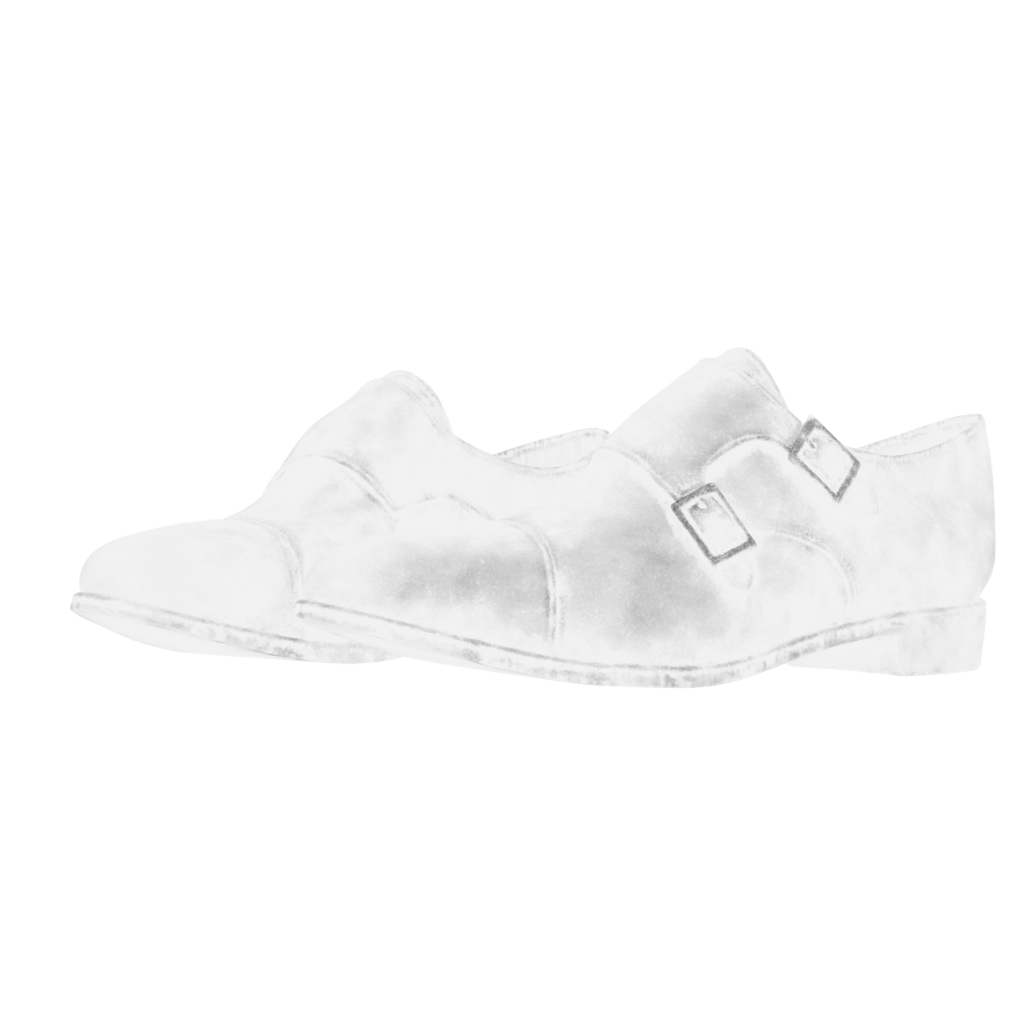}} &
	\raisebox{-0.5\height}{\includegraphics[width=0.093\textwidth]{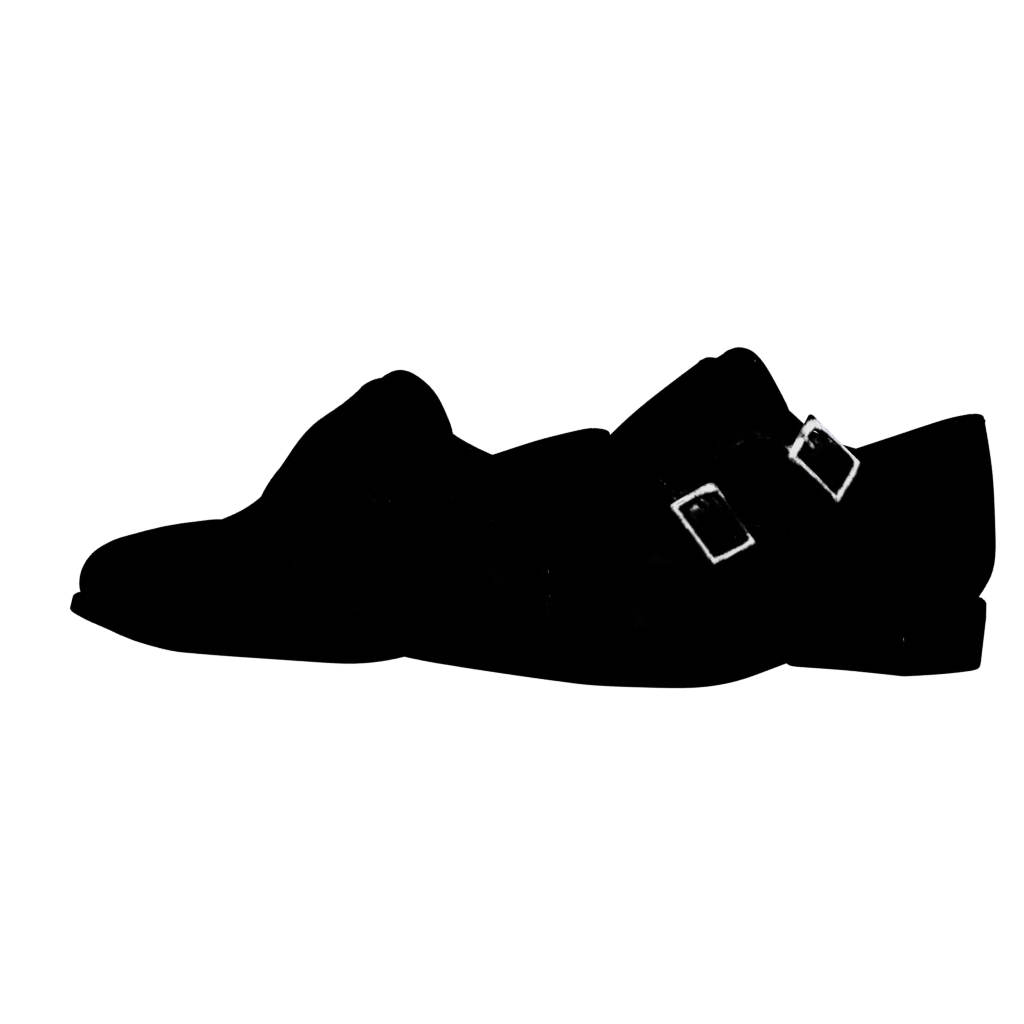}} \\

   	& Geometry & & Relit & Base color & Roughness & Metallic & & Relit & Base color & Roughness & Metallic
\end{tabular}
\vspace*{-2mm}
\caption{
	Text-to-material generation. We compare against Paint-it~\cite{youwang2024paintit} and DreamMat~\cite{Zhang2024dreammat} on three example meshes from the BlenderVault~\cite{litman2025materialfusion} dataset. We encourage the reader to compare the quality of the intrinsics (base color, roughness, metallic) to the reference. While significant deviations are expected in purely text-guided methods, we note that the base color predicted by our method is significantly more demodulated, or ``flat'' than the competing work. This is particularly noticeable for the \textsc{Gramophone} model. Our roughness and metallic guides are also more faithful to the reference, though with a slight bias towards lower roughness.
}
\label{fig:main_quality_results}
\end{figure*}
}

%%%%%%%%%%%%%%%%%%%%%%%%%%%%%%%%%%
% Relighting
%%%%%%%%%%%%%%%%%%%%%%%%%%%%%%%%%%

\newcommand{\figRelighting}{
\begin{figure}
    \small
    \centering
    \includegraphics[width=0.99\columnwidth]{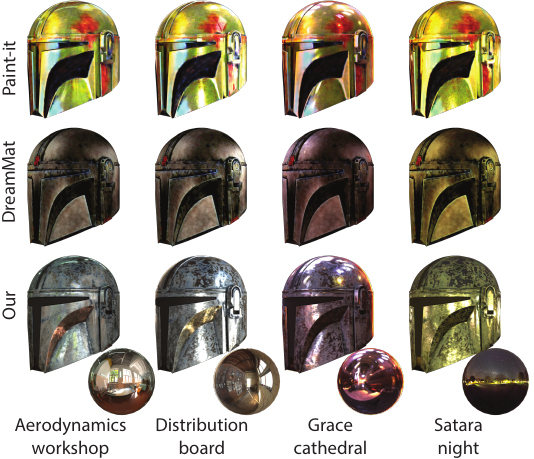}
    \vspace*{-2mm}
    \caption{
    	The \textsc{Helmet} model rendered with four light probes chosen to stress-test different lighting conditions. 
        Our result show accurate specular reflections. DreamMat reproduces the global color tone of the probe, 
        but the baked highlights are static and do not reflect the primary light direction.    	
    }
    \label{fig:relighting}
    \end{figure}
} 

\newcommand{\tinyrow}[2]{\rotatebox{90}{\centering\tiny #2}}

%%%%%%%%%%%%%%%%%%%%%%%%%%%%%%%%%%
% Variety
%%%%%%%%%%%%%%%%%%%%%%%%%%%%%%%%%%

\newcommand{\figVariety}{
\begin{figure}
    \centering
    \setlength{\tabcolsep}{1pt}
    \begin{tabular}{ccccc}
       \includegraphics[width=0.19\columnwidth]{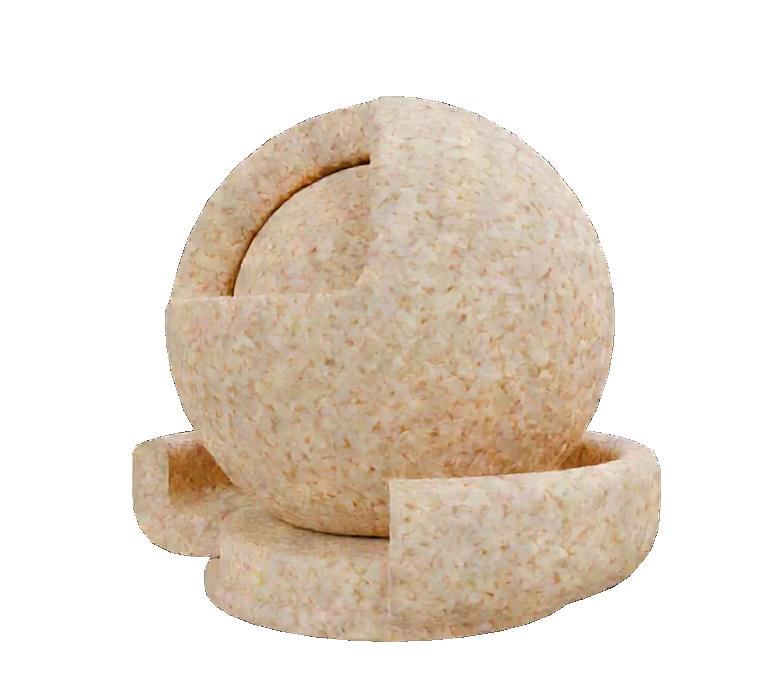} &
       \includegraphics[width=0.19\columnwidth]{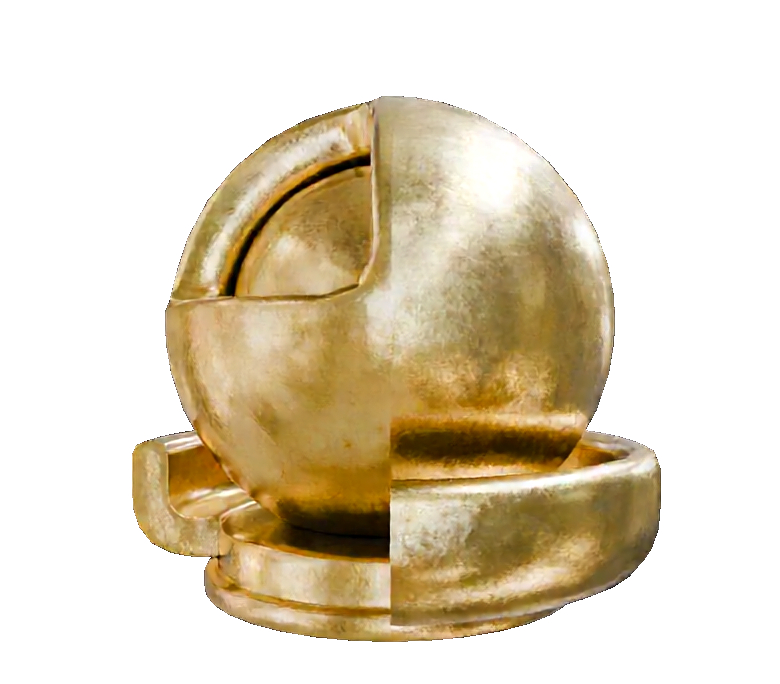} &
       \includegraphics[width=0.19\columnwidth]{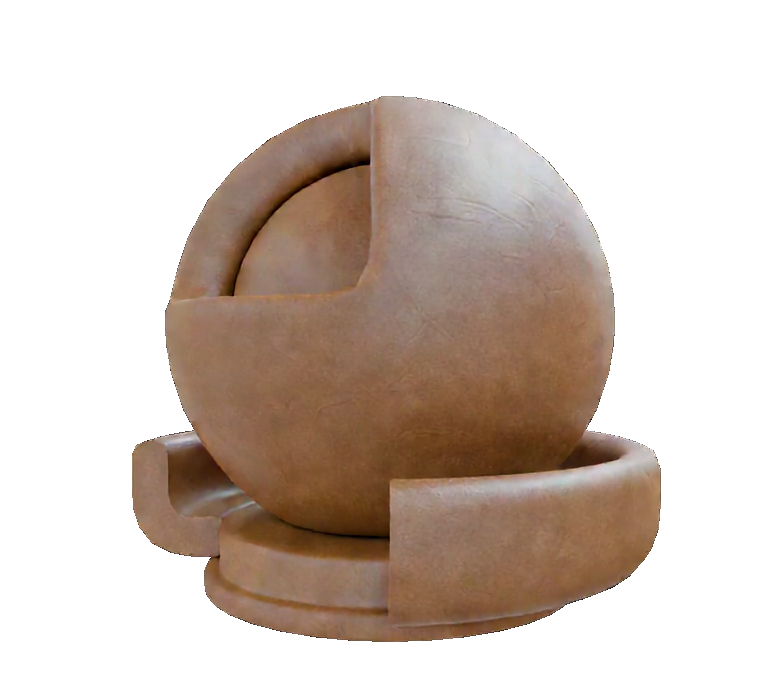} &
       \includegraphics[width=0.19\columnwidth]{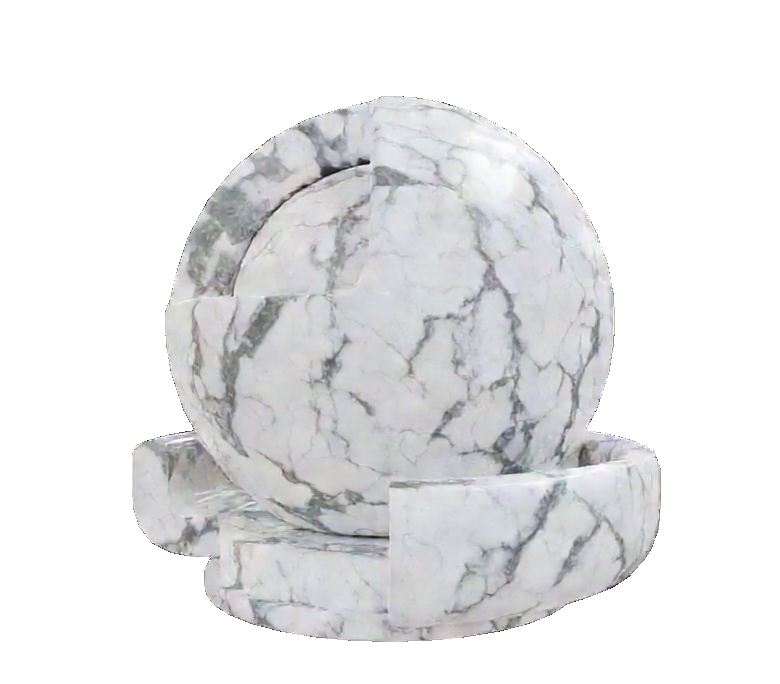} &
       \includegraphics[width=0.19\columnwidth]{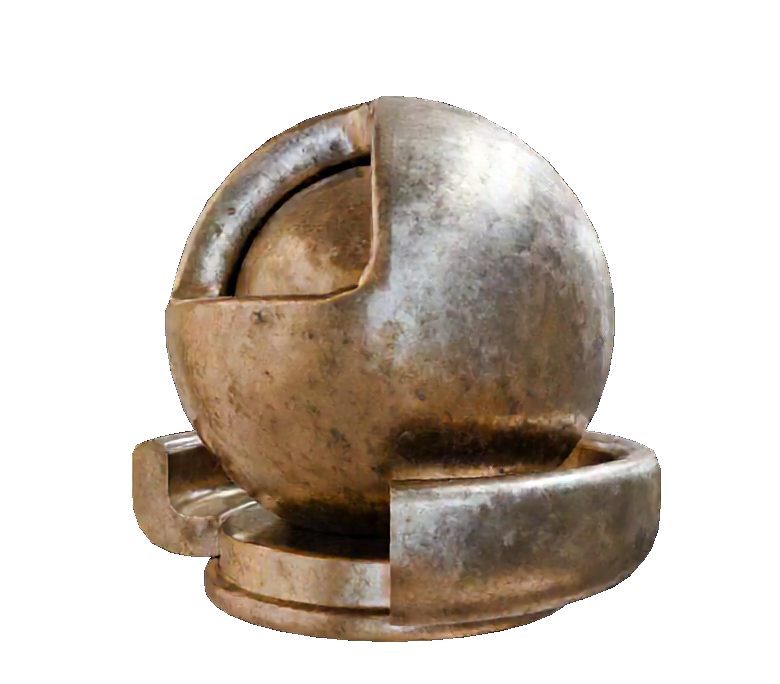} \\

       \includegraphics[width=0.19\columnwidth]{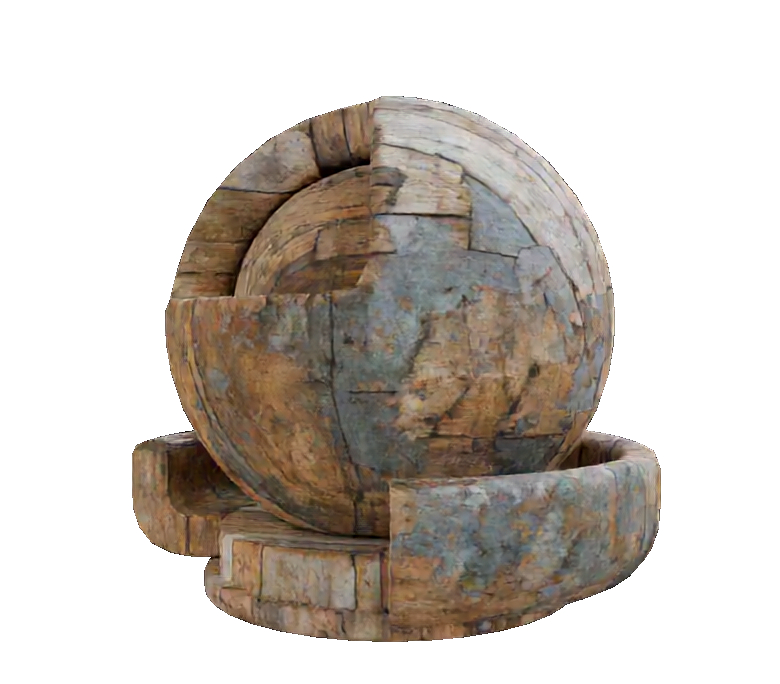} &
       \includegraphics[width=0.19\columnwidth]{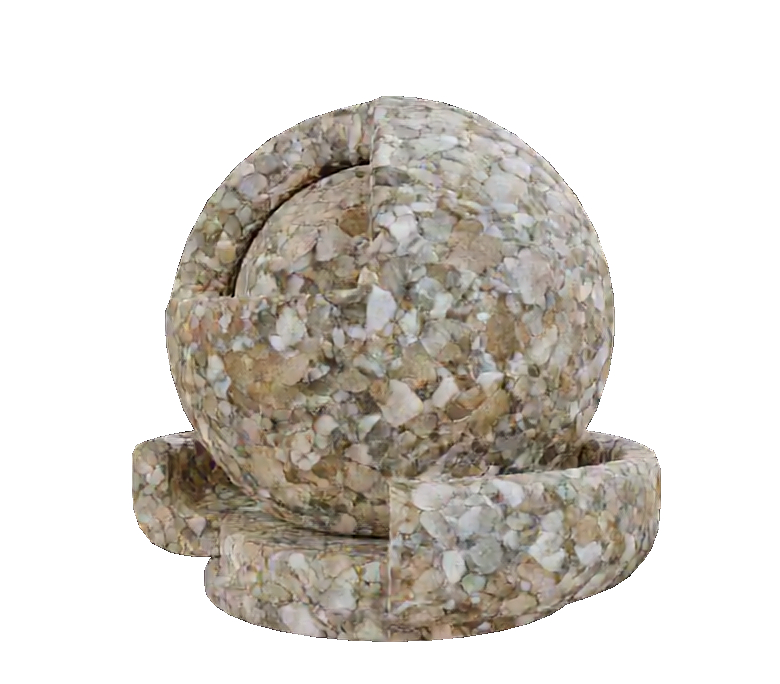} &
       \includegraphics[width=0.19\columnwidth]{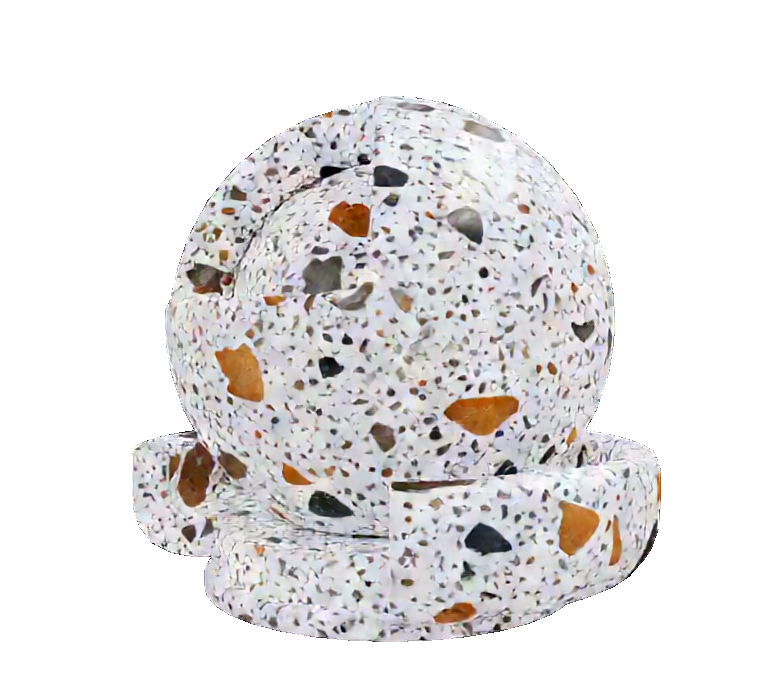} &
       \includegraphics[width=0.19\columnwidth]{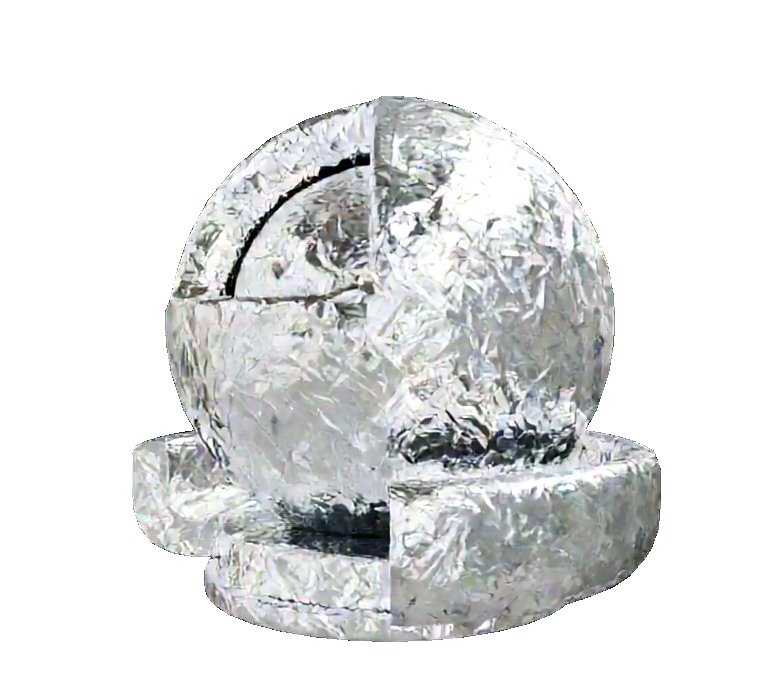} &
       \includegraphics[width=0.19\columnwidth]{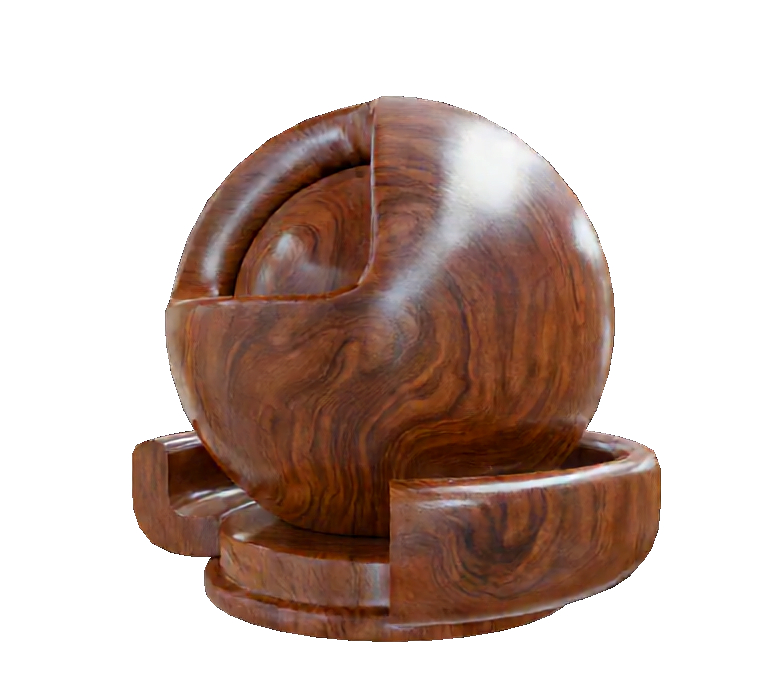} 
    \end{tabular}
    \vspace*{-2mm}
    \caption{Our finetuned video model generates view-consistent multi-view images of 
    diverse materials from text prompts, while closely respecting the input geometry and lighting.}
    \label{fig:variety}
 \end{figure}
}

%%%%%%%%%%%%%%%%%%%%%%%%%%%%%%%%%%
% Guides
%%%%%%%%%%%%%%%%%%%%%%%%%%%%%%%%%%

\newcommand{\figGuides}{
\begin{figure}
    \centering
    \includegraphics[width=1.0\columnwidth]{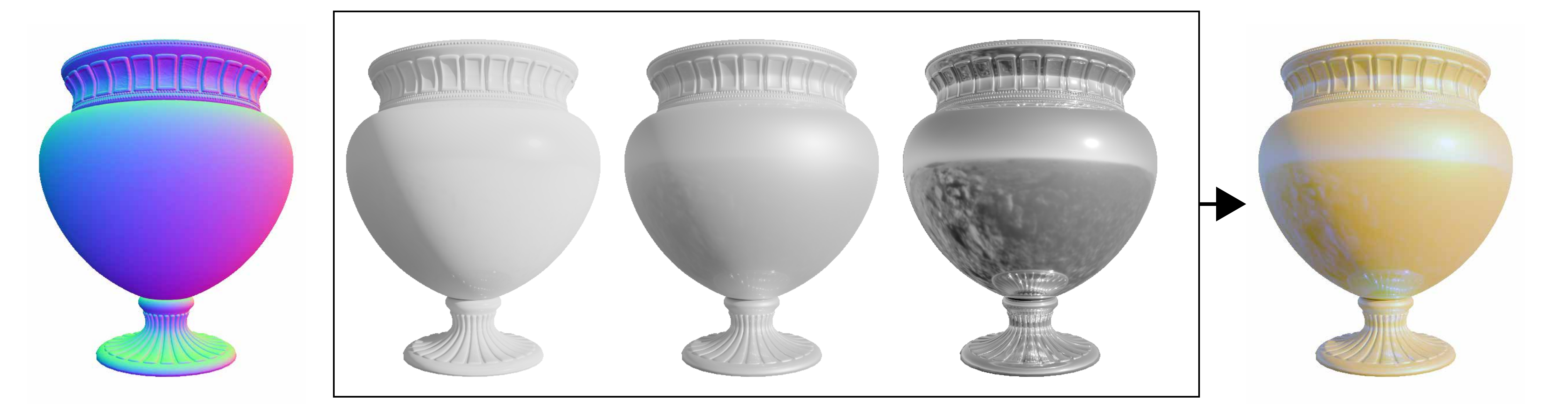}
    \setlength{\tabcolsep}{6pt}
    \begin{tabular}{ccccc}
    \small{Normal} & \small{$r\!\!=\!\!1,m\!\!=\!\!0$} & \small{$r\!\!=\!\!0.5,m\!\!=\!\!0.5$} & \small{$r\!\!=\!\!0,m\!\!=\!\!1$} & \small{Combined} \\   
    \end{tabular}
    \vspace*{-2mm}
    \caption{We condition the video model with normal maps and 
    three shading conditions, all with a uniform base color $k_d = (0.7,0.7,0.7)$
    and varying roughness ($r$) and metallic ($m$) parameters. The shading conditions are combined into an RGB image.}
    \label{fig:guides}
 \end{figure}
}

%%%%%%%%%%%%%%%%%%%%%%%%%%%%%%%%%%
% Impact of regularizers
%%%%%%%%%%%%%%%%%%%%%%%%%%%%%%%%%%

\newcommand{\figRegularizers}{
	\begin{figure}
		\centering
		\small
		\setlength{\tabcolsep}{1pt}
		\def\arraystretch{1.0}
	    \begin{tabular}{ccccc}
			& \multicolumn{2}{c}{Relit image} & \multicolumn{2}{c}{Base color}\\
			
			\rotatebox[origin=c]{90}{\textsc{Shed}} &
			\raisebox{-0.5\height}{\makecropbox{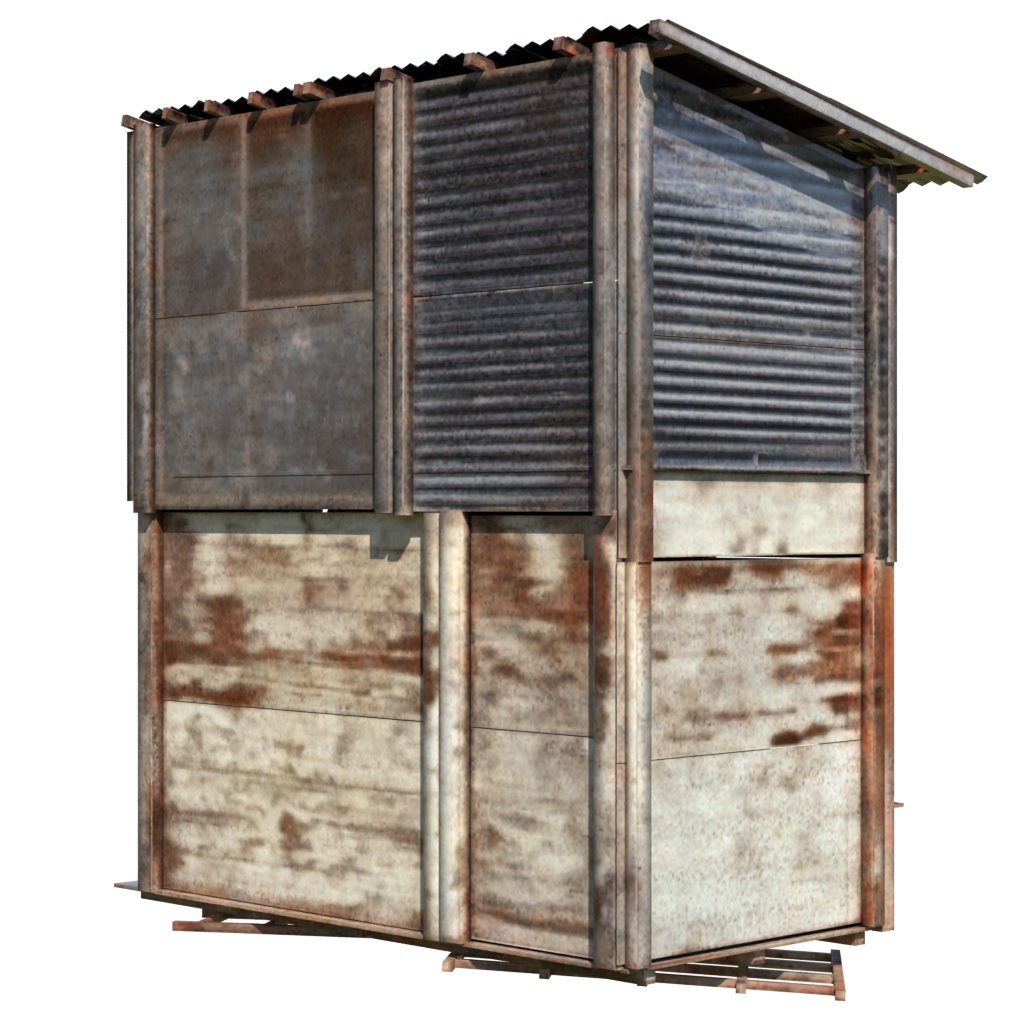}{0.23\columnwidth}{0.52}{0.52}{1.0}{1.0}} &
			\raisebox{-0.5\height}{\includegraphics[height=0.23\columnwidth]{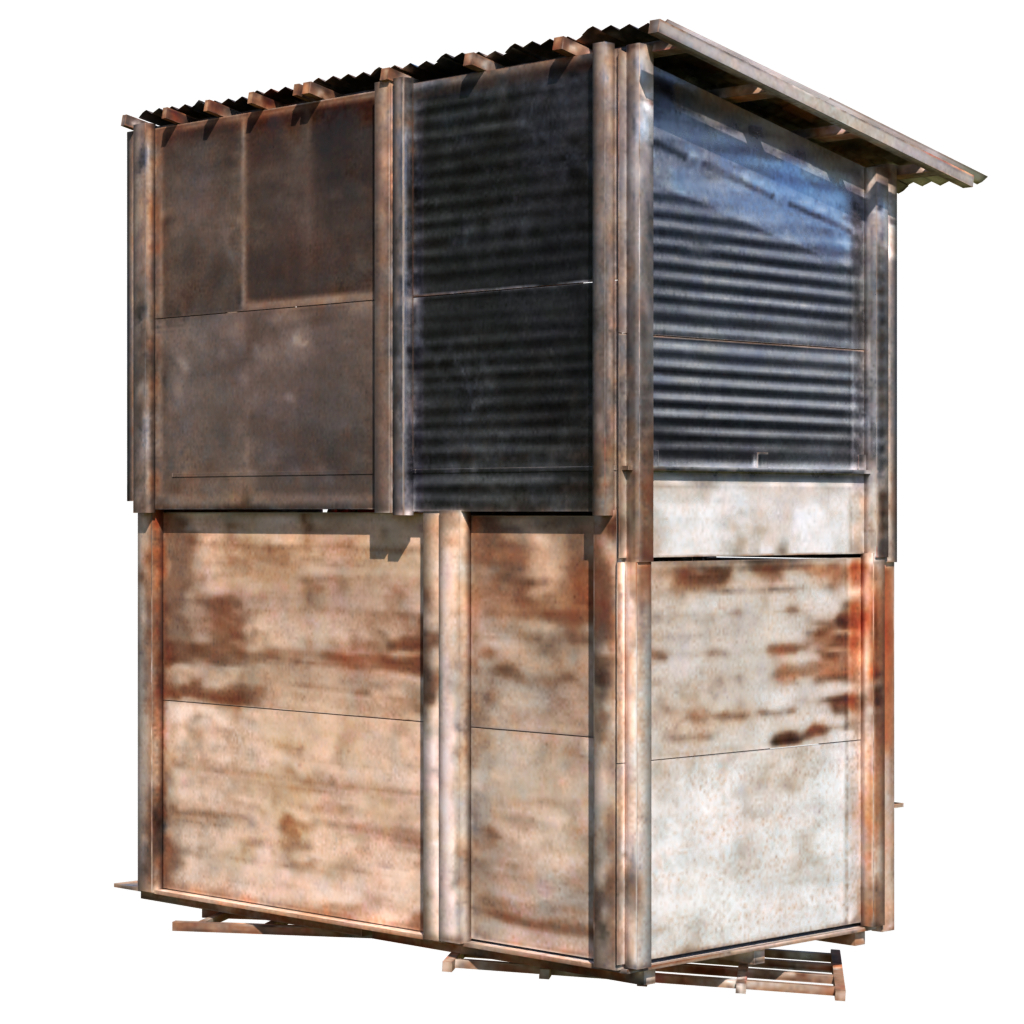}} &
			\raisebox{-0.5\height}{\includegraphics[height=0.23\columnwidth]{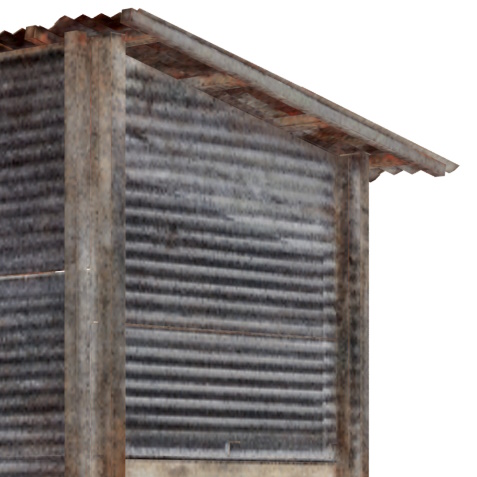}} &
			\raisebox{-0.5\height}{\includegraphics[height=0.23\columnwidth]{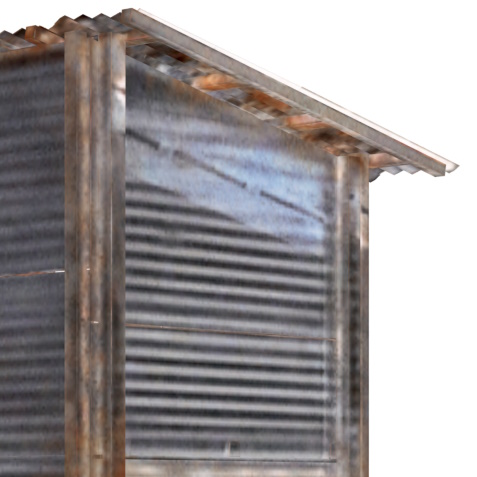}} \\
            \\
			\rotatebox[origin=c]{90}{\textsc{Gramophone}} &			
			\raisebox{-0.5\height}{\makecropbox{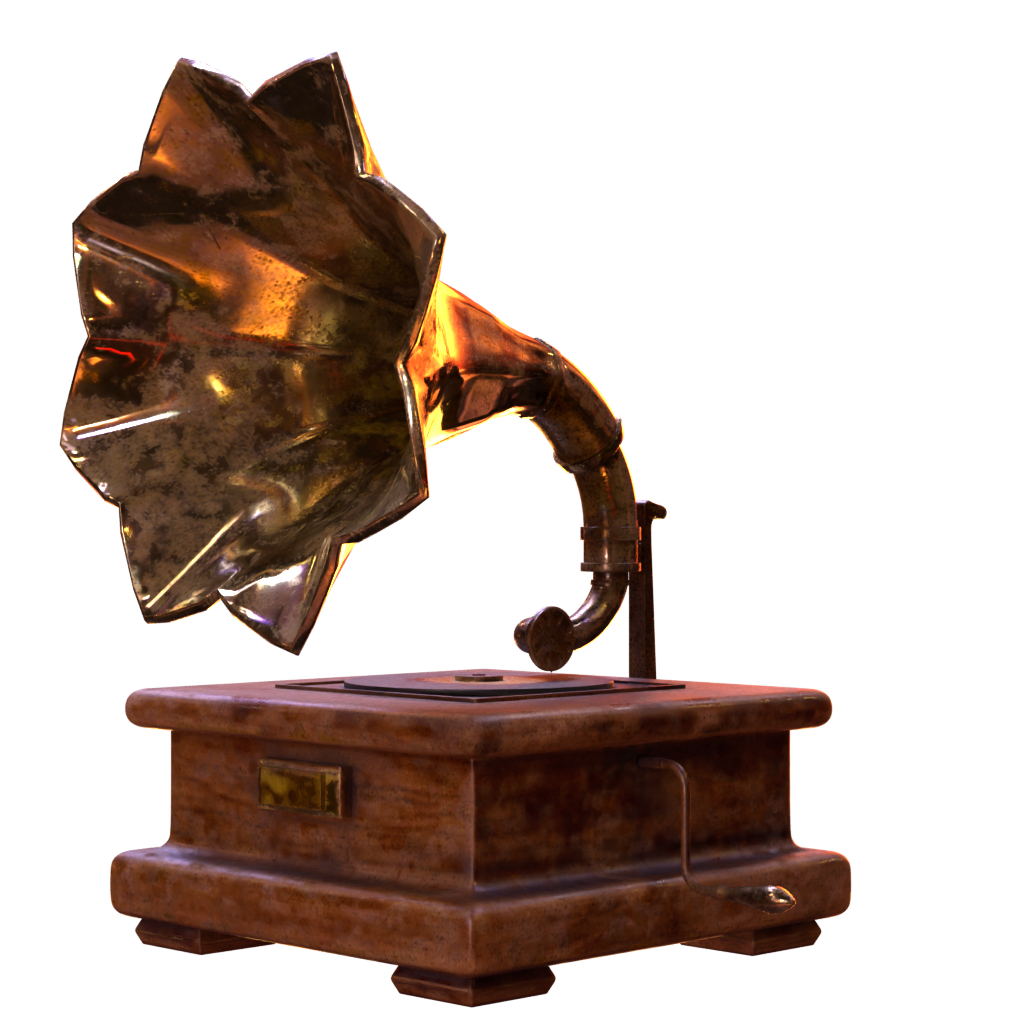}{0.23\columnwidth}{0.05}{0.45}{0.50}{0.90}} &
			\raisebox{-0.5\height}{\includegraphics[height=0.23\columnwidth]{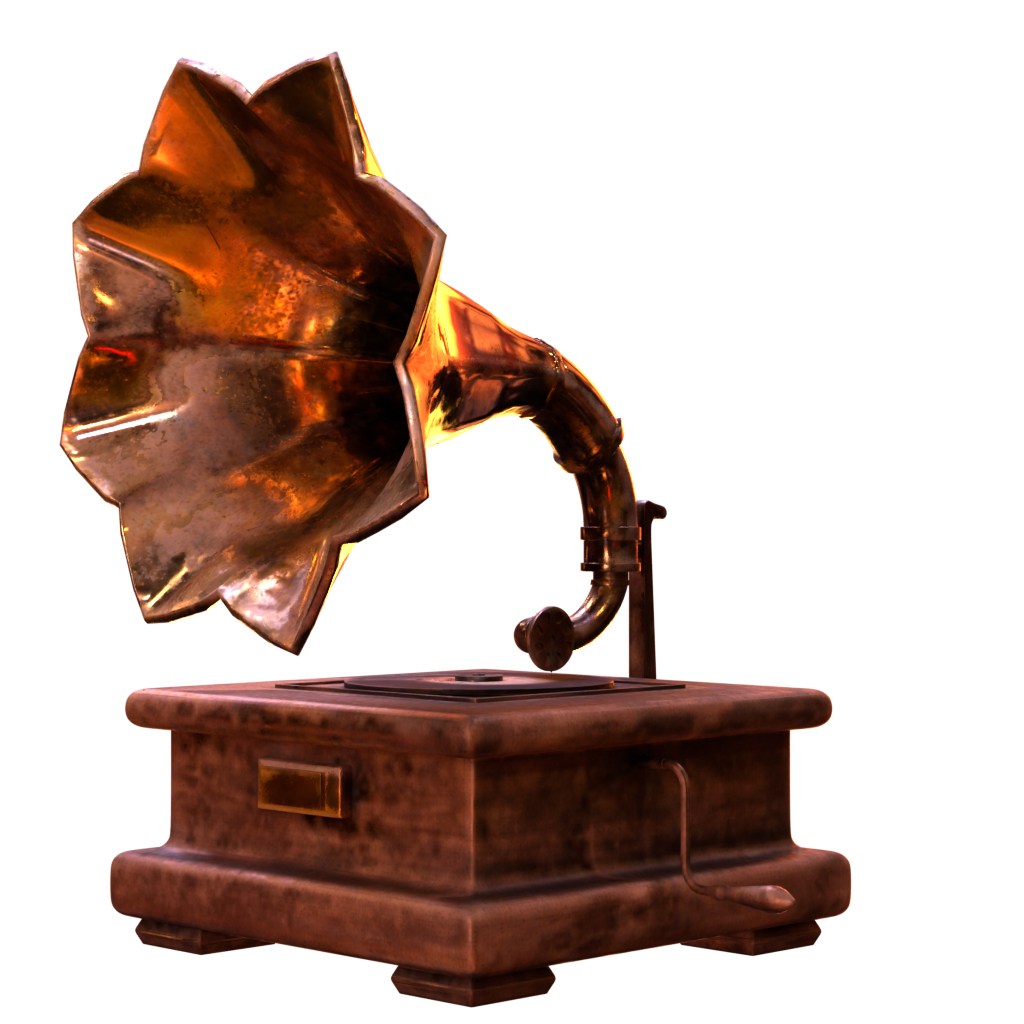}} &
			\raisebox{-0.5\height}{\includegraphics[height=0.23\columnwidth]{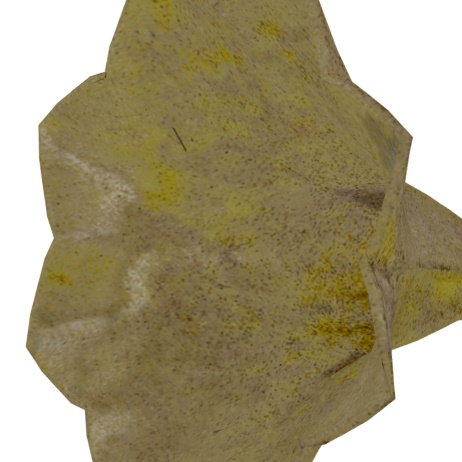}} &
			\raisebox{-0.5\height}{\includegraphics[height=0.23\columnwidth]{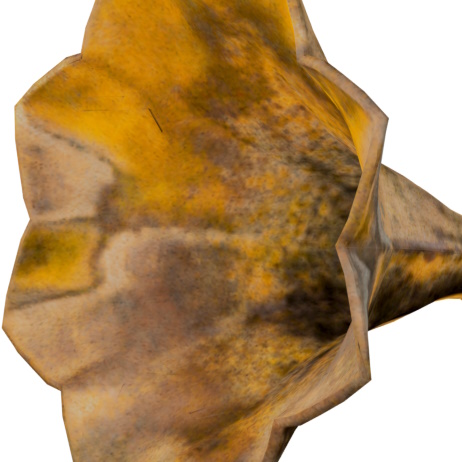}} \\			
			& Our & w/o regularizer & Our & w/o regularizer \\
		\end{tabular}
		\caption{The impact of using intrinsic guides for regularization. Differentiable rendering sometimes struggles with 
        disentangling lighting from materials. Without regularizer, the strong shadow from the roof is not demodulated in the 
        \textsc{Shed} scene. Similarly, the \textsc{Gramophone} example exhibits clear base color darkening in the horn, which 
        creates an overly shadowed region when lit.}
		\label{fig:regularizer_ablation}
	\end{figure}
}

%%%%%%%%%%%%%%%%%%%%%%%%%%%%%%%%%%
% Warp
%%%%%%%%%%%%%%%%%%%%%%%%%%%%%%%%%%

\newcommand{\figWarp}{
\begin{figure}
    \centering
    \setlength{\tabcolsep}{1pt}
    \begin{tabular}{lcccc}
       \tinyrow{0.3}{Warped} &
       \includegraphics[width=0.23\columnwidth]{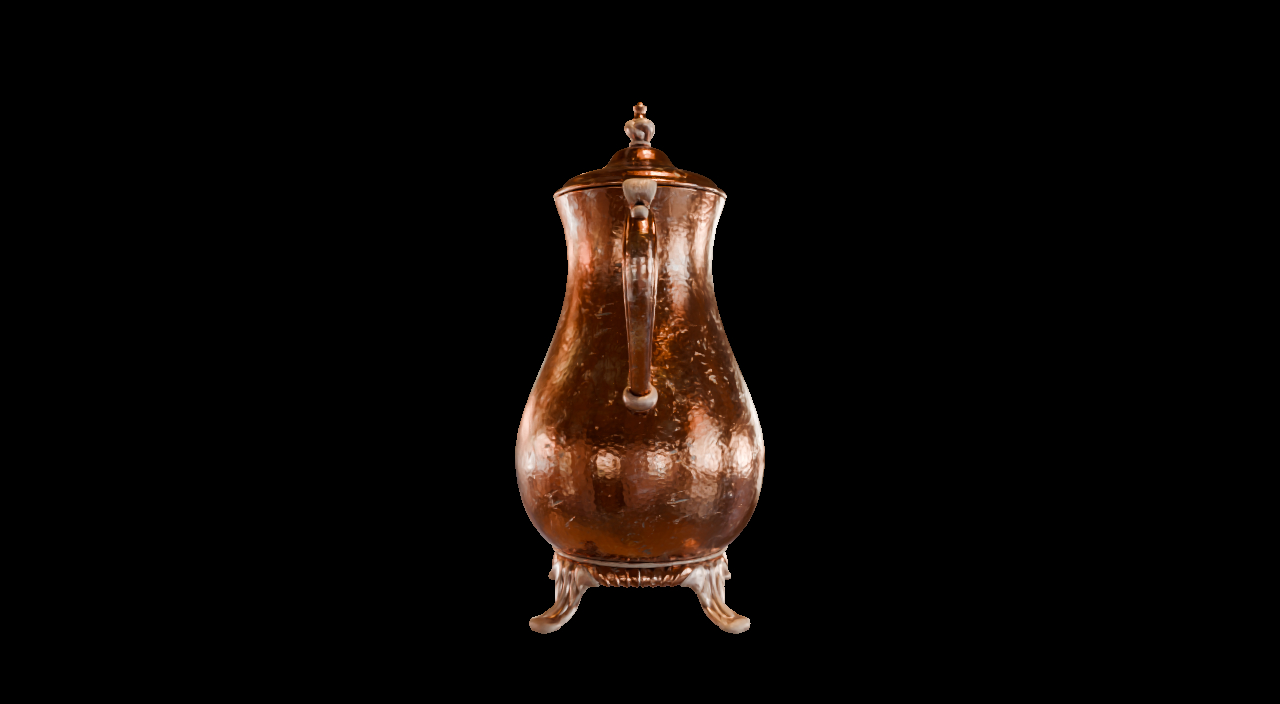} &
       \includegraphics[width=0.23\columnwidth]{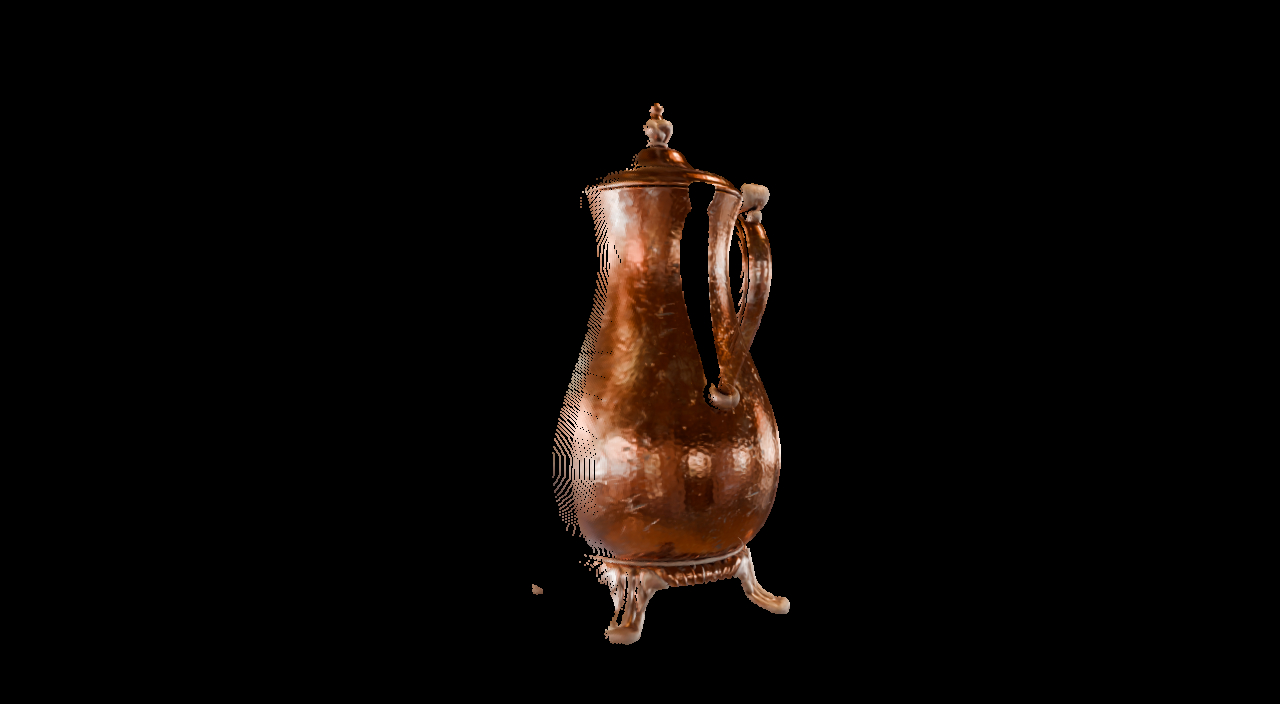} &
       \includegraphics[width=0.23\columnwidth]{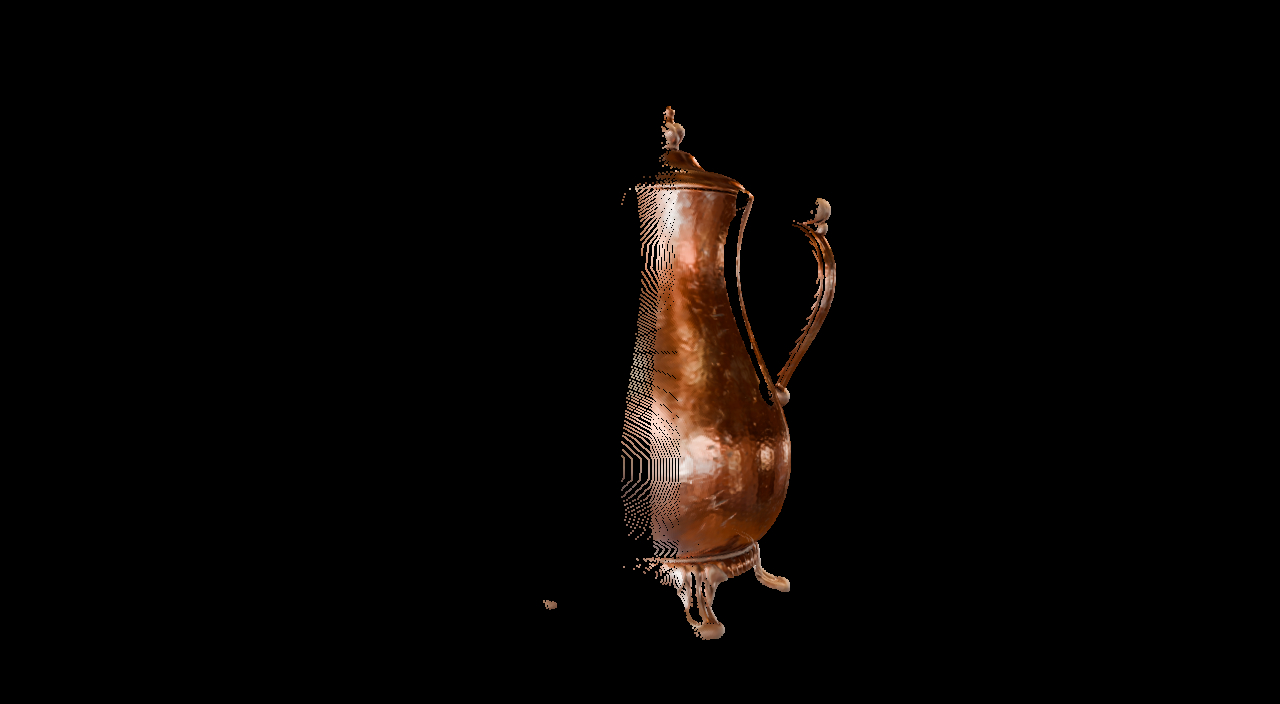} &
       \includegraphics[width=0.23\columnwidth]{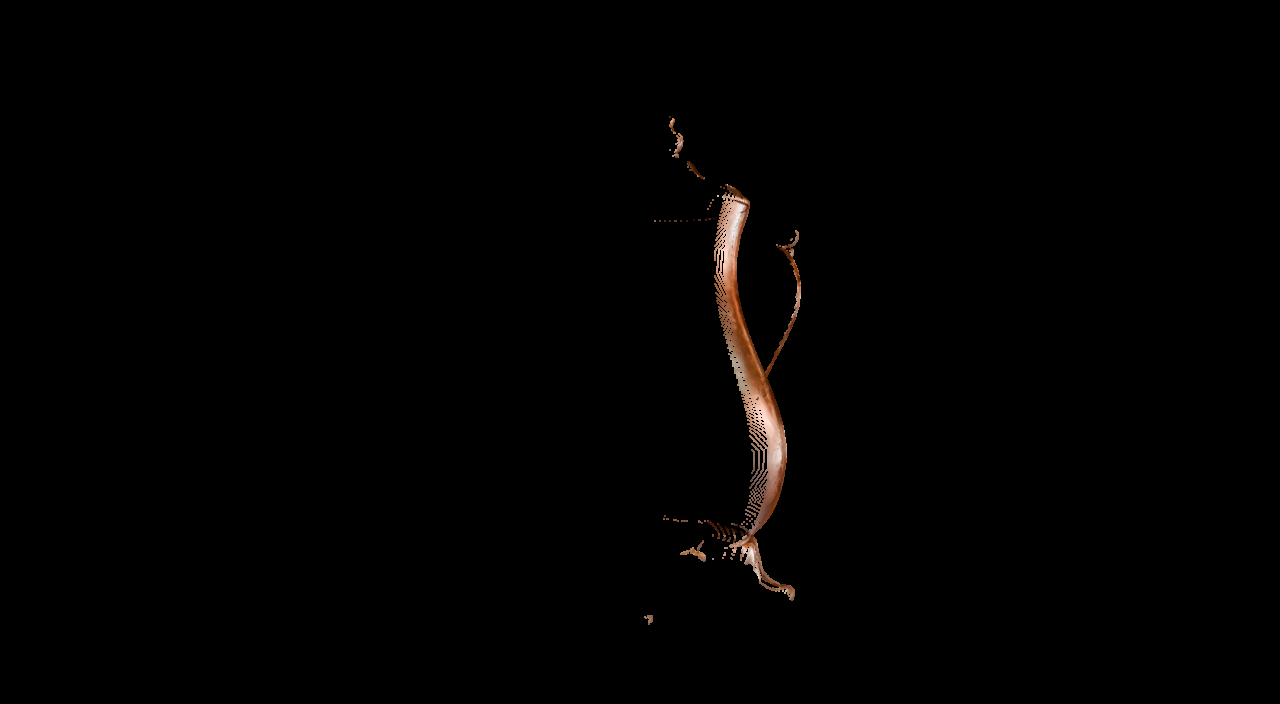} \\
       
       \tinyrow{0.3}{Output} &
       \includegraphics[width=0.23\columnwidth]{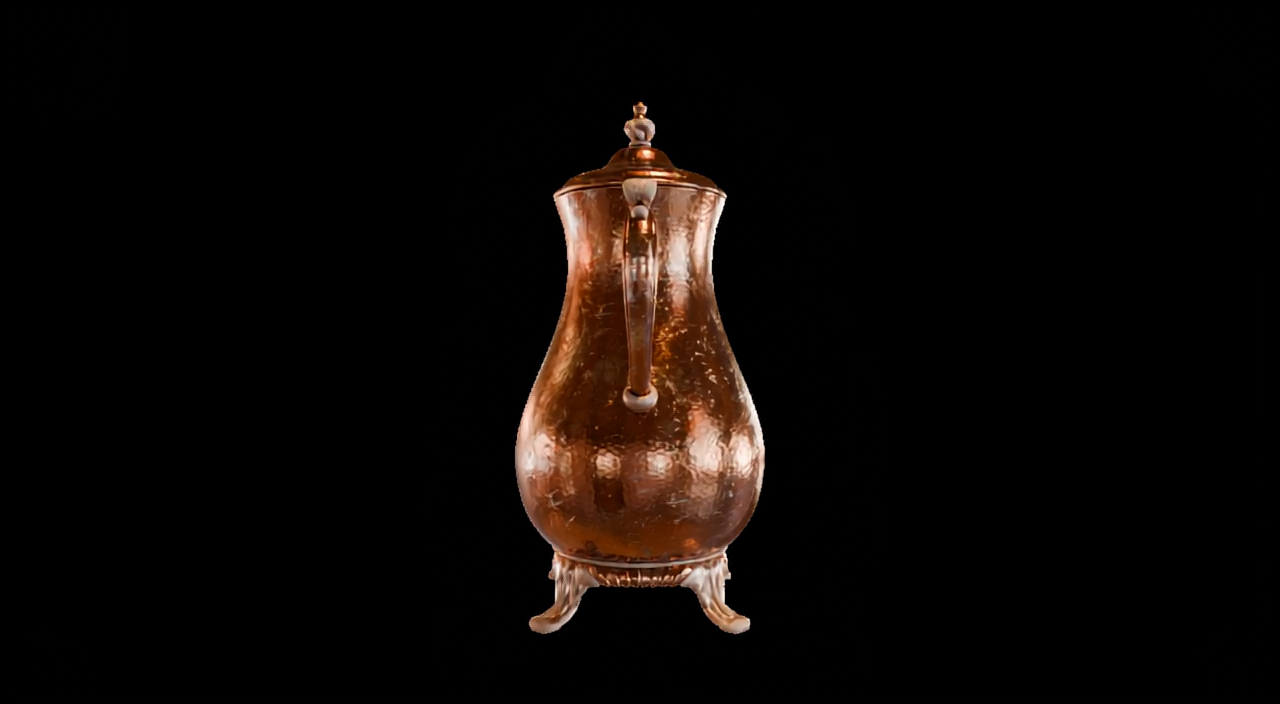} &
       \includegraphics[width=0.23\columnwidth]{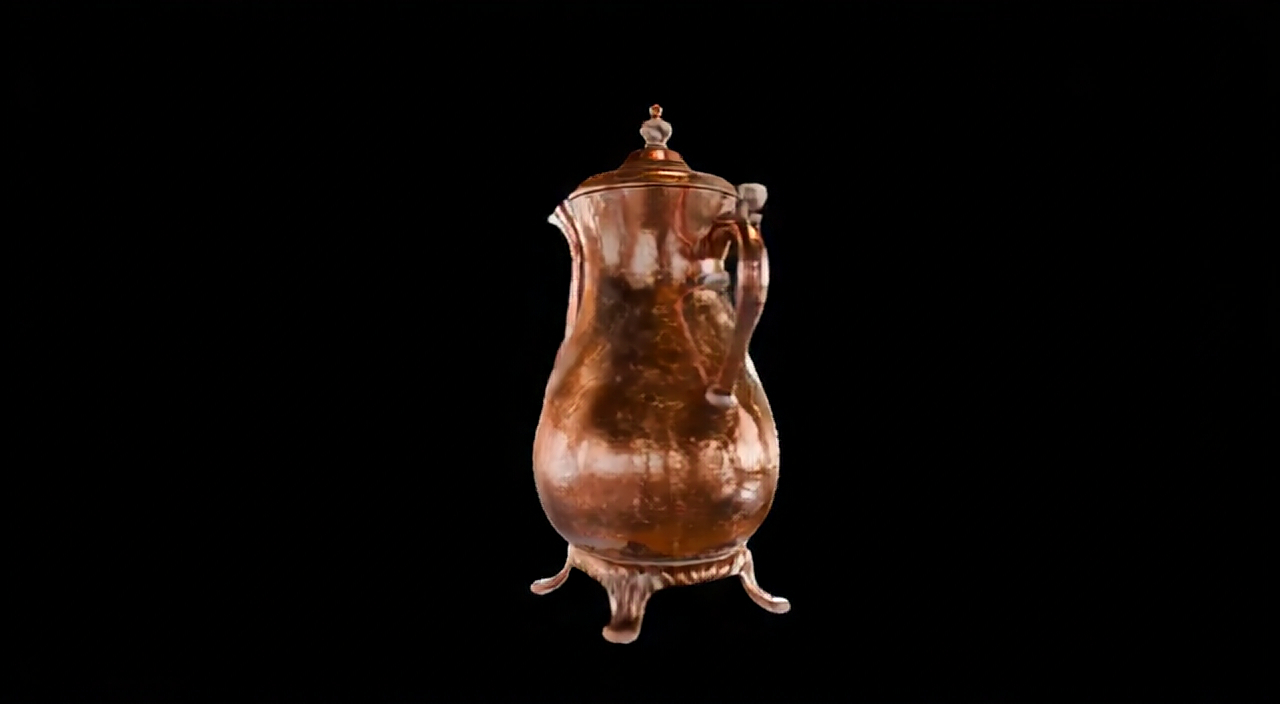} &
       \includegraphics[width=0.23\columnwidth]{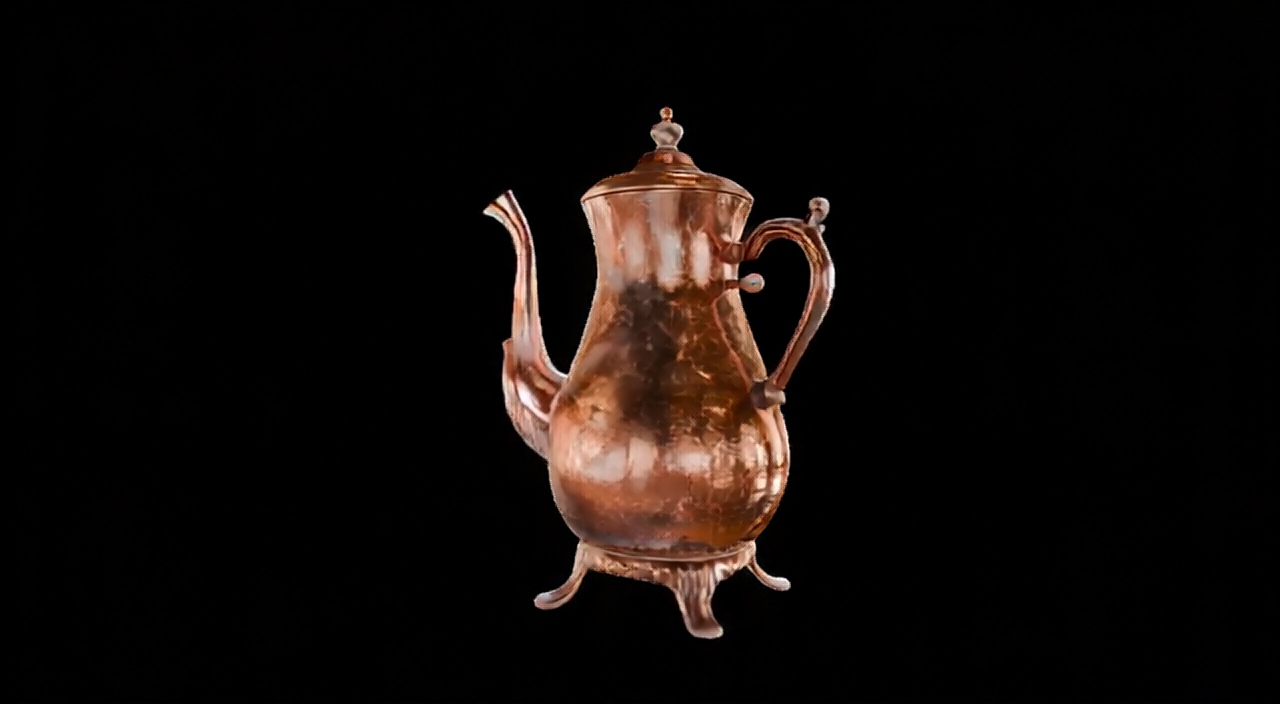} &
       \includegraphics[width=0.23\columnwidth]{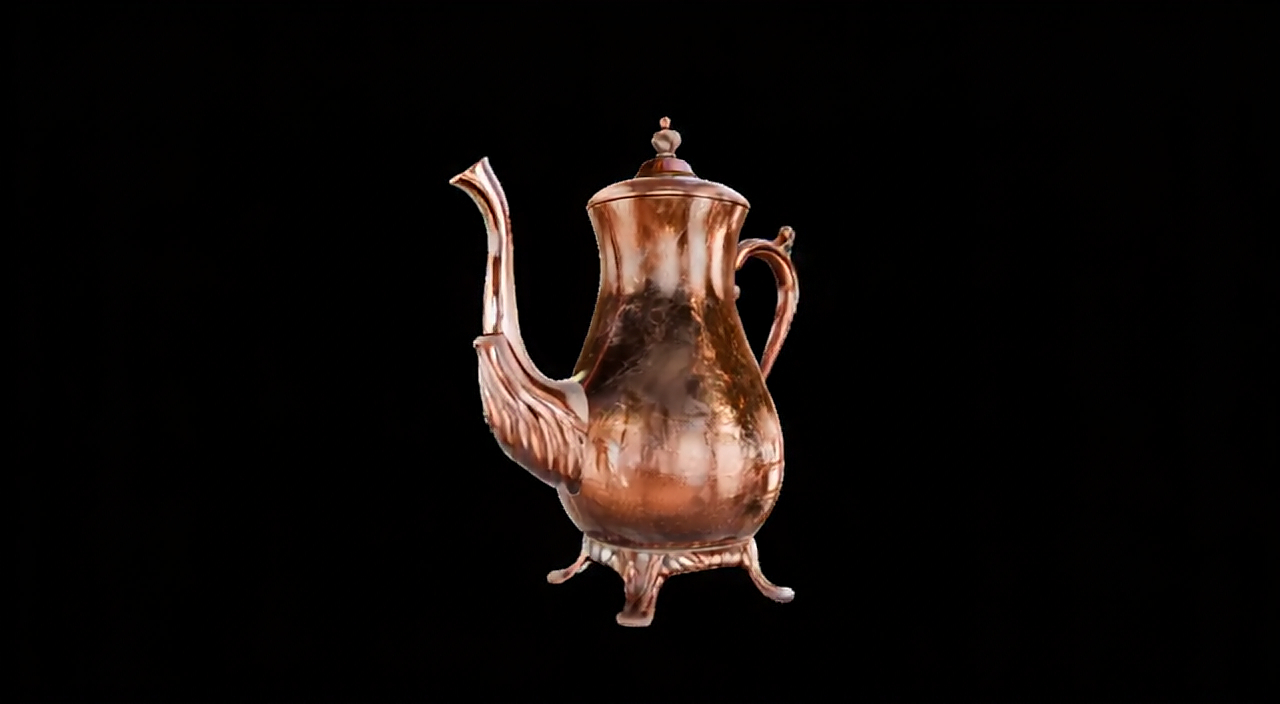} \\
       
       \tinyrow{0.3}{Reference} &
       \includegraphics[width=0.23\columnwidth]{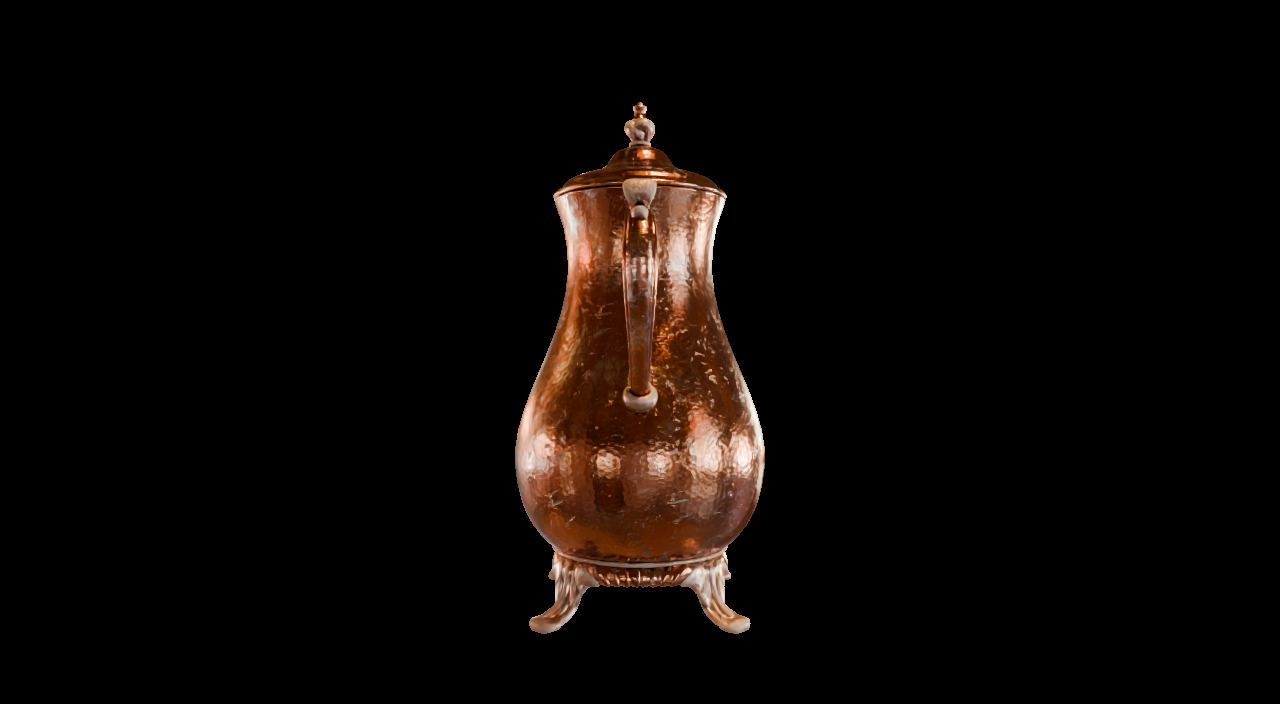} &
       \includegraphics[width=0.23\columnwidth]{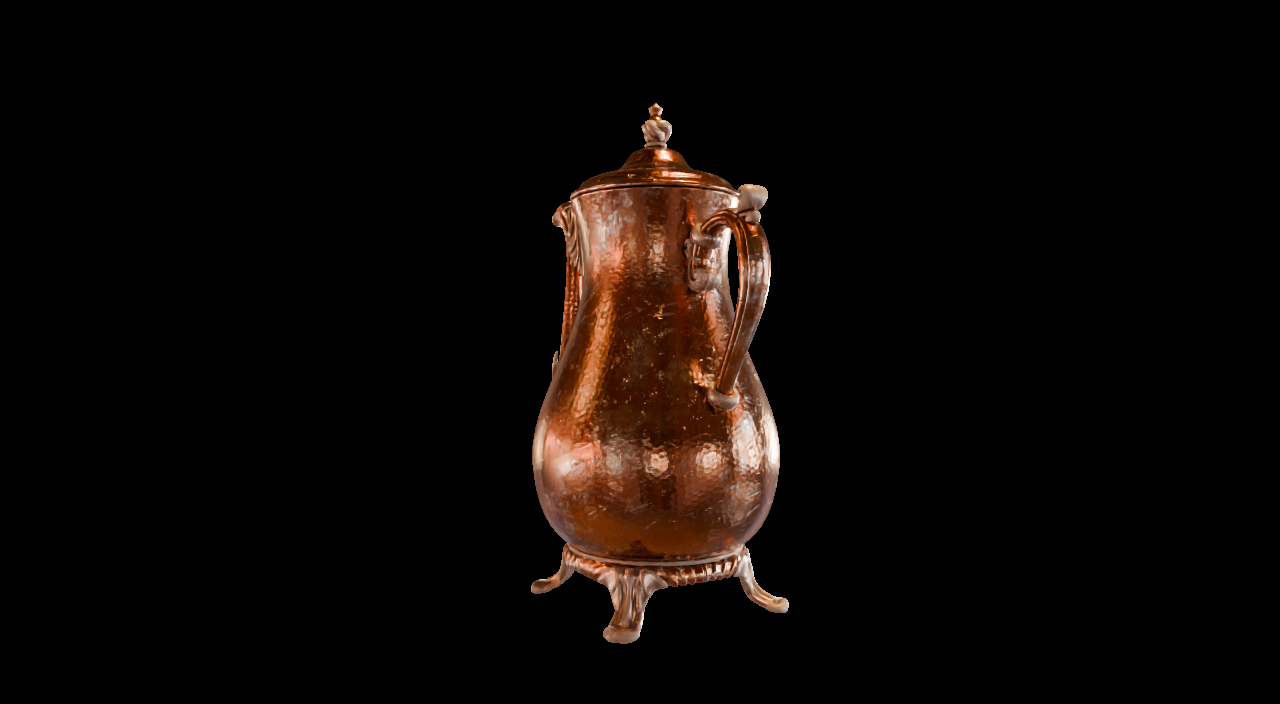} &
       \includegraphics[width=0.23\columnwidth]{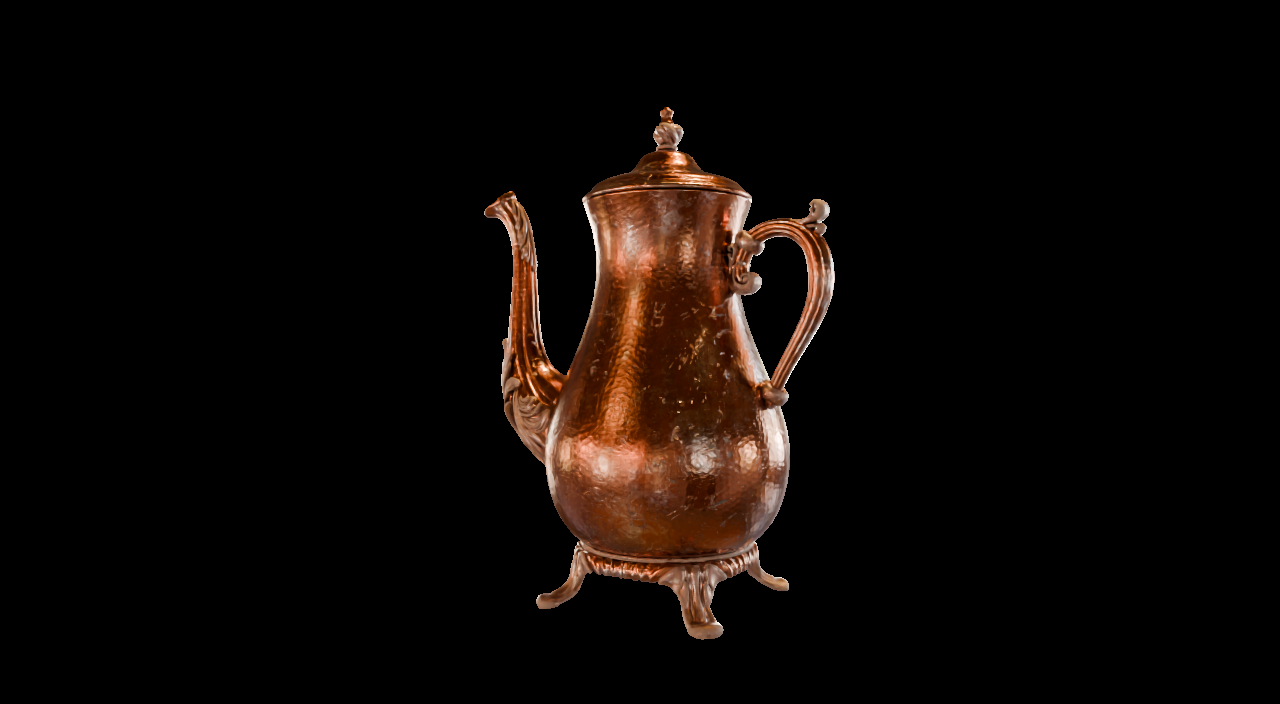} &
       \includegraphics[width=0.23\columnwidth]{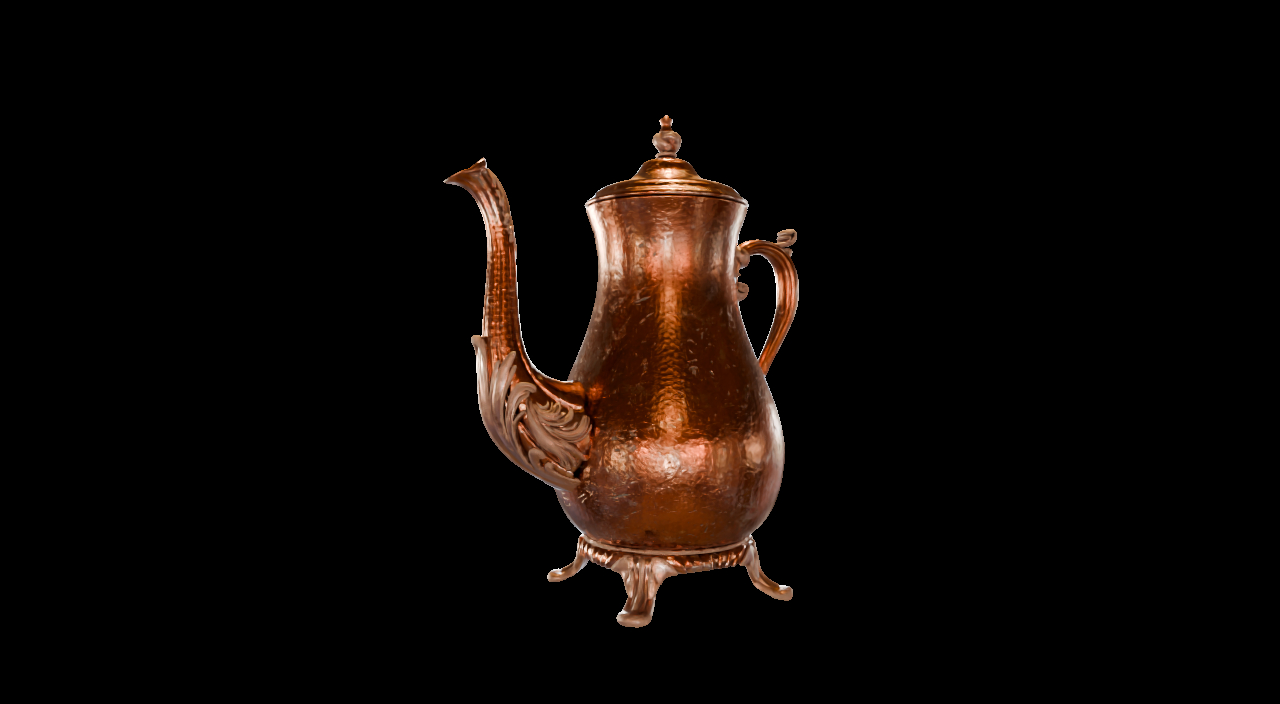} \\
              
       & \small{Frame~0} & \small{Frame~10} & \small{Frame~20} & \small{Frame~40} 

    \end{tabular}
    \vspace*{-2mm}
    \caption{Our image-to-material pipeline is conditioned on a single image,
    warped using a depth map to align with the current camera pose. The warping introduces disocclusion artifacts and 
    incorrectly attaches specular highlights
    to the surface. The video model successfully inpaints the missing regions and generates plausible specular highlights}
    \label{fig:warp}
 \end{figure}
}

%%%%%%%%%%%%%%%%%%%%%%%%%%%%%%%%%%
% Image-2-video model
%%%%%%%%%%%%%%%%%%%%%%%%%%%%%%%%%%

\newcommand{\figItoV}{
	\begin{figure}
		\centering
		\small
		\setlength{\tabcolsep}{1pt}
		\def\arraystretch{1.0}
	    \begin{tabular}{cccc}		
            \rotatebox[origin=c]{90}{\textsc{Shed}} &			
			\raisebox{-0.5\height}{\includegraphics[height=0.32\columnwidth]{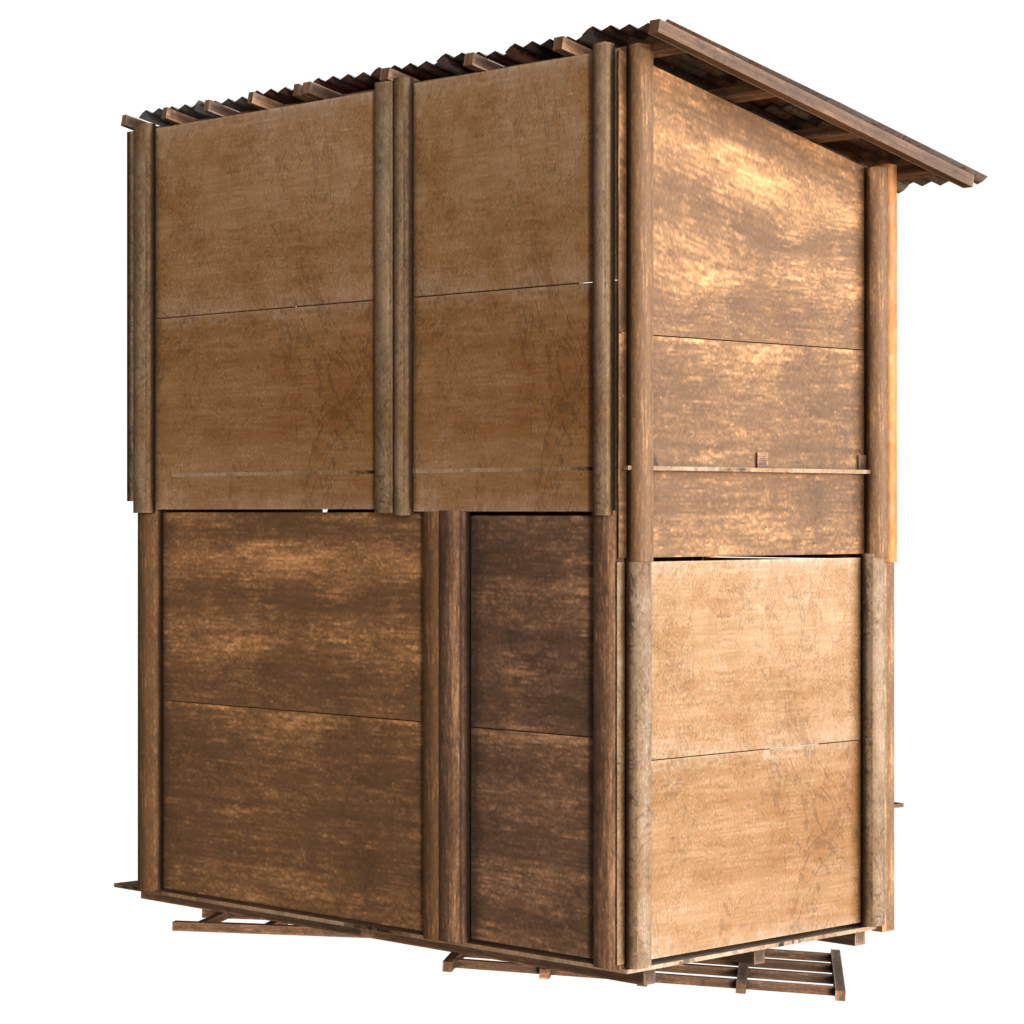}} &
			\raisebox{-0.5\height}{\includegraphics[height=0.32\columnwidth]{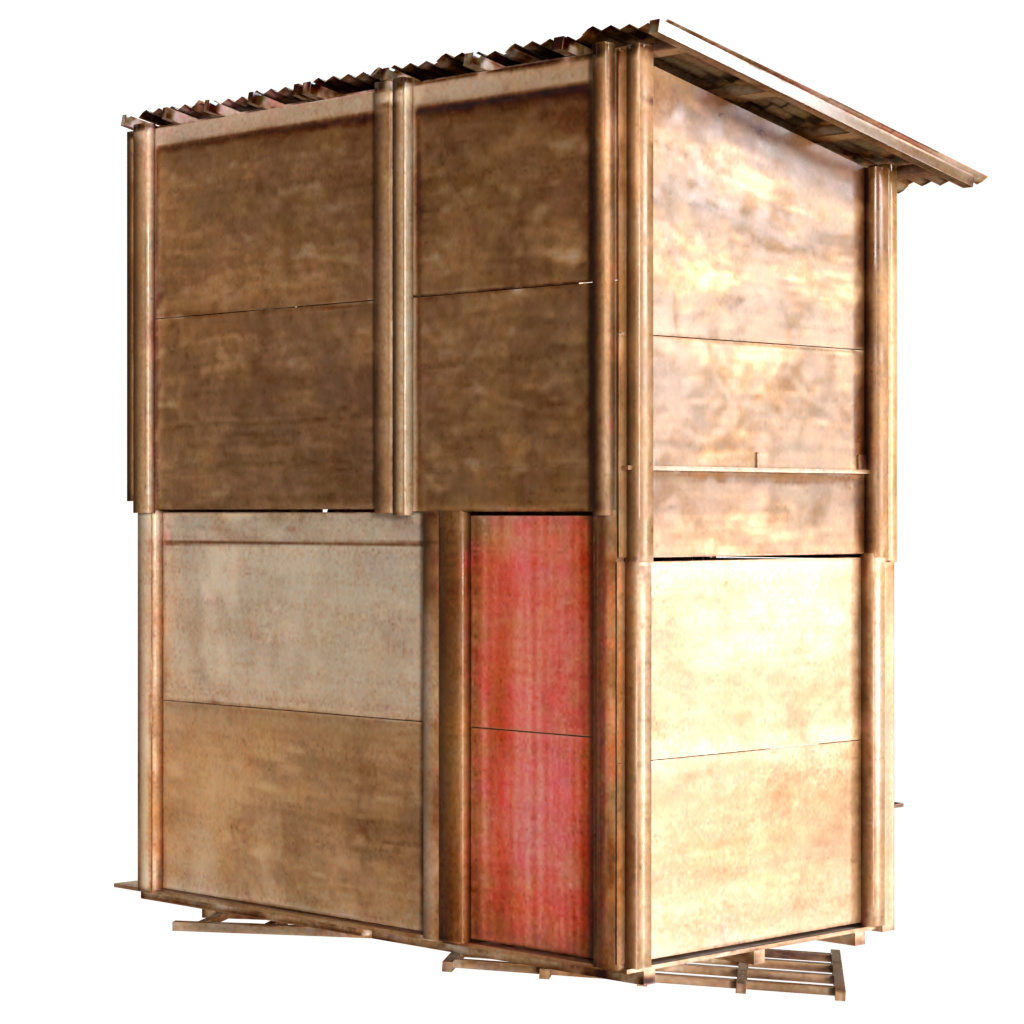}} &
			\raisebox{-0.5\height}{\includegraphics[height=0.32\columnwidth]{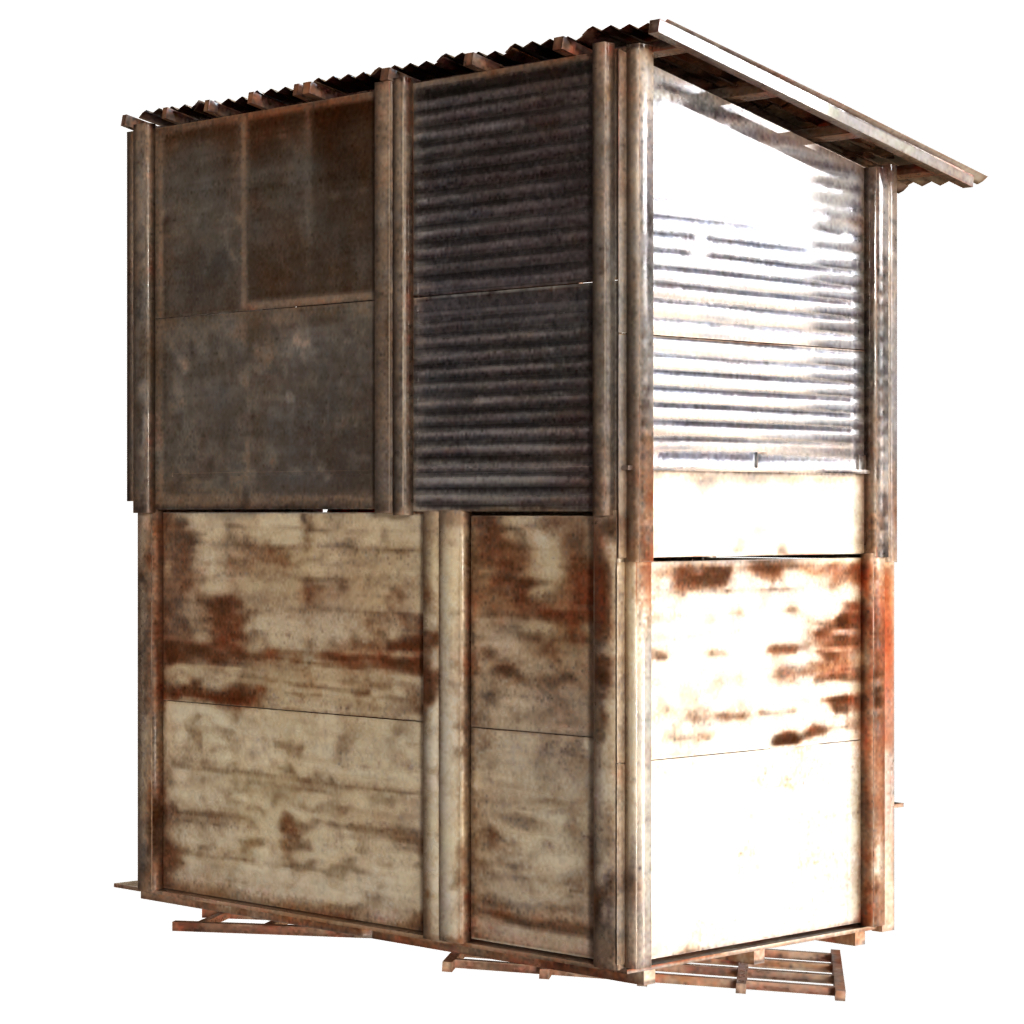}} \\

            \rotatebox[origin=c]{90}{\textsc{Helmet}} &
			\raisebox{-0.5\height}{\includegraphics[height=0.32\columnwidth]{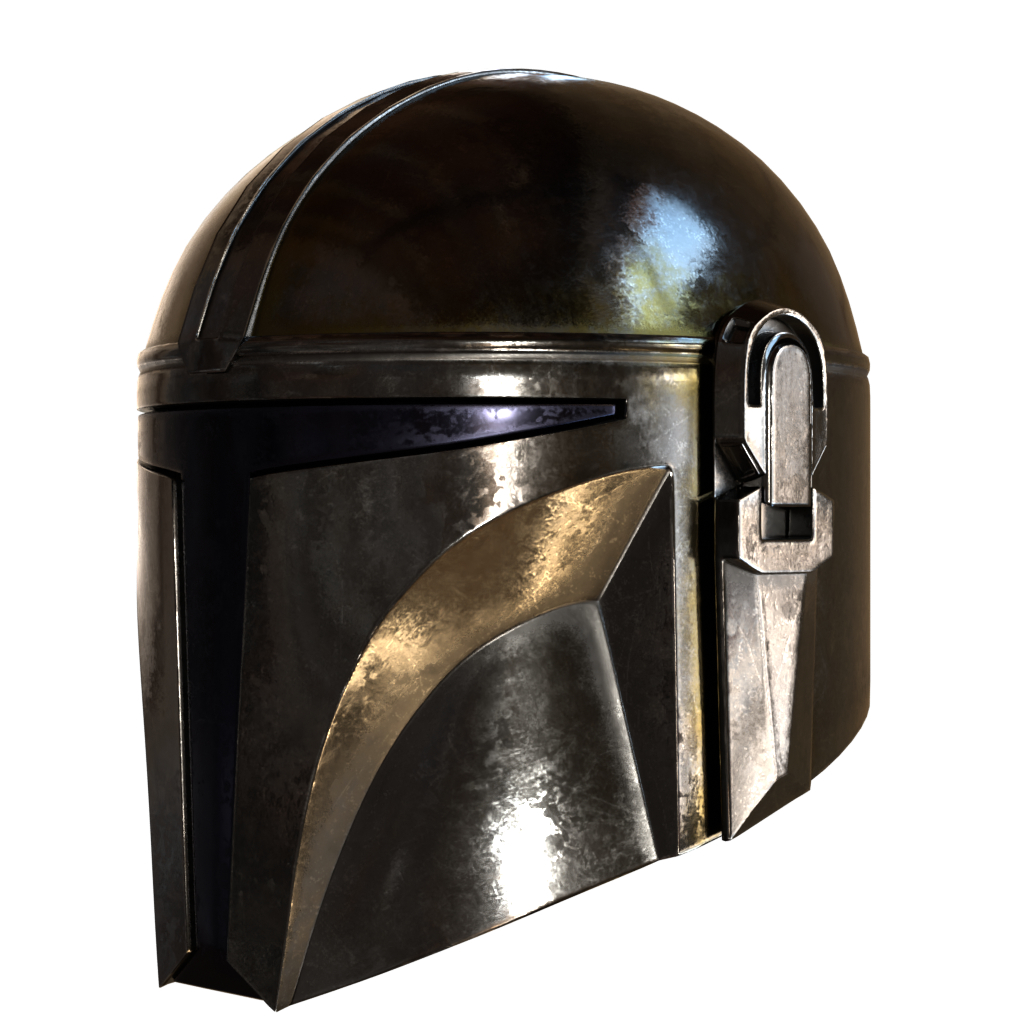}} &
			\raisebox{-0.5\height}{\includegraphics[height=0.32\columnwidth]{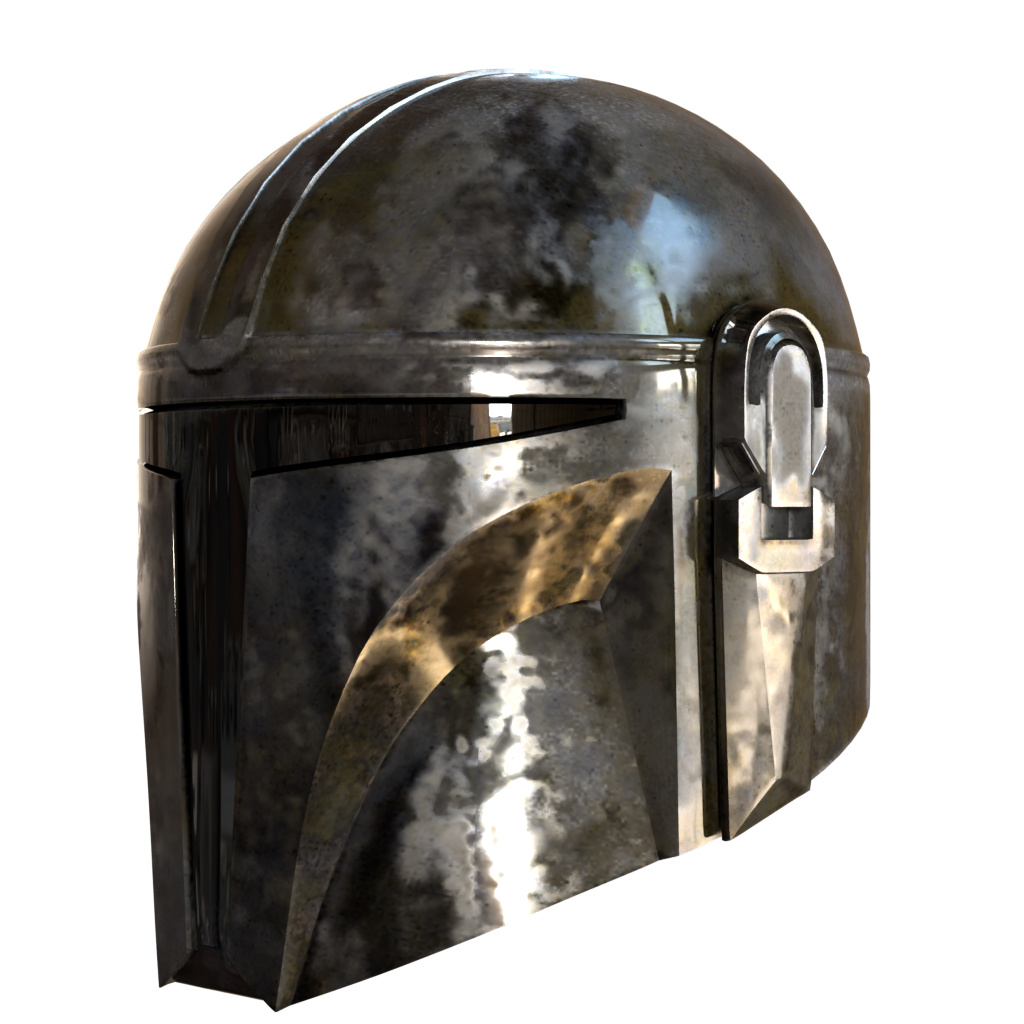}} &
			\raisebox{-0.5\height}{\includegraphics[height=0.32\columnwidth]{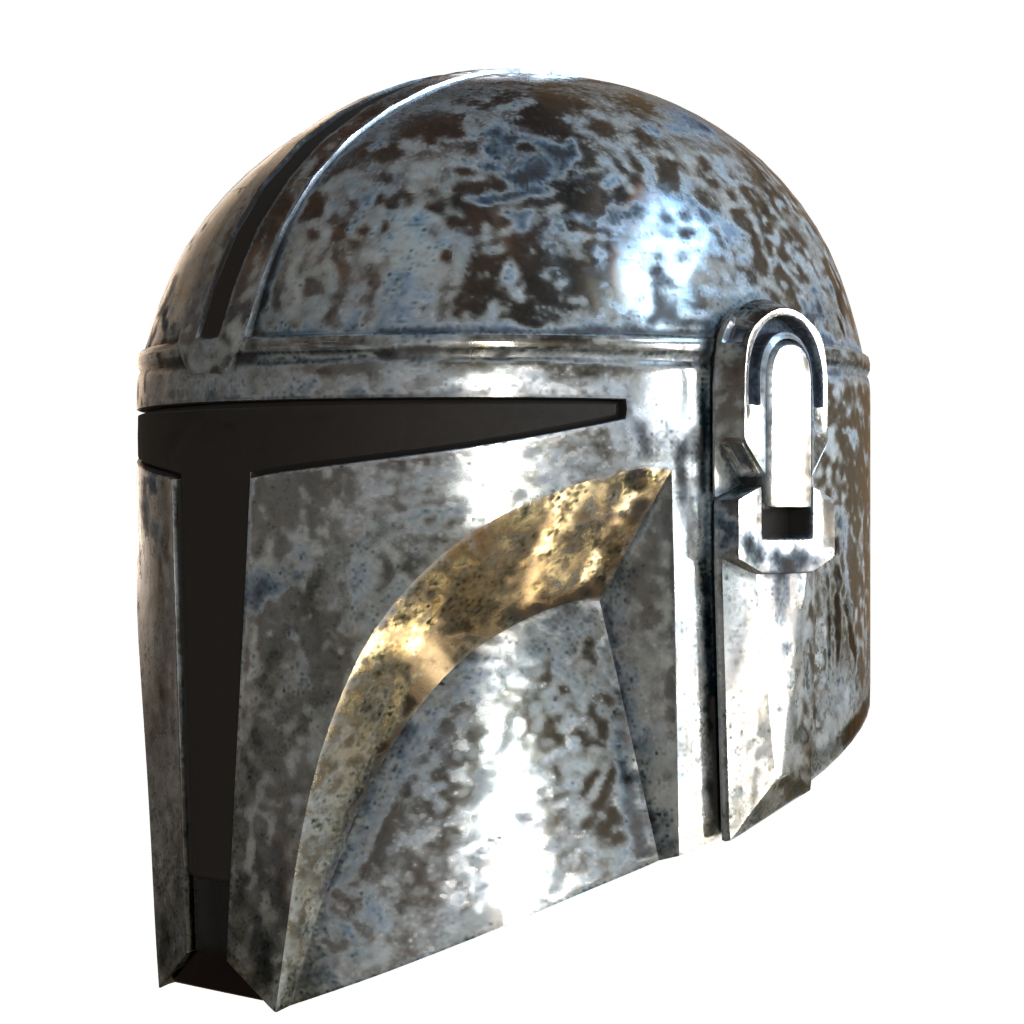}} \\

        	& Reference & Image-to-material & Text-to-material \\
        \end{tabular}
		\vspace*{-2mm}
		\caption{
            Two examples comparing our text- and image-guided models. 
            The image-to-material model is guided by a single shaded view from the reference scene, in addition to the text prompt, 
            and consequently respects the aesthetics of the reference scene better.
        }
		\label{fig:i2v_results}
	\end{figure}
}

\maketitle

\section{Introduction}

DreamFusion~\cite{poole2022dreamfusion} kickstarted a new research field of leveraging 
diffusion models to generate 3D models from text prompts.
Most approaches extract data from image diffusion models through differentiable rendering, typically using neural radiance fields~\cite{Mildenhall2020} (NeRFs) or 3D Gaussian splats (3DGS)~\cite{kerbl3Dgaussians}.
While results look impressive, e.g., for novel view synthesis, 
extracting materials for physically-based rendering (PBR)~\cite{Burley12,Walter2007}
remains challenging.
The lack of view-consistency in image diffusion models typically introduces blur and 
washes out material details. This is particularly obvious when extracting parameters for PBR materials models 
which relies on consistency of specular reflections.
Furthermore, multi-view optimization in a surface-based renderer which disentangles materials and lighting, e.g., 
a physically-based path tracer, is substantially more challenging than optimizing the volumetric approaches of 
NeRFs and 3DGS which bake lighting and material in common appearance representations. 

Video diffusion models are beneficial for extracting spatially varying and specular materials. 
They far exceed image diffusion models in handling specular highlights and provide increased view consistency. 
This greatly helps when extracting material parameters and for intrinsic decomposition~\cite{DiffusionRenderer}. 

We present a method for text-to-material extraction based on video diffusion models. 
Our method starts from a known 3D geometry and a text prompt describing the desired material.
We condition a diffusion model on G-buffer guides (normals, uniform shading),  similar to previous 
work~\cite{Zhang2024dreammat,deng2024flashtex}, which we extend to the video domain.
By finetuning a recent video model, Cosmos~\shortcite{cosmos_short}, with these conditions,
we generate video sequences of known 3D objects with synthesized materials and controlled lighting. 
We then apply the inverse rendering pipeline of DiffusionRenderer~\cite{DiffusionRenderer} to extract 
\emph{albedo}, \emph{roughness}, and \emph{metallic} material maps from the generated video. 
Finally, materials are baked to texture maps through a multi-view optimization using a differentiable path tracer regularized on the guides
extracted by DiffusionRenderer. 

As shown in Figure~\ref{fig:teaser}, we produce spatially varying, detailed materials
that are robust to varying lighting conditions. In comparison to related 
work~\cite{Zhang2024dreammat,youwang2024paintit,fang2024makeitreal}, we show higher 
quality results and improved separation of lighting and materials.

\section{Related Work}

\subsection{Video Diffusion Models}
Diffusion models construct a Markov chain that gradually adds random noise to data through a sequence of diffusion steps. They are trained to reverse this process, enabling sample generation by a denoising process starting from Gaussian noise. 
Several diffusion-based generative models have been developed based on similar principles~\cite{sohl2015deep, ho2020denoising, dhariwal2021diffusion}.
Recently, video diffusion models~\cite{blattmann2023videoldm, blattmann2023svd, hong2023cogvideo, yang2024cogvideox, cosmos_short} extend image-based diffusion approaches to the temporal domain, enabling video generation from inputs such as text or an initial frame.
In this work, we build upon the Cosmos~\shortcite{cosmos_short} video diffusion model. 

\subsection{Differentiable Rendering}

Fuzzy scene representations such as NeRFs~\cite{Mildenhall2020} and 
Gaussian splatting~\cite{kerbl3Dgaussians} are commonly used in optimization setups, 
and generate impressive novel-view synthesis results. However, disentangling materials and lighting is non-trivial. 
In this paper, we focus on surface geometry with PBR materials~\cite{Burley12}. 
Previous work include differentiable rasterization~\cite{Laine2020diffrast}, 
which has low run-time cost and has been successfully applied to photogrammetry~\cite{Munkberg_2022_CVPR}. 
Differentiable path tracing~\cite{Zhang:2020:PSDR,Mitsuba3} approaches are considerably more costly, and introduce Monte-Carlo noise 
in the training process, which can make gradient-based optimization more challenging. However, path tracing accurately simulates global 
illumination effects, and has higher potential reconstruction quality. In this work we use a custom path tracer similar to publicly 
available solutions~\cite{Mitsuba3} but with inline support for querying a hashgrid~\cite{mueller2022instant} representation in the ray tracing loop. 

\subsection{Material Extraction from Diffusion Models}

DreamFusion~\cite{poole2022dreamfusion} introduces a \emph{score distillation sampling} (SDS) loss, and generates 3D assets from pre-trained text-to-image diffusion models. This approach has since been refined~\cite{zhu2023hifa,wang2023prolificdreamer} and integrated in many text-to-3D pipelines~\cite{chen2023fantasia3d,wang2023prolificdreamer} 
to extract PBR materials. 
Paint-it~\cite{youwang2024paintit} proposes  
representing material texture maps with randomly initialized convolution-based neural kernels. This regularizes the optimization 
landscape, improving material quality.

TextureDreamer~\cite{yeh2024texturedreamer} finetunes the diffusion model using Dreambooth~\cite{ruiz2022dreambooth} with a few 
images of a 3D object, and uses \emph{variational score distillation}~\cite{wang2023prolificdreamer} to optimize the material maps.
DreamMat~\cite{Zhang2024dreammat} and FlashTex~\cite{deng2024flashtex} improve on light and material disentanglement by finetuning 
image diffusion models to condition on geometry and lighting, allowing for optimization over many known lighting conditions. 
Various hybrid approaches combine SDS with inpainting, or coarse-to-fine texture 
refinement~\cite{richardson2023texture,chen2023text2tex,zeng2024paint3d}. 

A common limitation for most image models is lack of view consistency, which may show up as 
blur in the extracted textures. SV3D~\cite{voleti2024sv3d} and Hi3D~\cite{yang2024hi3d} improve on this aspect by 
finetuning video models for object rotations, and extract 3D models from the generated views. However, these approaches 
have limited resolution and do not provide PBR materials. Trellis~\cite{xiang2024structured} and TEXGen~\cite{yu2024texgen} 
avoid the view consistency problem altogether by having the diffusion model operate directly in 3D space~\cite{xiang2024structured} 
and texture space~\cite{yu2024texgen} respectively. 
These methods show great promise, but they are limited by training on relatively small, synthetic datasets of 3D assets.

MaPa~\cite{zhang2024mapa}, MatAtlas~\cite{ceylan2024matatlas}, and Make-it-Real~\cite{fang2024makeitreal} start from a database of known 
high-quality materials, and learn to project the input (image or text) onto the known representation. MaPa relies on \
material graphs and optimize parameters of known graphs, while Make-it-Real uses a database of PBR-textures, and 
MatAtlas a database of procedural materials. These methods are limited by the expressiveness of their material databases, but 
benefit from much improved regularization.

There are also several approaches for 3D asset creation through large-scale generative models. 
CLAY~\cite{zhang2024clay} generates 3D geometry and materials from text and image inputs. 
Their material generation model uses a finetuned multi-view image diffusion model~\cite{shi2023MVDream} conditioned on normal maps. 
The material model generates four canonical views of the PBR texture maps (albedo, roughness, metallic), which are then projected 
into texture space. 3DTopia-XL~\cite{chen2024primx} proposes a novel
3D representation, which encodes the 3D shape, textures, and materials in volumetric primitive anchored to the surface of the object. 
Their denoising process jointly generates shape and PBR materials.
In concurrent work, MCMat~\cite{zhu2024mcmat} leverages Diffusion Transformers (DiT) to extract multi-view images of PBR material maps, 
combined with a second DiT to enhance details in UV space. 

\figSystem

Another related line of research is intrinsic decomposition of images. IntrinsicAnything~\cite{chen2024intrinsicanything}
decomposes images into diffuse and specular components, and leverages these components as priors using physically-based inverse
rendering to extract material maps. MaterialFusion~\cite{litman2025materialfusion} introduces
a 2D diffusion model prior to help estimate material parameters in an multi-view reconstruction pipeline.
RGB$\leftrightarrow$X~\cite{zeng2024rgb} shows both intrinsic decomposition into G-buffers
and neural rendering from G-buffers using diffusion models. DiffusionRenderer~\cite{DiffusionRenderer} extends RGB$\leftrightarrow$X
to videos, and also support relighting. NeuralGaffer~\cite{jin2024neural_gaffer} and DiLightNet~\cite{zeng2024dilightnet} leverage 
diffusion models for relighting single views. 
IllumiNerf~\cite{zhao2024illuminerf} relights each view in a multi-view dataset, then reconstructs a NeRF model with these relit images.

In concurrent work, IntrinsiX~\cite{kocsis2025intrinsix} combines intrinsic predictions for PBR G-buffers for a single view 
from text (using image diffusion models) with a rendering loss. Our approach is related, in that we combine intrinsic predictions 
and differentiable rendering. However, our goal is extracting PBR maps in \emph{texture space}, which requires consistent predictions 
from many views. To that end, we leverage video diffusion models for view-consistent multi-view predictions and
a more accurate renderer and lighting model in our rendering loss.

\section{Method}

Our pipeline, as shown in Figure~\ref{fig:system}, combines two video models with a differentiable renderer to extract PBR
material textures. We assume a given 3D model with a unique texture parametrization, 
and a lighting condition specified by an HDR light probe.
We first generate frames for a $360^\circ$ orbit from a video model, conditioned on geometry and lighting.  
Next, we run multi-view reconstruction from the generated views using differentiable rendering to extract the material maps. 
To regularize the reconstruction process, we leverage image intrinsics extracted from the generated video using 
a second video model. Below, we describe each step in detail.

\subsection{Video Model I: Multi-view generation}
\label{sec:video_model1}
In a first step, we generate videos of a $360^\circ$ camera revolution around the 3D model, 
based on a text prompt describing the object's material. We finetune a recent Diffusion Transformer (DiT) video model, 
Cosmos~\shortcite{cosmos_short}, with additional conditions to respect the lighting and 
input geometry. We use the \texttt{Cosmos-1.0-Diffusion-7BVideo2World}\footnote{https://github.com/NVIDIA/Cosmos} base 
model which operates in a latent space with 8$\times$
compression in the spatial and temporal domain. Encoding and decoding to and from latent space is done by the pretrained VAE
\texttt{Cosmos-1.0-Tokenizer-CV8x8x8}.
The base model supports text- and image guided video generation at a resolution of 1280$\times$704 pixels and 121 frames.

Similar to DreamMat~\cite{Zhang2024dreammat} and FlashTex~\cite{deng2024flashtex} we condition the diffusion models 
with intrinsic guides: normals in camera space, and a \emph{simple shading} guide which provides scene lighting information. 
The shading guide is rendered using three uniform 
materials with constant albedo, $k_d = 0.7$, roughness, $r=\{1,0.5,0\}$ and metallic, $m = \{0,0.5,1\}$,
as shown in Figure~\ref{fig:guides}. 
We compress the three shading guides into a single RGB-image by encoding the luminance of each image as a separate channel
and provide these conditions per-frame. 

\figGuides

We increase the input feature count of the input embedding layer of the diffusion model to account for our additional inputs
(normals and simple shading), encoded into latent space using the frozen Cosmos tokenizer. 
We concatenate the encoded features with the noisy video in latent space along the channel dimension,
and finetune the embedding layer and all DiT layers using a custom dataset for 10k iterations on 64 GPUs. 
We use the the denoising score matching loss from Cosmos~\shortcite{cosmos_short} unmodified.

Our dataset consists of 40k videos of object-centric $360^\circ$ orbits of 3D models from 
Objaverse~\cite{objaverse}, BlenderVault~\cite{litman2025materialfusion}, ShaderBalls~\cite{mazzone2023shaderball} (50 materials), 
ABO~\cite{collins2022abo}, and HSSD~\cite{khanna2023hssd}. For each object, we render a video with 121 frames
at a resolution of 1280x704 using a path tracer with three bounces. For lighting, we use a randomly selected probe from 670 HDR 
maps from Poly Haven~\cite{polyhaven}. 
We also render the intrinsic normal and simple shading guides from Figure~\ref{fig:guides}. 
Video captions are automatically generated using the CogVLM2~\cite{hong2024cogvlm2} model.
Since our videos are cyclical, we augment the dataset by randomly reversing the video, and randomly offsetting the video 
start frame on each training iteration.

\subsection{Video Model II: Intrinsic Decomposition}

To extract intrinsic channels---base color, roughness, and metallic maps---from the generated video, we re-implement the 
inverse renderer from DiffusionRenderer~\cite{DiffusionRenderer}. 
The inverse renderer is a video diffusion model that estimates material attributes conditioned on the input video. 
The same model is used to generate all material attributes. It generates one material map at a time, and uses optimizable 
domain embeddings to specify the target attributes to generate. 

We finetuned the inverse rendering model from the Cosmos~\shortcite{cosmos_short} base model
\texttt{Cosmos-1.0-Diffusion-7BVideo2World}, which provides higher quality results than the 
SVD-based~\cite{blattmann2023svd} version from the DiffusionRenderer paper. 
We follow the training pipeline from DiffusionRenderer closely, and finetune the base model 
using 150k videos with objects from Objaverse~\cite{objaverse}.

\subsection{Differentiable Rendering}

To extract materials from the generated videos, we run multi-view reconstruction with a 
differentiable path tracer. 
Our renderer is similar to publicly available solutions, e.g., Mitsuba3~\cite{Mitsuba3}, but with added support for querying a 
hash grid+MLP~\cite{mueller2022instant} representation inline in the ray tracing loop. We leverage this representation 
as a multi-scale material representation. As noted in in methods using SDS loss~\cite{chen2023fantasia3d,wang2023prolificdreamer}, 
the  hierarchical hash grid+MLP representation is more robust to errors and noisy loss functions than 2D texture maps. By inlining 
the hash grid+MLP evaluation in the path tracer, we avoid 1) the renderer-PyTorch roundtrip for each evaluation and 2) logic to 
gather a large batch of shading requests for the PyTorch evaluations to be efficient.
After optimization, we sample out the hash grid into standard 2D texture maps.
We leverage CUDA WMMA instructions for efficient evaluation of the MLP linear layers,
Slang~\cite{bangaru2023slang} shaders (for autodiff) and GPU-accelerated OptiX~\cite{parker2010optix} ray tracing kernels.

The differentiable renderer takes a 3D model and HDR probe, alongside camera poses and 121 frames from 1) the generated video 
(Video Model I) and 2) intrinsics (Video Model II). We run multi-view reconstruction to optimize the material maps for 1k 
iterations with a batch size of eight, using 1-3 bounces, 128~spp in the forward pass and 4~spp in the backward pass. 

\paragraph*{Shading model}

Our goal is all-frequency lighting including shadows, reflections, and indirect illumination. The outgoing radiance $L(\omega_o)$ in 
direction $\omega_o$ can be expressed using the rendering equation~\cite{kajiya1986rendering}:
\begin{equation}
	L(\omega_o) = \int_\Omega L_i(\omega_i)f(\omega_i,\omega_o) (\omega_i \cdot \mathbf{n}) d\omega_i.
    \label{eq:ibl}
\end{equation}
This is an integral of the product of the incident radiance, $L_i(\omega_i)$ from  direction $\omega_i$ and the BSDF 
$f(\omega_i, \omega_o)$. The integration domain is the hemisphere $\Omega$ around the surface intersection normal, $\mathbf{n}$. 

We use the physically-based (PBR) material model from Disney~\cite{Burley12}
which combines a Lambertian term and a Cook-Torrance microfacet specular shading model~\cite{cook1982reflectance}:
\begin{equation}
f(\omega_i,\omega_o) = \frac{D G F }{4 (\omega_o \cdot \mathbf{n}) (\omega_i \cdot \mathbf{n})},
\end{equation}
where $D$, $G$ and $F$ are functions representing the GGX~\cite{Walter2007} normal distribution (NDF), geometric attenuation and 
Fresnel term, respectively. This BSDF is parameterized with an RGB base color, $c_{\textrm{base}}$, and scalar roughness, $r$, and 
metallic, $m$, values at each shading point.
The specular lobe is described by the roughness value for the GGX normal 
distribution function, $D$. 
The diffuse albedo is given by $k_d = c_{\textrm{base}} \cdot (1-m)$, and 
specular albedo $k_s = 0.04 \cdot (1-m)   + c_{\textrm{base}} \cdot m$~\cite{karis2013real}. 

The rendering equation is evaluated using Monte Carlo integration, and 
we apply \emph{multiple importance sampling}~\cite{veach1998robust}, 
to reduce variance. We use three sampling techniques: light importance sampling~\cite{PBRT}, cosine sampling for the diffuse lobe, and 
GGX importance sampling~\cite{Heitz2018GGX} for the specular lobe.

\paragraph*{Optimization task}
Let $\phi$ denote our optimization parameters (i.e., spatially varying material maps).
For a given camera pose, $c$, the differentiable renderer produces an 
image $I_{\phi}(c)$. The reference image $I_{\mathrm{ref}}(c)$ is a view from the same camera,
generated by Video Model I. 
Given a loss function $L$, we minimize the empirical risk
\begin{equation}
\underset{\phi}{\mathrm{argmin}}\ \mathbb{E}_{c}\big[L\big(I_{\phi}(c), I_{\mathrm{ref}}(c)\big)\big]
\end{equation}
using Adam~\cite{kingma2017adam} based on gradients w.r.t.~the optimization parameters,
$\partial L/\partial\phi$.

Our image loss, $L_\mathrm{image}$, is $L_1$ norm on tonemapped and gamma corrected 
sRGB colors~\cite{Munkberg_2022_CVPR}. This is combined with a \emph{scale invariant}
regularizer loss:
\begin{equation}
	L_{\mathrm{reg}} = L_1\left(x / \overline{x}, \  y / \overline{y}\right),
\end{equation}
applied to each of the predicted intrinsics (base color, roughness, metallic) from Video Model II, 
where we divide each 
image by its mean value before computing $L_1$ loss. 
The scale invariant formulation is more robust against biases in the predicted 
intrinsics. All loss terms are masked, applied only to pixels covered by the object. 
Our total loss function is $L = L_\mathrm{image} + \lambda L_{\mathrm{reg}}$.
We use $\lambda = 0.2$ for all three regularizer terms, which empirically 
works well over our set of test scenes. We also experimented with adding an 
LPIPS~\cite{zhang2018unreasonable} loss term, but did not see improved results. 

\figWarp

\figMainQualityResults

\subsection{Image-conditioned Video Generation}
\label{sec:image_conditioned}

While our primary focus is on material generation from text, our pipeline can also accommodate
material generation from an image example since the Cosmos base model inherently supports both
text and image conditioning. We adopt an approach similar to Gen3C~\cite{ren2025gen3c}
where the video model is conditioned on an input image, which is warped (using a provided depth buffer) according to the camera trajectory. 
Please refer to Figure~\ref{fig:warp} for an example.
As in our text-to-video setting, we additionally condition the video model on normals and three shading conditions. 
As described in Section~\ref{sec:video_model1} we start from the Cosmos~\shortcite{cosmos_short} model, and increase the 
input feature count of the input embedding layer of the diffusion model to account for our additional inputs
(\emph{warped image}, normals, and simple shading) and finetune the video model using our custom dataset.
The reference image needs to (ideally) respect shape and lighting conditions of the additional inputs. We only tried this 
setting, and have not ablated the robustness of an arbitrary reference image.

\section{Results}

We evaluate the quality of our results primarily against Paint-it~\cite{youwang2024paintit} and 
DreamMat~\cite{Zhang2024dreammat}, which use image diffusion models and score distillation sampling~\cite{poole2022dreamfusion}
for text-guided PBR material generation for a given 3D model.  
Paint-it re-parameterizes material textures as network weights of a U-Net, to regularize the noisy SDS-loss.
We picked it as a representative comparison, as they compare favorably to recent SDS-based methods for texture and material 
generation~\cite{chen2023fantasia3d, richardson2023texture,chen2023text2tex}.
DreamMat finetunes an image diffusion model to condition 
on geometry and lighting and generates shaded views of an object with the same lighting. 
This disentangles lighting and appearance and improves PBR material quality. 
This is closely related to our approach, with the main differences being that we leverage video diffusion models for multi-view 
generation and material priors. 
We additionally compare to Make-it-Real~\cite{fang2024makeitreal}, as a representative 
method retrieving materials from a database. We initialize Make-it-Real from the base color generated by DreamMat. 
The visual results are similar to DreamMat, and are left out for brevity. Please refer to Table~\ref{tab:metrics} and 
our supplemental material for the full comparison.

In text-to-material applications, the reference is not clearly defined. To compute quantitative metrics,
we took the following approach: We selected six 3D models with PBR materials from BlenderVault 
(held out from our training set), rendered out reference videos and generate captions for each example with CogVLM2~\cite{hong2024cogvlm2}. 
The generated prompts (see supplemental) and geometry for each model are used as input in our text-to-material evaluations. 
We use these six 3D models as ground truth when computing metrics for intrinsics and relighting evaluations.

Figure~\ref{fig:main_quality_results} shows qualitative results.
We note that our method outperforms previous work in terms of intrinsics and relighting quality. This is 
apparent in the base color of the \textsc{Gramophone} scene, where our 
method produces a demodulated texture, similar to the reference, while the base color produced by
Paint-it and DreamMat both show baked specular highlights. Similar behavior is also clearly noticeable in the \textsc{Diver}
and \textsc{Shoes} examples. 
Our roughness and metallic guides are noisy, and somewhat biased towards lower roughness (more specular reflections). 
However, they are considerably more faithful to the reference than the competing methods.
Please refer to the supplement for additional scenes and videos. 

\figRelighting

In Figure~\ref{fig:relighting}, we relight the extracted materials with four different probes. DreamMat captures the overall 
color tone of each probe, but the baked highlights make the lighting appear to always come from the same direction. 
This is apparent in the \emph{Grace Cathedral} probe, where the main light source is placed straight above the object. 
We present error metrics in Table~\ref{tab:metrics}.
\begin{table}[tb]
{
	\centering
	\small
	\setlength{\tabcolsep}{4pt}
	\begin{tabularx}{\columnwidth}{lcccc|lcc}
		PSNR (dB) $\uparrow$ & relit & $k_d$ &   $r$ &   $m$ & LPIPS$\downarrow$ & relit & $k_d$ \\
		\hline
		Our &  20.1 &  20.1 &  16.9 &  14.1 &               Our &  0.15 &  0.18 \\
		DreamMat &  16.1 &  16.7 &  13.1 &  13.1 &          DreamMat &  0.21 &  0.23 \\
		Make-it-Real &  17.3 &  16.7 &  17.5 &  13.2 &      Make-it-Real &  0.20 &  0.23 \\
		Paint-it &  14.0 &  13.8 &  14.1 &  10.0 &          Paint-it &  0.22 &  0.26 \\
	\end{tabularx}
}
\vspace*{2mm}
\caption{Image metrics for text-to-materials methods. Each material 
is compared to a reference 3D model from which we generated the prompt and used the geometry. 
Scores for intrinsics are averaged over six test scenes, 
and relighting results are averaged over six test scenes, each scene lit with 
eight light probes.}
\label{tab:metrics}
\setlength{\tabcolsep}{6pt}
\end{table}
Considering base color and relighting results, we note that our method scores highest or on par with the competing methods for 
all scenes. As an outlier, we score lower than DreamMat on the roughness and metallic guides for the \textsc{Shed} scene (see 
supplement), which is explained by our method adding metallic features, which, while plausible, are not present in the reference example.

\subsection{User study} 
To evaluate the perceptual quality of the generated texture and materials, we conducted a user study where participants were 
presented with paired video rendering in random order--one generated by our method and the other by a baseline method. 
Participants were asked to compare the two results, assess the quality of texture and material realism, and choose the 
image they perceived as more visually appealing. 

We render each asset with four different lighting conditions, resulting in a set of 24 samples to evaluate. We invite 11 users 
to make a binary selection for each video pair, and then apply majority voting to determine the preferred method for each video sample. 

Additionally, inspired by recent work on using VLMs as perceptual evaluators~\cite{wu2023gpteval3d}, we randomly sample seven 
frames per video and conduct the same evaluation on image pairs using GPT-4o~\cite{openai2024gpt4o} as the perceptual evaluator. 
In Table~\ref{tab:user_study}, we report the percentage of samples where our method was preferred over the baselines. 
The average user preference for our method is above 50\% for both human and VLM evaluators, showing our method outperforms 
baselines with higher perceptual quality. 

\begin{table}[htbp]
{
	\centering
	\small
	\begin{tabularx}{\columnwidth}{l|cc}
	Setting & Human Eval. & VLM Eval. \\
	\hline
	Our over DreamMat~\cite{Zhang2024dreammat}   & $62\%$  & $62\%$  \\ 
	Our over Paint-it~\cite{youwang2024paintit}  & $58\%$  & $100\%$  \\ 
	\end{tabularx}
}
\vspace*{2mm}
\caption{\textbf{User study.} We report the percentage of samples where our method is preferred over baselines, 
based on evaluations from both human annotators and a VLM (GPT-4o). 
A preference score above 50\% indicates that our method is favored.
} 
\label{tab:user_study}
\end{table}

\subsection{Ablations}

\figVariety

\paragraph*{Material diversity}
Our finetuned video model leverages the powerful base model of Cosmos, and can generate a diverse set of materials. 
In Figure~\ref{fig:variety} we prompt the model with ten different material descriptions
while keeping the lighting and geometry constant. The video model produces view-consistent, high-quality 
video frames, while respecting the input geometry and lighting.

\figItoV

\paragraph*{Image-conditioning}

We evaluate our image-conditioned video model by providing a single rendered view of the reference 3D model in addition to the text prompt, effectively performing material inpainting.
As expected, Figure~\ref{fig:i2v_results} shows that the resulting materials match the style of the reference more closely, and this is also reflected in the errors metrics below.
\begin{center}
	\small
	\setlength{\tabcolsep}{5pt}
	\begin{tabular}{lcccc|lcc}
	
		PSNR$\uparrow$ & relit & $k_d$ & $r$ & $m$ & LPIPS$\downarrow$ & relit & $k_d$ \\
		\hline
		Our-text  & 20.1 & 20.1 & 17.0 & 14.1 & Our-text  & 0.15 & 0.18 \\
        Our-image & 21.1 & 21.0 & 17.1 & 15.4 & Our-image & 0.13 & 0.15 \\
	\end{tabular}
	\setlength{\tabcolsep}{6pt}
\end{center}

\paragraph*{Impact of regularization with intrinsics}

\figRegularizers

We ablate the intrinsic regularization in Figure~\ref{fig:regularizer_ablation}. While 
multi-view reconstruction from the generated video frames with our differentiable path tracer works reasonably well, we note that the intrinsic guides help demodulate the albedo, and reduce discontinuities in metallic and roughness parameters, resulting in a more natural look. 
This is further corroborated by a small but significant quality improvement, as shown in the metrics below.
\begin{center}
	\small
	\setlength{\tabcolsep}{5pt}
	\begin{tabular}{lcccc|lcc}
		PSNR$\uparrow$ & relit & $k_d$ & $r$ & $m$ & LPIPS$\downarrow$ & relit & $k_d$ \\
		\hline
		Our     & 20.1 & 20.1 & 17.0 & 14.1 & Our     & 0.15 & 0.18 \\
		No reg. & 19.2 & 18.6 & 18.5 & 12.8 & No reg. & 0.16 & 0.18 \\
	\end{tabular}
	\setlength{\tabcolsep}{6pt}
\end{center}

\paragraph*{Impact of differentiable path tracing}

The base video model we leverage is trained on natural images and captured videos. 
To reduce the domain gap between computer generated images and photographs,
we opted for physically-based rendering with path tracing, incorporating reflections, shadows, and global 
illumination effects, both when generating our training data and for inverse rendering. Path tracing is more 
expensive and introduce higher noise level due to 
extensive use of Monte Carlo integration to approximate lighting. However, it produces 
more realistic images and helps disentangle lighting from material parameters. 
Figure~\ref{fig:raydepth} is an illustrative example of reduced light baking
by more accurate simulation of multiple light bounces.

\figRayDepth

\section{Conclusion and Limitations}

We conclude that modern video diffusion models combined with high-quality differentiable rendering show great promise for text-guided PBR 
material generation. While we limit ourselves to material generation, we believe our method can be a valuable component in larger systems 
targeting text-to-3D, single-image reconstruction, and for material augmentation for 3D datasets.

A limitation of our current prototype is multi-view coverage. The base model we leverage expects smooth camera movements, and we use 
$360^\circ$ orbits with a fixed elevation. Unseen areas of the mesh will not be accurately reconstructed, and textures will contain 
data extrapolated by the hash grid+MLP~\cite{mueller2022instant} representation. Ideally, we would like to uniformly distribute our 
views on a sphere around the object to maximize coverage. Recent works~\cite{gao2024cat3d,wu2024cat4d} show promising signs of this 
type of camera control, but the models are currently not publicly available.

Furthermore, current video models target natural videos and use aggressively compressed latent spaces, which shows up in generated 
videos as motion blur and video compression artifacts. This breaks the pinhole camera 
and instant shutter assumptions of our differentiable renderer. Optimization with an $L_1$ loss recovers some image detail, but
that a more principled solution is desired to reduce blur. We are hopeful that our method will greatly benefit from future video 
models, which is a very active research field.

We currently do not consider normal mapping and rely on geometric normals for shading. Normal mapping can be trivially enabled 
in our differentiable renderer, and DiffusionRenderer outputs a normal intrinsics map, but we note that previous work struggles 
with the fundamentally under-determined problem of normal map optimization~\cite{Munkberg_2022_CVPR}. Robust priors, or 
regularizers, would be required for high quality results, and we leave this to future work.

An interesting avenue for future work is to leverage image-con\-di\-ti\-ned models (Section~\ref{sec:image_conditioned}), to generate 
localized coverage and slow down camera movement, by stitching together multiple videos. We would also like to expand
the material extraction to handle more rendering effects, e.g., refractions and subsurface scattering. While our early 
experiments are encouraging, these experiments require novel datasets.

\begin{acks}
The authors are grateful to Aaron Lefohn, Sanja Fidler, Ming-Yu Liu, and Jun Gao for
their support and discussions during this research, as well as the
anonymous reviewers for their valuable comments and feedback.
\end{acks}

\bibliographystyle{ACM-Reference-Format}
\bibliography{paper}

\end{document}